\documentclass[10pt, pre, amsmath, onecolumn, showpacs, superscriptaddress, nofootinbib]{revtex4-1}
\usepackage{graphicx,color}
 \graphicspath{{./}{./figures/}}
\usepackage{ams math}
\usepackage{enumerate}
\usepackage{mathbbol}
\usepackage{mathtools}
\usepackage{amsfonts}
\usepackage{natbib}

\usepackage{bm}
\usepackage{hyperref}
\hypersetup{
    colorlinks=true,       
    linkcolor=red,          
    citecolor=blue,        
    filecolor=magenta,      
    urlcolor=cyan           
}

\usepackage{color}
\usepackage[dvipsnames]{xcolor}

\usepackage{cancel}

\usepackage[caption=false]{subfig}
\usepackage{bbm}

\usepackage{soul}
\setstcolor{red}
\usepackage{multirow}

\newcommand{\nn}{\nonumber}

\newcommand{\sgn}{{\mathrm{sgn}}}

\newcommand{\magenta}{\color{magenta}}

\newcommand{\be}{\begin{equation}}
\newcommand{\ee}{\end{equation}}
\newcommand{\bea}{\begin{eqnarray}}
\newcommand{\eea}{\end{eqnarray}}
\newcommand{\beq}{\begin{eqnarray}}
\newcommand{\eeq}{\end{eqnarray}}

\newlength{\bilderlength}

\usepackage{subfig}

\begin{document}

\title{Density and Current Fluctuations for Run-and-Tumble Particles With and Without Interactions
}

\author{L\'eo \surname{Touzo}}
\affiliation{University of Chicago, James Franck Institute, 929 E 57th Street, Chicago, IL 60637}
\author{Pierre Le Doussal}
\affiliation{Laboratoire de Physique de l'Ecole Normale Sup\'erieure, CNRS, ENS and PSL Universit\'e, Sorbonne Universit\'e, Universit\'e Paris Cit\'e,
24 rue Lhomond, 75005 Paris, France}
\author{Gr\'egory \surname{Schehr}}
\affiliation{Sorbonne Universit\'e, Laboratoire de Physique Th\'eorique et Hautes Energies, CNRS UMR 7589, 4 Place Jussieu, 75252 Paris Cedex 05, France}

\date{\today}

\begin{abstract} 

We study density and current fluctuations in a one-dimensional gas of run-and-tumble particles (RTPs), whose intrinsic velocities switch between two opposite values at a constant rate. We consider particles with and without pair interactions, including regularized Riesz gases, and also obtain corresponding results for Brownian particles. By linearizing the Dean–Kawasaki (DK) equations around a homogeneous stationary state, we compute density correlations and the variance of the integrated current through the origin. For repulsive Riesz interactions decaying as $1/|x|^s$ with $s>0$, this variance grows at long times as $t^{1/2}$ for $s>1$, but as $t^{s/(1+s)}$ for $0<s<1$. Starting from the same DK description, but retaining the Poisson statistics of direction reversals, we derive a macroscopic fluctuation theory (MFT) for RTPs with smooth pair interaction potential, which allows to compute the large time large deviation of the time integrated current. 
We show that, in the absence of interactions, the MFT saddle point equations can be solved explicitly using
a generalized Cole–Hopf transform. This allows to recover and generalize previous results for the current distribution obtained by microscopic calculations,
for RTPs with annealed or quenched step initial conditions.
Beyond these distributions, MFT allows to compute the optimal density histories producing a prescribed current through the origin. We find that these density profiles, for both RTPs and Brownian particles, exhibit a nontrivial space-time structure. 
We also study stationary current fluctuations on an interval connected to reservoirs. Finally, we analyze the MFT equations
perturbatively in the interaction. In both the infinite-line and reservoir geometries, we obtain explicit first-order corrections to the current generating functions for short-range interacting RTPs and Brownian particles.

\end{abstract}

\maketitle

\tableofcontents

\bigskip

\newpage

\section{Introduction and main results}
\label{sec:introduction}

\subsection{Overview}
\label{sec:intro_overview}

Run-and-tumble particles (RTPs) provide one of the simplest models of
persistent active motion. A particle moves ballistically with a finite
self-propulsion velocity during a ``run'', while random tumbling events
renew its direction of motion. This dynamics was introduced originally in
connection with bacterial motion and chemotaxis
\cite{Schnitzer1993,TailleurCates2008,Berg2004}, and has since become a paradigmatic
model of active matter~\cite{Ramaswamy2010,CatesTailleur2015}. In one
dimension the internal state reduces to a dichotomous variable
$\sigma=\pm1$, with velocities $\pm v_0$ and tumbling rate $\gamma$.
A free particle then crosses over, on scales large compared with the
persistence time $\gamma^{-1}$ and persistence length $v_0/\gamma$, to an
effective Brownian motion with diffusion coefficient
$D_a=v_0^2/(2\gamma)$ \cite{Weiss2002}. At shorter scales, however, the finite propagation
velocity and the internal state remain essential and lead to fluctuation
properties which can differ strongly from those of Brownian particles~\cite{Ang2017,Malakar2018,Dhar2019,BenDorWoillezKafriKardarSolon2019,LMS2020,Fry22,GMS26}.

 Interactions make this difference considerably richer. For instance, due to the persistent motion of RTPs, excluded volume interactions may lead to jamming and clustering, or on larger scales to motility-induced phase separation \cite{TailleurCates2008,CatesTailleur2013,CatesTailleur2015,
SlowmanEvansBlythe2016,CatesNardini2025}. Due to the out-of-equilibrium nature of these systems, they cannot be described by a Gibbs stationary measure, which makes understanding these collective behaviors quite challenging.
Furthermore, persistence introduces a microscopic time and length scale which cannot always be eliminated in favor of an effective temperature or diffusion coefficient. Several classes of interacting active particles
nevertheless admit useful coarse-grained descriptions. These include lattice models of run-and-tumble dynamics with exclusion interactions
\cite{ThompsonTailleurCatesBlythe2011,KEBT18,Erignoux2021,Agranov2021,Agranov2022,Dandekar2023,MasonErignouxJackBruna2023,ChakrabortyPradhan2024,MukherjeeSahaSadhuDharSabhapandit2025}, models where the motility of the active particles is regulated by the density
\cite{CatesTailleur2013,DinelliObyrneTailleur2024}, as well as active particles with long-range repulsive interactions 
described by a two-body potential $W(x) \propto 1/|x|^s$,
such as the active Dyson Brownian motion ($s=0$) and other active Riesz gases
\cite{ADBM1,RieszFluct,Santra_Calogero}. These studies also illustrate an important distinction between a hydrodynamic regime, in which persistence can sometimes be absorbed into effective transport coefficients, and shorter-scale regimes in which the internal active dynamics remains explicit. 

A natural microscopic starting point for fluctuating hydrodynamics is the Dean--Kawasaki (DK) equation. For interacting Brownian particles the DK equation gives an exact stochastic evolution equation for the microscopic density, with a multiplicative conserved noise \cite{Kawasaki1994,Dean1996}. Generalizations of this viewpoint to active particles are less straightforward because, in addition to translational currents, one must describe fluctuations of the internal state. DK or closely related fluctuating-hydrodynamic approaches have been
used for active lattice gases~\cite{Erignoux2021,Agranov2021,MasonErignouxJackBruna2023,MukherjeeSahaSadhuDharSabhapandit2025}, for the active Dyson
Brownian motion and active rank diffusions~\cite{ADBM1,activeRD}, for active particles with taxis or quorum-sensing interactions
\cite{DinelliObyrneTailleur2024}, as well as in the more general framework of actively switching particles \cite{Bressloff2024Switching}, and more recently for active Brownian particles with pairwise interaction potentials \cite{Soriani2026}. For strictly 1D systems where the interactions prevent particle crossings, it was found however that the jamming caused by the persistent motion of RTPs prevents the derivation of closed hydrodynamic equations
\cite{ADBM1,MukherjeeSahaSadhuDharSabhapandit2025,Bressloff2024Switching}.
In one dimensional systems of RTPs for which a DK description is applicable, e.g. when the particle can cross \cite{ADBM1},
the natural hydrodynamic variables are the densities $\rho_\pm$ of particles with orientation $\pm$. Contrary to the translational diffusive noise which is Gaussian by nature, the noise due to tumbling events is Poissonian. This distinction
is relatively unimportant at the level of typical fluctuations, where the Poisson noise can be expanded to quadratic order, but it becomes essential for large deviations.

In diffusive systems, the dynamical large deviations are described by macroscopic fluctuation theory (MFT)
\cite{Derrida2007,BertiniMFT2015}. In its standard one-component form, MFT is expressed in terms of a density-dependent diffusion coefficient and mobility, and has led to many exact results for current and density
fluctuations, both on the infinite line and in boundary-driven systems
\cite{Derrida2009,BodineauDerrida2004,BertiniCurrent2005,
KrajenbrinkLeDoussal2021,GrabschEtAl2022,MallickMoriyaSasamoto2022,
MallickMoriyaSasamoto2024,KrajenbrinkLeDoussalWASEP}. 
For interacting Brownian particles with arbitrary short-range pair forces, this framework has recently been connected explicitly to the microscopic Dean--Kawasaki description and to equilibrium statistical mechanics in
Ref.~\cite{Grabsch2025,Grabsch2026} (see also \cite{SahaJangidArnoulxKlamserSadhu2026}). 
By contrast, systems with long-range interactions are sub-diffusive on the macroscopic scale and thus their dynamics cannot be reduced to a diffusion coefficient and mobility as in standard MFT \cite{RieszGas2023,DysonCurrent2024}.
Finally, MFT methods have recently been extended to ballistic systems, such as hard rods~\cite{DeNardisBallistic2025,DeNardisMFTHardRods,Powdel2026HardRods,Mondal2026}.  

For diffusive systems, there is also a substantial multi-component literature. The hydrodynamics of particles carrying internal ``colors''
goes back to Ref.~\cite{QuastelColor1992}, with dynamical large deviations developed subsequently in
Ref.~\cite{QuastelRezakhanlouVaradhan1999}. 
Multi-species MFT with cross-diffusion has been considered in Refs.~\cite{Vanicat2017,MasonJackBruna2023}, while systems combining conservative transport with local reactions or interconversion are described by reaction--diffusion large-deviation theories
\cite{Elgart2004,JonaLandimVares1993,BodineauLagouge2010,
BodineauLagouge2012,BouchaudBiroli2006,BiroliLefevre2007}.
For active lattice gases, these ingredients were combined in Ref.~\cite{Agranov2022}: conservative current fluctuations coexist there with the Poisson statistics of tumbling events.

Recent large-deviation results for RTPs and related active particle models have also been obtained through other methods.
In \cite{FreeRTPs2020}, the large deviations of the current through the origin for non-interacting RTPs with a step initial condition were derived from a microscopic calculation. These results were later extended to other types of initial conditions in  \cite{BJC22,JoseRossoRamola2023,JoseRossoRamola2024}, and more general independent Gaussian processes \cite{DMS23} (see also \cite{LeDoussal2024}), including more recently active Ornstein-Uhlenbeck particles \cite{JKKS26}.
Stationary fluctuations of the density, including in the presence of an exclusion interaction, were studied
in Ref.~\cite{RedigVanWiechen2024}, while
Ref.~\cite{PulvirentiRedigVanWiechen2026} established a hydrodynamic limit
and trajectory large deviations.
These developments show that the large-deviation theory of active particles is a rapidly expanding field.
Nevertheless, to our knowledge, a general continuum MFT derived directly from microscopic RTP dynamics in the presence of smooth external forces and smooth spatial pair interactions, while keeping explicitly both the orientation-resolved densities and the Poissonian tumbling fluctuations,
has not yet been formulated. Existing active MFT constructions concern primarily lattice or exclusion models.

One aim of the present work is to provide such a construction. We consider
one-dimensional particles whose microscopic dynamics is given in
\eqref{eq:Langevin}, with internal velocities $\pm v_0$, thermal diffusion
$D$, a smooth external potential $V(x)$, and an orientation-independent
pair potential $W(x)$. We assume throughout the $D=0$ interacting
calculations that the force is sufficiently regular and weak at contact
for particles to be able to cross, as specified in \eqref{condition1}. Starting from
the DK dynamics for the densities $\rho_+$ and $\rho_-$, we use two complementary levels of
description. For typical fluctuations, we coarse grain and linearize the
stochastic equations, replacing the tumbling noise by its Gaussian
limit. For large deviations we instead retain the underlying Poisson
statistics of the tumbling events exactly and derive the corresponding
MFT action. Brownian particles are treated systematically as a benchmark:
they allow us to check the construction against standard MFT and, in the
interacting case, against the recent continuum formulation of
Ref.~\cite{Grabsch2025,Grabsch2026}. The free RTP problem then provides an exactly
solvable two-state theory, while interactions are incorporated
perturbatively.

The paper is organized as follows. In
Section~\ref{sec:typical} we introduce the DK equations for interacting RTPs,
linearize them around a homogeneous state and obtain the space-time
density and orientation correlations. We then apply these results in
Section~\ref{sec:typical_current} to the variance of the integrated
current, with particular emphasis on short- and long-ranged Riesz
interactions and on the crossover between persistent and diffusive
regime. In Section~\ref{sec:MFT_gen} we return to the underlying
Poissonian tumbling dynamics and construct the full MFT path integral and
its saddle-point equations for a general observable. We apply this theory
in Section~\ref{sec:MFT_step} to the current through the origin on the
infinite line for a step initial density. After treating Brownian and free
RTPs, including different annealed and quenched preparations, we add weak
interactions in Section~\ref{sec:step_interaction}. Finally,
Section~\ref{sec:MFT_reservoirs} considers the stationary current on an
interval coupled to reservoirs. We obtain exact results for free
Brownian particles and free RTPs and perturbative results for weakly
interacting particles. The Appendices contain
technical derivations of the perturbative corrections and details on the boundary
conditions, as well as a study of the optimal histories of the density leading to the current large deviations discussed in Section~\ref{sec:MFT_step}.

\subsection{Observables and summary of main results}
\label{sec:intro_results}

We consider RTPs with intrinsic velocities $\pm v_0$ and tumbling rate $\gamma$.
To remain fully general we also introduce an additional thermal noise of amplitude $D$ [see \eqref{eq:Langevin}]. All of the results for RTPs presented in this section are however obtained for $D=0$. In some cases we also present results for purely Brownian particles of diffusion coefficient $D$. 

We investigate current fluctuations in two geometries: (i) on the full line and (ii) on the interval $[0,L]$ with open boundary conditions. 
\\

\noindent {\bf Typical fluctuations on the real line.} On the infinite line, studied in Sections \ref{sec:typical} and \ref{sec:MFT_step}, the basic observable
is the integrated current through the origin,
\be \label{Q_intro}
Q(t_f)
=
\int_0^{t_f}dt\,J_s(0,t)
=
\int_0^{+\infty}dx\,
\left[
\rho_s(x,t_f)-\rho_s(x,0)
\right].
\ee
The initial mean density is chosen to be a step,
$r(x)=\rho_0\theta(-x)+\rho_1\theta(x)$. In the annealed case
the initial density $\rho(x,0)$ can fluctuate around its mean $r(x)$,
while in the quenched case it is fixed and equal to $r(x)$. 

In Section~\ref{sec:typical} we study the  
stationary (i.e annealed) and homogeneous initial condition with $\rho_0=\rho_1= \rho$, in which case
$\langle Q(t_f)\rangle=0$. 
We consider RTPs interacting
via a two-body potential $W(x)$, such that the particles
can cross [see Eq. \eqref{eq:Langevin} and condition \eqref{condition1}].
We also assume $W(\pm \infty)=0$. One example is the Riesz gas 
$W(x) \propto 1/|x|^s$ regularized with a short scale cutoff $a_0$
[see \eqref{Riesz_regularized}]. We use the linearized Dean Kawasaki (DK) equations to compute the variance $\langle Q(t_f)^2\rangle_c$ of the integrated current, for which we obtain an exact expression at all times [see Eq.~\eqref{Qvar_general}].  
At short times $t_f \to 0$ the current is controlled by ballistic
motion and is independent, at leading order, of the interaction,
\be
\left\langle Q(t_f)^2\right\rangle
\simeq
\rho v_0t_f  \;.
\ee
At long times $t_f \to \infty $ the behavior is instead determined by the small-wavevector
relaxation rates. For a short-ranged interaction,
the slow mode is diffusive with effective diffusion coefficient
$D_{\rm eff}=D_a+\rho W_0$ with
$W_0=\int_{-\infty}^{+\infty}dx\,W(x)$, and where 
$D_a=v_0^2/(2 \gamma)$ is the effective diffusion coefficient of a single RTP.
One finds
\be \label{Qvar_intro_SR}
\left\langle Q(t_f)^2\right\rangle_c
\simeq
\frac{2\rho D_a}{\sqrt{\pi D_{\rm eff}}}\sqrt{t_f} \;.
\ee
For a (regularized) Riesz interaction, 
with $0<s<1$, the relaxation is superdiffusive and
as $t_f \to +\infty$ one has
\be \label{Qvar_intro_LR}
\left\langle Q(t_f)^2\right\rangle_c
\simeq
\frac{
2\Gamma\left(\frac1{1+s}\right)\rho D_a
}{
\pi s(C_s\rho g)^{1/(1+s)}
}
t_f^{s/(1+s)} \quad , \quad  C_s=\frac{\sqrt{\pi}}{2^s} \frac{\Gamma(\frac{1-s}{2})}{\Gamma(1+\frac{s}{2})} \;.
\ee
The log-gas is the limiting case $s=0$ and gives the much slower
growth
$\langle Q(t_f)^2\rangle_c
\simeq2D_a(\pi^2g)^{-1}\log(t_f/t_0)$, where $t_0$ is a microscopic time, while at the marginal value
$s=1$ an additional logarithmic correction appears.
For long-ranged interactions these large time asymptotics agree, including the leading amplitudes,
with results obtained for passive and active Riesz gases by rather
different microscopic or MFT methods
\cite{RieszGas2023,RieszFluct,DysonCurrent2024}, after the replacement of
the passive diffusion constant by $D_a$. Note that the RTPs considered here are allowed to cross for $D=0$,
whereas for the standard Riesz interaction ($a_0=0$ for $s>0$)
they behave as single-file.
\\

\noindent {\bf Large deviations on the real line: non-interacting case.} In Section \ref{sec:MFT_gen} we develop a macroscopic fluctuation theory for RTPs directly in the continuum, extending the approach of \cite{Agranov2022} in the discrete setting, see also \cite{MasonErignouxJackBruna2023,MukherjeeSahaSadhuDharSabhapandit2025,KEBT18}. It allows in principle to compute the full cumulant generating function
\be 
\Sigma(\lambda,t_f)=\ln\langle e^{\lambda Q(t_f)}\rangle =  \sum_{n \geq 1} 
\langle Q(t_f)^n \rangle_c \frac{\lambda^n}{n!} \;,
\ee
{both for free and interacting RTPs. We start with free, i.e., non-interacting RTPs.}

In Section~\ref{sec:MFT_step} we consider a step initial condition on the infinite line, with arbitrary mean densities $\rho_0$ and $\rho_1$ on the two sides. More precisely, denoting by $r_\pm(x)$ the mean densities of particles with each orientation, we take $r_\pm(x)=\rho^\pm_0\theta(-x)+\rho^\pm_1\theta(x)$, with $\rho_0=\rho^+_0+\rho^-_0$ and $\rho_1=\rho^+_1+\rho^-_1$. We first consider the annealed, unpolarized preparation, in which particle positions are independently Poisson distributed on each side and each particle is assigned either orientation $\pm$ with equal probability, so that $\rho^\pm_{0,1}=\rho_{0,1}/2$. For this preparation, we obtain the exact finite-time result
\be \label{Sigma_RTP_intro}
\Sigma_{\rm an}(\lambda,t_f)
=
\left[
(e^\lambda-1)
+
\frac{\rho_1}{\rho_0}(e^{-\lambda}-1)
\right]\mu(t_f),
\qquad
\mu(t_f)
=
\frac{\rho_0v_0t_f}{2}
e^{-\gamma t_f}
\left[
I_0(\gamma t_f)+I_1(\gamma t_f)
\right] \;,
\ee
where $I_\nu(x)$ denotes the standard Bessel function of index $\nu$. For $\rho_1=0$, this is the cumulant generating function of a Poisson variable and coincides with the result of Ref.~\cite{FreeRTPs2020}, obtained there from an independent single-particle approach rather than from MFT. For $\rho_1>0$, it is the cumulant generating function of the difference of two independent Poisson variables with respective means $\mu(t_f)$ and $(\rho_1/\rho_0)\mu(t_f)$. Taking the Legendre transform, 
the PDF of $Q(t_f)=Q$ takes the large deviation form
\be \label{Ian_intro}
P_{\rm an}(Q)\asymp e^{-{\cal I}_{\rm an}(Q)},
\quad  
{\cal I}_{\rm an}(Q)
=
\mu(t_f)
\Psi_{\rm an}
\left(\frac{Q}{\mu(t_f)}\right) , 
\quad 
\Psi_{\rm an}(q)
=
q\ln\left(
\frac{q+\sqrt{q^2+4 \frac{\rho_1}{\rho_0}}}{2}
\right)
-\sqrt{q^2+4 \frac{\rho_1}{\rho_0}}+1+\frac{\rho_1}{\rho_0}.
\ee
{Here, and in the following, we use the symbol $\asymp$ to indicate a logarithmic equivalent.} 
For $\rho_1=0$, this reduces to the Poisson rate function $\Psi_{\rm an}(q)=q\ln q-q+1$, $q\geq0$. The single-particle RTP dynamics enters the full annealed current statistics only through $\mu(t_f)$, which crosses over from the ballistic short-time behavior $\mu(t_f)\simeq\rho_0v_0t_f/2$ for $\gamma t_f\ll1$ to the diffusive long-time behavior $\mu(t_f)\simeq\rho_0\sqrt{D_at_f/\pi}$ for $\gamma t_f\gg1$, with $D_a=v_0^2/(2\gamma)$. Thus the large-deviation speed crosses over from $t_f$ at short times to $\sqrt{t_f}$ at long times.

We then consider quenched particle positions, with the initial density fixed to $r(x)=\rho_0\theta(-x)+\rho_1\theta(x)$, while the initial orientations remain annealed and are chosen independently with equal probability, so that $\rho_{0,1}^\pm=\rho_{0,1}/2$. 
At long times, two different current regimes appear: a diffusive one,
$Q=O(\rho_0\sqrt{D_at_f})$, and a ballistic one,
$Q=O(\rho_0v_0t_f)$.  In the diffusive regime, both densities $\rho_0$
and $\rho_1$ contribute and the RTP dynamics becomes effectively Brownian,
with diffusion constant $D_a=v_0^2/(2\gamma)$.  Hence
\be \label{Pqu_diffusive_intro_new}
P_{\rm qu}(Q)
\asymp
\exp\left[-\rho_0\sqrt{\frac{D_at_f}{\pi}}\,
\Psi_{\rm qu}\left(\sqrt{\frac{\pi}{D_at_f}}\frac{Q}{\rho_0}\right)\right],
\qquad Q=O(\rho_0\sqrt{D_at_f}),
\ee
where $\Psi_{\rm qu}$ is precisely the quenched Brownian rate function given in
Eq.~\eqref{LD_free_Br_quenched}, with its dependence on $\rho_1/\rho_0$.
For $\rho_1=0$, this reproduces the diffusive regime of
Ref.~\cite{FreeRTPs2020} obtained by a very different method. 
Much larger (positive) currents probe instead the ballistic regime.  In this case,
\be \label{Pqu_ballistic_intro_new}
P_{\rm qu}(Q)
\asymp
\exp\left[-\rho_0v_0\gamma t_f^2\,
\Psi_{\rm RTP}\left(\frac{Q}{\rho_0v_0t_f}\right)\right],
\qquad 0<Q\leq \rho_0v_0t_f,
\ee
with
\be \label{Psi_RTP_intro_new}
\Psi_{\rm RTP}(q)
=
q-\frac{q}{2}\sqrt{1-q^2}-\frac12\sin^{-1}q,
\qquad 0\leq q\leq1.
\ee
For $\rho_1=0$, this is the ballistic large-deviation regime of
Ref.~\cite{FreeRTPs2020}.  For $\rho_1>0$, the particles initially on the
right give only a subleading contribution to a large positive current, so the
leading ballistic rate function is unchanged.  The negative ballistic tail for $Q <0$ is
obtained by exchanging $\rho_0$ and $\rho_1$.

Finally, we consider the fully quenched case, in which both the initial
positions and orientations are fixed.  The exact finite-time CGF is given by
Eq.~\eqref{Sigma_or} for arbitrary $\rho_{0,1}^\pm$.  For
$\rho_1^\pm=0$ and $\rho_0^+=\rho_0^-=\rho_0/2$, it reduces to the generating
function obtained in Refs.~\cite{JoseRossoRamola2023,JoseRossoRamola2024}, using a microscopic approach,
where it was used mainly to study the variance and its short- and long-time
behaviors. At long times, the fully quenched problem again displays the same two current
regimes: diffusive and ballistic.  In the diffusive regime the memory of the initial orientation is
lost, and one recovers the quenched Brownian large-deviation function in
Eq.~\eqref{Pqu_diffusive_intro_new}.  
In the ballistic regime, one recovers the same leading large-deviation form as for quenched positions and annealed initial orientations given in Eqs. (\ref{Pqu_ballistic_intro_new}).
Thus, the initial polarization affects finite-time behaviors of the average and the variance of the integrated current~\cite{JoseRossoRamola2023,JoseRossoRamola2024} at short times, but not the leading long-time diffusive and ballistic rate functions.
\\

\begin{figure}[t]
\centering
\includegraphics[width=0.59\textwidth]{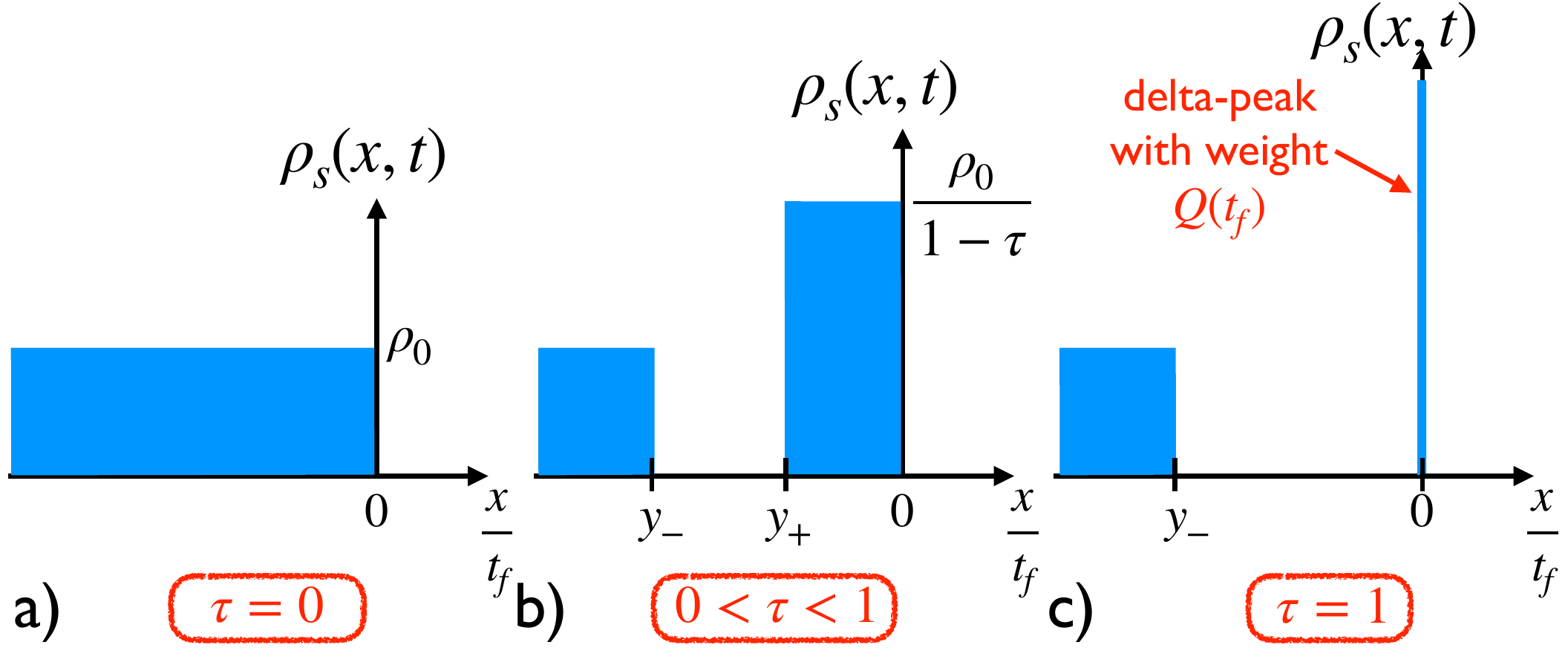}
\includegraphics[width=0.4\linewidth]{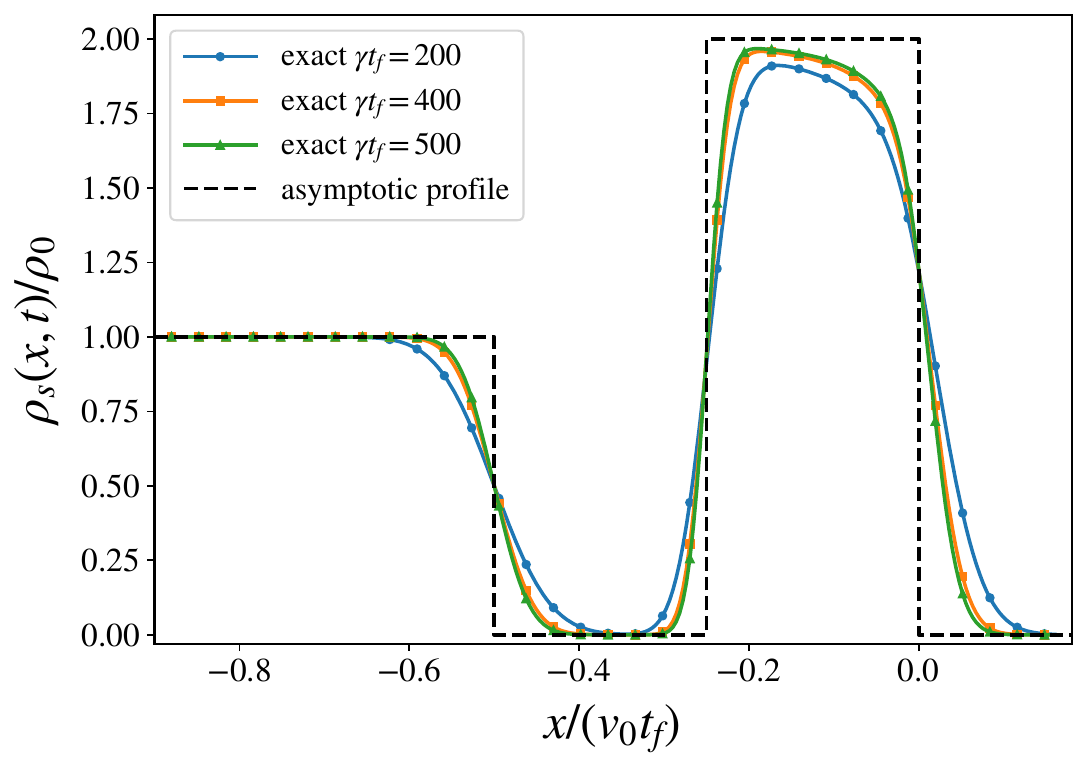}
\caption{{\bf Left:} Optimal density history in the quenched ballistic regime for large $t_f$ with fixed total current $Q=q\rho_0v_0t_f$. 
We denote $\tau=t/t_f$ and $y=x/t_f$.
The initially fixed density develops a macroscopic region
of vanishing density, for $y_- < y < y_+$, where $y_-=-q v_0$ and $y_+ = - q v_0(1-\tau)$, while the particles producing the rare current are compressed near the origin.  As $t\to t_f$, this compressed region collapses
onto the origin and carries a total mass $Q$. {{\bf Right:} Numerical check of the optimal macroscopic density profile for $q=1/2$ and $\tau=1/2$. The symbols and solid lines are obtained by evaluating Eqs.~\eqref{Gs_sol_freeRTP}, \eqref{Gd_sol_freeRTP} and \eqref{Fs_quenched} numerically and using \eqref{rho_reconstruction_app}, with $\lambda$ fixed by the exact current relation \eqref{Q_free_rtp_qa_step2}. We show three values of
$\gamma t_f$, namely $200$, $400$ and $500$. The dashed line is the large-$t_f$ result \eqref{rho_ballistic_profiles_intro_new}.}}
\label{fig:ballistic_outer_history}
\end{figure}

\noindent{\bf Optimal evolution of the density profile.} The MFT solution gives, in addition, the optimal density history associated
with these current fluctuations. To our knowledge, this quantity has received little attention, even though it can exhibit interesting behavior.
We analyze it here for {\it quenched positions}
and {\it annealed orientations}, taking $\rho_1=0$.  This is a genuinely new
result.  We describe here first the ballistic regime, which is the most striking, but the diffusive regime also exhibits similar, although smoother behavior, see below. For a prescribed ballistic current $Q=q\rho_0v_0t_f$ and
$0<t<t_f$, the density profile at leading order at large $t_f$ is
\be \label{rho_ballistic_profiles_intro_new}
\frac{\rho(x,t)}{\rho_0}\longrightarrow
\begin{cases}
1,&x<-Q/\rho_0,\\[1mm]
0,&-Q/\rho_0<x<-(1-t/t_f)Q/\rho_0,\\[1mm]
\dfrac{t_f}{t_f-t},&-(1-t/t_f)Q/\rho_0<x<0,\\[2mm]
0,&x>0.
\end{cases}
\ee
The particles initially contained in $[-Q/\rho_0,0]$ are thus selected to
produce the rare current.  At time $t$ they occupy the shorter interval
$[-(1-t/t_f)Q/\rho_0,0]$, leaving a macroscopic region of vanishing density
behind them.  As $t\to t_f$, this interval collapses onto the origin while
carrying a total mass $Q$, as shown in {the left panel of} Fig.~\ref{fig:ballistic_outer_history}. The right panel of the same figure shows the optimal density profile for finite values of $t_f$. As $\gamma t_f$ increases, the finite-time rounding near the three edges becomes narrower and the exact profile approaches
the piecewise limiting form of \eqref{rho_ballistic_profiles_intro_new}.

\begin{figure}[t]
\centering
\includegraphics[width=0.6\linewidth]{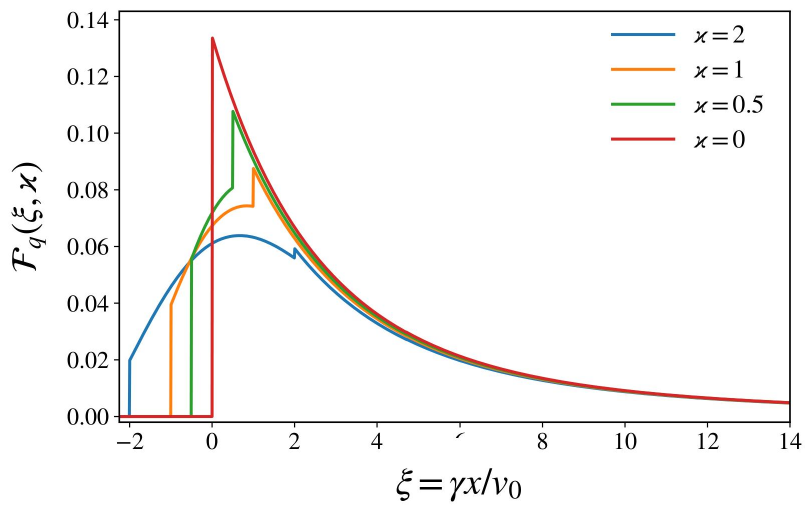}
\caption{Final-time scaling function ${\cal F}_q(\xi,\kappa)$ for $q=1/2$ and
$\kappa=2,1,0.5,0$.  For $\kappa>0$ the leading $O(t_f)$ contribution has support
only for $\xi\geq-\kappa$ and has a second discontinuity at $\xi=+\kappa$.  The
curves for different $\kappa$ share the same ${\cal F}_q(\xi,\kappa)\sim\xi^{-2}$ large-$\xi$ tail.}
\label{fig:Fq_terminal_profiles}
\end{figure}

Very close to the final time, the macroscopic profile exhibits an interesting boundary layer. Indeed, defining
\be \label{terminal_scaling_intro_new}
\kappa=\gamma(t_f-t),\qquad \xi=\frac{\gamma x}{v_0},
\ee
with $\kappa = O(1)$ and $\xi = O(1)$ fixed, one finds
\be \label{Fq_scaling_intro_new}
\rho_s(x,t)\simeq \rho_0\gamma t_f\,{\cal F}_q(\xi,\kappa).
\ee
A plot of the scaling function ${\cal F}_q(\xi,\kappa)$ as a function of $\xi$ is shown in
Fig.~\ref{fig:Fq_terminal_profiles}.  Finite propagation speed gives
${\cal F}_q(\xi,\kappa)=0$ for $\xi<-\kappa$, while the discontinuities at
$\xi=\pm \kappa$ are the direct signature of trajectories with no tumble during
the remaining time. Interestingly, it exhibits an algebraic tail at large $\xi$, namely
\be \label{Fq_tail_intro_new}
{\cal F}_q(\xi,\kappa)\simeq \frac{1}{\xi^2},\qquad \xi\to+\infty,
\ee
independently of $q$. The explicit form of ${\cal F}_q(\xi,\kappa)$ is derived in Appendix~\ref{app:density_RTP}.
In summary, we find that in the optimal history the particles wait until about the final time $t_f$ to cross the origin, and they do within a time window of order $O(1)$ near $t_f$, and on a spatial scale $O(v_0/\gamma)$.
The macroscopic profile \eqref{rho_ballistic_profiles_intro_new} and the
final-time profile \eqref{Fq_scaling_intro_new} match near the origin when
$1\ll\kappa=\gamma(t_f-t)\ll\gamma t_f$.
At fixed $0<q<1$ and $z=\xi/\kappa=x/[v_0(t_f-t)]$, away from
$z=-q$ and $z=0$, Eq.~\eqref{h:terminal} gives
\be \label{RTP_macro_terminal_matching_intro}
\kappa\,{\cal F}_q(\kappa z,\kappa)\longrightarrow
\begin{cases}
1,&-q<z<0,\\
0,&z<-q\ \hbox{or}\ z>0,
\end{cases}
\qquad \kappa\to\infty\;.
\ee
Thus the final-time density is
$\rho_s\simeq\rho_0\gamma t_f/\kappa=\rho_0t_f/(t_f-t)$
on $-qv_0(t_f-t)<x<0$, reproducing the height and width in
Eq.~\eqref{rho_ballistic_profiles_intro_new}.
This limit concerns $x=O(v_0(t_f-t))$; the far-left region
$x<-qv_0t_f$ is outside this scale.

Interestingly, for non-interacting Brownian particles with diffusion coefficient $D$, one can also observe similar non-trivial density profiles, for quenched initial conditions in the limit of large integrated currents. We thus
consider the same quenched initial density $r(x)=\rho_0\theta(-x)$ and a
prescribed positive current $Q\gg\rho_0\sqrt{Dt_f}$. At fixed
$0<\tau=t/t_f<1$, the optimal density profile on the scale $Q/\rho_0$ takes the same form \eqref{rho_ballistic_profiles_intro_new} as in the RTP case on macroscopic scales.
Very close to the final time, we define
\be \label{brownian_terminal_variables_intro}
\kappa_{\rm B}=\frac{Q^2(t_f-t)}{4\rho_0^2Dt_f^2},
\qquad \xi_{\rm B}=\frac{xQ}{4\rho_0Dt_f}\;.
\ee
For fixed $\kappa_{\rm B}>0$ and $\xi_{\rm B}$ in the large-current
limit, the density takes the scaling form
\be \label{brownian_terminal_profile_intro}
\rho(x,t)\simeq\frac{Q^2}{4\rho_0Dt_f}\,
{\cal B}(\xi_{\rm B},\kappa_{\rm B})\;,
\ee
which is the equivalent of the late time profile found for the RTP in (\ref{Fq_scaling_intro_new}).
For $\kappa_{\rm B}>0$, the function ${\cal B}$ is smooth and nonzero
on both sides of the origin. Thus the finite support boundary and the
discontinuities of the RTP profile (seen clearly on Fig. \ref{fig:Fq_terminal_profiles}) are absent in the Brownian case.
For large positive $\xi_{\rm B}$, it has the algebraic tail
\be \label{brownian_terminal_tail_intro}
{\cal B}(\xi_{\rm B},\kappa_{\rm B})\simeq
\frac{1}{2\xi_{\rm B}^2},\qquad \xi_{\rm B}\to+\infty,
\ee
at fixed $\kappa_{\rm B}$. The explicit form of ${\cal B}$ is derived in
Appendix~\ref{app:density_br}.
A finite fraction of the prescribed current is therefore accumulated
within a time window of order $\rho_0^2Dt_f^2/Q^2$ near $t_f$, on a
spatial scale of order $\rho_0Dt_f/Q$.
\\

\vspace*{0.5cm}

\noindent{\bf Large deviations on the real line: weakly interacting case.}
We next include interactions perturbatively $W \to \varepsilon W$, to first order in $\varepsilon$, and we focus on short-range interactions, i.e., $s>1$ for the Riesz gas. On the infinite line, an
important issue is how the initial state is prepared. We first consider the case of Brownian particles, for which we distinguish between (i) a Poisson initial state in which the
interaction is switched on only during the observation time, and (ii) a
preparation where the initial state is the stationary equilibrium
in the presence of interactions. These preparations lead to different cumulants for the current: the explicit generating function is given in \eqref{Sigma_ann0_SR} for preparation (i), 
and in \eqref{Sigma2eme} for preparation (ii). Extracting the second cumulant gives for large $t_f$ 
\bea 
\langle Q(t_f)^2 \rangle_c
\simeq \begin{cases} 
\left[
\rho_0+\rho_1
+
\varepsilon
\frac{W_0}{4D}
\left\{
(3-2\sqrt2)(\rho_0^2+\rho_1^2)
-
2\rho_0\rho_1
\right\} +
O(\varepsilon^2)
\right]
\sqrt{\frac{Dt_f}{\pi}} \quad, \quad \text{case (i)} \\
\left[
\rho_0+\rho_1
-
\varepsilon\frac{W_0}{4D}
(\rho_0+\rho_1)^2
+
O(\varepsilon^2)
\right]
\sqrt{\frac{Dt_f}{\pi}}  \quad, \quad \hspace*{3.1cm}  \text{case (ii)}
\end{cases} \;,
\eea
where we recall that $W_0=\int_{-\infty}^{+\infty}dx\,W(x)$. For a homogeneous interacting preparation (ii) with $\rho_0=\rho_1$, the variance
agrees with the independent calculation from the stationary linearized DK
theory presented in Section~\ref{sec:typical_current}.

For interacting RTPs on the full line we only consider the case (i). In the long-time diffusive regime
$\gamma t_f\gg1$, the complete first-order result again reduces to the
Brownian one with $D$ replaced by $D_a$, see \eqref{SigmaannRTPexpl}. In the
persistent regime $\gamma t_f \ll 1$ the situation is qualitatively different. The non-interacting optimal density and response fields, which enter in the calculation, contain ballistic fronts (see discussion above)
on which the local gradient expansion
breaks down, and the leading interaction correction is controlled by the
contact value $W(0)$ rather than by the integrated interaction $W_0$.
The complete formula for the correction to the cumulant generating function at large $t_f$ but for arbitrary fixed value of $\gamma t_f$
is given in Eq. \eqref{Sigma1_RTP_ballistic_scaling}. 
It simplifies in the strongly persistent limit where the correction, denoted here $\delta \Sigma(\lambda)$, becomes
\be \label{Sigma1_RTP_ballistic_tau0_intro}
\delta \Sigma(\lambda)
\simeq
-\frac{W(0)t_f}{4}
\left[
\rho_0^2(e^\lambda-1)^2
+
\rho_1^2(e^{-\lambda}-1)^2
\right],
\qquad
\gamma t_f\ll1 \;.
\ee
This provides a direct signature of
persistence in the interaction correction to a current large deviation.
\\

\noindent{\bf Open system in contact with reservoirs.}
For the open system on the interval $[0,L]$, treated in Section \ref{sec:MFT_reservoirs}, we consider the space-averaged integrated current
\be
{\cal Q}_{t_f}(L)
=
\frac1L
\int_0^{t_f}dt\int_0^Ldx\,J_s(x,t).
\ee
At fixed $L$ in the stationary long-time regime its fluctuations are
encoded by the scaled cumulant generating function (SCGF) and by the cumulants
\be \label{Sigma_intro_def}
\widetilde\Sigma(\lambda)
=
\lim_{t_f\to+\infty}
\frac1{t_f}
\ln\left\langle
e^{\lambda{\cal Q}_{t_f}(L)}
\right\rangle \quad , \quad \lim_{t_f\to\infty}
\langle{\cal Q}_{t_f}(L)^n\rangle_c/t_f = \widetilde\Sigma^{(n)}(0) \;.
\ee
Equivalently, the time-averaged current
$j={\cal Q}_{t_f}(L)/t_f$ obeys the large deviation principle
\be \label{current_LDP_intro}
P\left({\cal Q}_{t_f}(L)\simeq jt_f\right)
\asymp
e^{-t_f I(j)},
\qquad
I(j)
=
\sup_\lambda
\left[
\lambda j-\widetilde\Sigma(\lambda)
\right].
\ee

Here we compute these observables using the MFT of Section \ref{sec:MFT_gen} 
restricted to $[0,L]$ with appropriate boundary conditions (see below).
We solve the MFT saddle-point equations restricting to time independent solutions, i.e. assuming the analogue of
the Bodineau--Derrida additivity principle \cite{BodineauDerrida2004} -- possible time-dependent
optimal profiles are not considered here.

We consider a system with two reservoirs at each end, which fix the density at $x=0$ and $x=L$. For Brownian particles it reads
$\rho(0,t)=\rho_a$ and $\rho(L,t)=\rho_b$. For free RTPs
at $D=0$ we use reservoirs which fix the incoming populations, which fixes
$\rho_+(0,t)=\rho_a/2$ and $\rho_-(L,t)=\rho_b/2$. This corresponds to the infinitely fast reservoir limit, see e.g. \cite{SuzukiSasamoto2026}. 
We have obtained the stationary SCGF for free RTP's at $D=0$
\be \label{SCGF_free_RTP_intro}
 \widetilde\Sigma(\lambda)
=
\frac{v_0}{2}
\left[
\rho_a
\ln\left(
\frac{\alpha+e^\lambda}{\alpha+1}
\right)
+
\rho_b
\ln\left(
\frac{\alpha+e^{-\lambda}}{\alpha+1}
\right)
\right] \quad , \quad \alpha=\frac{\gamma L}{v_0}.
\ee
The corresponding mean current is at large time
\be
\frac{1}{t_f}  \langle {\cal Q}_{t_f}(L) \rangle \simeq  \widetilde\Sigma'(0)=
\frac{v_0}{2(\alpha+1)}(\rho_a-\rho_b)
=
\frac{D_a}{L+v_0/\gamma}
(\rho_a-\rho_b) \;.
\ee
Hence we see that the persistent motion of the RTPs induces a modification of the Fick's law, where $L \to L_{\rm eff} = L + v_0/\gamma$ where $v_0/\gamma$ is the so-called Milne length scale \cite{LMS2019}.
Equivalently, defining
$p_{\rm tr}=1/(1+\gamma L/v_0)$, Eq.~\eqref{SCGF_free_RTP_intro} is the
SCGF of the difference of two independent Bernoulli transmission
processes: each incoming particle crosses the interval with probability
$p_{\rm tr}$. For $L\gg v_0/\gamma$ we recover the open Brownian result
with $D\to D_a$, whereas in the strongly persistent limit the transmission
noise disappears (i.e. $\tilde \Sigma''(\lambda)\to 0$) within this ideal reservoir prescription. 

We also obtained the corrections to first order in the (short-range) interactions to the stationary
open-system SCGF for Brownian particles and for RTP's. For equal reservoirs, $\rho_a=\rho_b=\rho$, the
Brownian result reads
\be \label{Sigma_Brownian_intro}
\widetilde\Sigma_{\rm Br}(\lambda)
=
\frac{2D\rho}{L}(\cosh\lambda-1)
-
\varepsilon\frac{8\rho^2W_0}{3L}
\sinh^4\left(\frac{\lambda}{2}\right)
+
O(\varepsilon^2).
\ee
The interaction therefore leaves the variance unchanged at
$O(\varepsilon)$ but modifies the fourth and higher order even cumulants. This
result is reproduced by the standard scalar MFT using
$\sigma(q)=2Dq$ and the virial expansion which leads to 
$D_{\rm coll}(q)=D+\varepsilon W_0q+O(\varepsilon^2)$
\cite{Grabsch2025,Grabsch2026}. 

For RTPs the finite-persistence correction is more involved but remains
explicit.
For $\rho_a=\rho_b=\rho$, and for the ideal-contact prescription in which the
interaction acts in the bulk but is switched off in microscopic contact
layers at the reservoirs (see Section \ref{sec:MFT_reservoirs} for details), we obtain the SCGF in first order in perturbation theory
\be \label{Sigma_RTP_intro_compact}
\widetilde\Sigma_{\rm RTP}(\lambda)
=
\frac{v_0\rho}{2}
\ln\left[
1+\frac{2\alpha u_\lambda}{(1+\alpha)^2}
\right] -
\varepsilon\frac{2W_0\rho^2}{3L}
\frac{
\alpha^2u_\lambda
\left[
6(1+\alpha)^3+P(\alpha)u_\lambda
\right]
}{
(1+\alpha)^2
\left[
(1+\alpha)^2+2\alpha u_\lambda
\right]^2
}
+
O(\varepsilon^2) \;,
\ee
where we defined the shorthand notations
\be
\alpha=\frac{\gamma L}{v_0},
\qquad
u_\lambda=\cosh\lambda-1,
\qquad
P(\alpha)=\alpha^4+\alpha^3+10\alpha^2+9\alpha+3 \;.
\ee

Unlike the case of Brownian particles in
\eqref{Sigma_Brownian_intro}, the interaction term now starts at order
$\lambda^2$. Hence the variance is corrected as
\be \label{variance_RTP_intro}
\frac1{t_f}
\left\langle
[{\cal Q}_{t_f}(L)]^2
\right\rangle_c
\simeq
\frac{v_0\rho \alpha}{(1+\alpha)^2}
-
\varepsilon
\frac{4W_0\rho^2}{L}
\frac{\alpha^2}{(1+\alpha)^3}
+
O(\varepsilon^2) \;.
\ee
Thus interactions renormalize the stationary current variance already at
first order whenever persistence is finite. In the diffusive limit
$\alpha\gg1$ this correction vanishes and
\eqref{Sigma_RTP_intro_compact} reduces precisely to
\eqref{Sigma_Brownian_intro} with $D$ replaced by
$D_a=v_0^2/(2\gamma)$. At finite persistence the result also illustrates a feature
absent from the passive problem: if the pair
interaction is allowed to extend into the contact region, the additional
term \eqref{Sigma1_RTP_contact} must be included and is not universal
without specifying the reservoir model (see Section \ref{sec:MFT_reservoirs} for details).

\section{Density-Density two point correlation function with interactions} \label{sec:typical}

\subsection{General framework from the Dean-Kawasaki equation}

\subsubsection{The Dean-Kawasaki equations for interacting run-and-tumble particles}

We consider $N$ run-and-tumble particles in an external potential $V(x)$, which interact via a pairwise potential $W(x)$, which we take as an even function (of order $O(1)$). The equation of motion reads
\be \label{eq:Langevin}
\frac{dx_i(t)}{dt} = -V'(x_i(t)) - \sum_{j\neq i} W'(x_i(t)-x_j(t)) + v_0 \sigma_i(t)  +
\sqrt{2D}\,\xi_i(t) \;.
\ee
where the $\xi_i$'s are independent standard Gaussian white noises, and each $\sigma_i(t) = \pm 1$ flips its sign independently with rate $\gamma$. In Eq. \eqref{eq:Langevin} $D$ is the thermal noise amplitude, and $\pm v_0$ the intrinsic velocities.
Note that all particles interact identically, independently of their orientations, which, for a logarithmic interaction potential, corresponds to model II as defined in \cite{ADBM1}.
We will be interested in the densities defined as
\be
\rho_\sigma(x,t) = \sum_{i=1}^N \delta(x-x_i(t))\delta_{\sigma,\sigma_i(t)}  \;.
\ee
Throughout the paper, we assume that the interaction allows the particles to cross even when $D=0$, avoiding the formation of clusters. For two-particle collisions this condition reads
\be \label{condition1}
\max_{x \in \mathbb{R}} |W'(x)|<v_0 \quad , \quad {\rm for} \; D=0 \;. 
\ee
Under this assumption, together with smoothness of $W(x)$, which then implies $W'(0)=0$ by parity of $W(x)$, we can write a pair of coupled Dean-Kawasaki (DK) equations, as done in \cite{ADBM1} (note however that we have normalized the densities differently), 
\bea \label{dean1} 
\partial_t \rho_\sigma(x,t) &=& \partial_x [ \rho_\sigma(x,t) (- v_0 \sigma + V'(x) + \int dy \rho_s(y,t) W'(x-y)) ] - \gamma \rho_{\sigma}(x,t) + \gamma \rho_{-\sigma}(x,t) \nonumber
\\
&& + D \partial^2_x \rho_\sigma(x,t)  
+ \partial_x [ \sqrt{ 2 D \rho_\sigma(x,t) } \, \eta_\sigma(x,t) ] + \sigma \sqrt{\gamma \rho_s(x,t)} \, \zeta_K(x,t) \;,
\eea
where $\rho_s=\rho_++\rho_-$ and $\eta_+$, $\eta_-$ and $\zeta_K$ are independent standard Gaussian white noises. The first line contains the drifts from internal velocities, and from external and interaction forces, as well as the mean tumbling dynamics, while the second line contain the diffusion term and the passive and active noise terms. These equations are exact, except for the active noise term where we have made the replacement
\be \label{PoissonDK}
\sum_{i=1}^N
\left[
\delta_{\sigma,-\sigma_i}
-
\delta_{\sigma,\sigma_i}
\right]
\delta(x-x_i)\,d\widetilde N_i
\quad
\leadsto
\quad
\sigma\sqrt{\gamma\rho_s(x,t)}\,\zeta_K(x,t)\,dt ,
\ee
where $d\widetilde N_i(t)=dN_i(t)-\gamma dt$ and $N_i(t)$ is the Poisson process for particle $i$. This approximation is valid for typical fluctuations in the large $N$ limit, where each coarse-graining box contains many particles and many tumbling attempts (see discussion below). Rewriting these equations in terms of $\rho_s$ and $\rho_d= \rho_+ - \rho_-$ gives
\bea \label{dean2} 
\partial_t \rho_s(x,t) &=& \partial_x [ - v_0 \rho_d(x,t)  + V'(x) \rho_s(x,t) + \rho_s(x,t) \int dy \rho_s(y,t) W'(x-y) ] + D \partial^2_x \rho_s(x,t) + \partial_x [ \sqrt{ 2 D \rho_s(x,t) } \, \eta_s(x,t) ] \nonumber 
\\
\partial_t \rho_d(x,t) &=& \partial_x [ - v_0 \rho_s(x,t)  + V'(x) \rho_d(x,t) + \rho_d(x,t) \int dy \rho_s(y,t) W'(x-y) ] - 2\gamma \rho_d(x,t) + 2 \sqrt{\gamma \rho_s(x,t)} \, \zeta_K(x,t) \nonumber \\
&&+ D \partial^2_x \rho_d(x,t) + \partial_x [ \sqrt{ 2 D \rho_s(x,t) } \, \eta_d(x,t) ] 
\eea
with $\langle \eta_s(x,t) \eta_d(x',t') \rangle = \frac{\rho_d(x,t)}{\rho_s(x,t)} \delta(x-x')\delta(t-t')$. 
These equations coincide with those of \cite{Agranov2021}, except for the thermal noise terms, which differ because there is no exclusion constraint here.

Examples of relevant interactions for which the present method applies are regularized repulsive Riesz interactions
\be \label{Riesz_regularized}
-W'(x)=\frac{g \, x}{|x|^{2+s}+a_0^{2+s}} \;,
\ee
where $s>0$ is a real number, and where the characteristic length $a_0>0$ is chosen such that the condition \eqref{condition1} is satisfied. The case $s=0$ corresponds to a regularized version of the active log-gas, or active Dyson Brownian motion, introduced in \cite{ADBM1}. The case $s>1$ corresponds to short-range interactions, which includes the Calogero interaction for $s=2$ \cite{Santra_Calogero}.
Active Riesz gases with $a_0=0$ have been studied in \cite{RieszFluct}, which focused on microscopic quantities. A possible interpretation for the interaction \eqref{Riesz_regularized} is to consider particles confined in a narrow channel, such that the system is effectively 1d, but the particles can still pass each other.
\\

In the remainder of this section, we consider particles on the real line, such that at $t=-\infty$, the $N$ particles are contained in the interval $[-L/2,L/2]$, and we take the joint limit $N,L\to+\infty$ with fixed average density $\rho=N/L$. The mean density is thus equal to $\rho$ at all times and on the whole real axis {(with equal proportions of $+$ and $-$ particles on average)}. Alternatively, we could place the $N$ particles on a circle of size $L$ and consider a periodized interaction potential, as explained in \cite{RieszFluct}. This would give the same results in the thermodynamic limit $N,L\to+\infty$ with fixed $\rho=N/L$. Throughout this section, we focus on the case where there is no external potential, $V(x)=0$, and no thermal noise, $D=0$.

\subsubsection{Coarse-grained density and linearized equations}
\label{sec:linearized_RTP}

We are interested in the small fluctuations of the density around the average constant density profile. This requires us to coarse grain the density by defining
\be \label{coarse_graining}
\tilde \rho_\sigma(x,t) = \frac{1}{\ell} \int_{-\ell/2}^{\ell/2} dz \, \rho_\sigma(x+z,t) \;.
\ee
The coarse-graining distance $\ell$ should be large compared with the mean interparticle spacing, so that each coarse-graining box contains a macroscopic number of particles, but small compared with the scale over which the interaction potential $W(x)$ varies. For instance, for an interaction of the form \eqref{Riesz_regularized}, we require
\be \label{conditioncoarsegraining}
\rho^{-1} \ll \ell \ll a_0 \;.
\ee
Under this scale separation, the coarse-grained fields obey an effective Dean--Kawasaki equation of the same form as the microscopic one. Indeed, the interaction term can be closed as
$\overline{\rho_\sigma(W'*\rho_s)}\simeq \tilde\rho_\sigma(W'*\tilde\rho_s)$, where {$\overline{\cdots}$ denotes the coarse-graining operation \eqref{coarse_graining} and} $f * g$ denotes a convolution between the two functions $f$ and $g$, while the ballistic drift, diffusion, and mean tumbling terms commute with the coarse graining. The coarse-grained noises are spatially correlated over a distance of order $\ell$, but for Fourier modes satisfying $k\ell\ll1$ these correlations can be replaced by spatial white noise. For free Brownian particles, the same condition can be expressed in real space: the smoothing is irrelevant at spatial separations $|x-x'|\gg\ell$, and for correlations at a time separation $\tau=|t-t'|$ when $\sqrt{D\tau}\gg\ell$. For free RTPs, the corresponding condition is $L_{\rm RTP}(\tau)\gg\ell$, where
$L_{\rm RTP}(\tau)$ is the  
root-mean-square displacement of a single RTP over a time interval $\tau$.
This reduces to $L_{\rm RTP}(\tau) \simeq v_0\tau\gg\ell$ for $\gamma\tau\ll1$ and to $L_{\rm RTP}(\tau) \simeq \sqrt{D_a\tau}\gg\ell$ for $\gamma\tau\gg1$, with $D_a=v_0^2/(2 \gamma)$. 
In the presence of interactions, the general condition remains $k\ell\ll1$, together with $\ell\ll a$, where $a$ denotes the characteristic scale of variation of $W$. The corresponding temporal condition depends on the long-wavelength relaxation rate, denoted $\lambda_-(k)$. This rate will be determined below: for short-range interactions, $\lambda_-(k)\simeq D_{\rm eff} k^2$,
{where $D_{\rm eff}$ is an effective diffusion coefficient defined in \eqref{defDeff},} 
so that the coarse graining is irrelevant at time separations satisfying $\sqrt{D_{\rm eff} \tau}\gg\ell$. Independently of these spatial coarse-graining conditions, replacing the Poisson tumbling noise by Gaussian noise requires many tumbling events within each resolved mesoscopic space-time cell, $\gamma\tilde\rho_s\ell\,\Delta t\gg1$, or equivalently, for fluctuations probed at frequency $\omega$, $|\omega|\ll\gamma\tilde\rho_s\ell$. 
To simplify the notation, and since from now on we will only use the coarse-grained density \eqref{coarse_graining}, we omit the tilde in the remainder of this paper.

This coarse-graining allows us to linearize the Dean--Kawasaki equations around the homogeneous stationary state
\be
\rho_s(x,t)=\rho+\delta\rho_s(x,t),
\qquad
\rho_d(x,t)=\delta\rho_d(x,t) \;.
\ee
We assume that the homogeneous state is linearly stable and that the fluctuations $\delta\rho_s$ and $\delta\rho_d$ remain small.
Since $W'(x)$ is the derivative of a pair potential, its integral over the real axis vanishes,
\be
\int dy\,W'(x-y)=0 \;.
\ee
Taking this into account, the Dean-Kawasaki equations simplify to
\bea \label{dean_delta2} 
\partial_t \delta\rho_s(x,t) &=& -\partial_x [ v_0 \delta \rho_d(x,t) + \rho \mathcal{H}[\delta\rho_s](x,t) ] \;, \nonumber 
\\
\partial_t \delta \rho_d(x,t) &=& - v_0 \partial_x \delta \rho_s(x,t) - 2\gamma \delta \rho_d(x,t) + 2 \sqrt{\gamma \rho} \, \zeta_K(x,t) \;,
\eea
where we introduce the notation
\be \label{defconvH}
\mathcal{H}[f](x,t) = -\int dy \, W'(x-y) f(y,t) \;.
\ee

We will now use these equations to calculate the two-point correlations of the density fluctuations $\delta\rho_s$ and $\delta\rho_d$. Below we derive these correlation functions in Fourier space and real time. The results expressed for both Fourier space and time are given in Appendix~\ref{app:two_point_fourier_time}.

\subsection{Space-time correlation functions} \label{sec:correlations}

\subsubsection{Linearized equations in Fourier space}

In the following we will be working directly in the thermodynamic limit. Hence, we define the Fourier transform in space as
\be
\hat f(k,t) = \int dx \; e^{-ikx} f(x,t) \;.
\ee
Taking the Fourier transform of the equations \eqref{dean_delta2} leads to
\bea \label{dean_fourierx_step1} 
\partial_t \delta \hat\rho_s(k,t) &=& - ik v_0 \delta \hat \rho_d(k,t) - i k \rho \mathcal{\hat H} [ \delta \hat\rho_s](k,t) \;,
\\
\partial_t \delta \hat \rho_d(k,t) &=& - ik v_0 \delta \hat \rho_s(k,t) - 2\gamma \delta \hat \rho_d(k,t) + 2 \sqrt{\gamma \rho} \, \hat \zeta_K(k,t) \;, \nonumber
\eea
with
\be \label{Fourier_etacorr}
\langle \hat \zeta_K(k,t) \hat \zeta_K^*(k',t')\rangle = 2\pi \delta(k-k') \delta(t-t') \;.
\ee
Since the interaction term in real space takes the form of a convolution, in Fourier space it can be simply written
\be \label{fourierHgeneral}
\widehat{\mathcal{H}}[\delta \rho_s](k,t) = -i h(k) \delta \hat \rho_s(k,t) \;,
\ee
with
\be \label{fourierHgeneral2}
h(k)= -i\int dx \, e^{-i k x} W'(x) = - \int dx \sin(k x) W'(x) \;,
\ee
where we have used the fact that $W'(x)$ is odd, and where $h(k)$ is real. For interactions of the form \eqref{Riesz_regularized}, we have in the limit $k\ll a_0^{-1}$ (i.e. on scales where the regularization can be neglected), for $0<s<1$ (see \cite{RieszGas2023} and references therein),
\be \label{fourierH}
h(k) \simeq  C_s g \, {\rm sgn}(k)|k|^{s} \quad , \quad C_s=\frac{\sqrt{\pi}}{2^s} \frac{\Gamma(\frac{1-s}{2})}{\Gamma(1+\frac{s}{2})} \quad , \quad 0 <s < 1 \;. 
\ee
In the particular case of the regularized log-gas $s=0$, the Fourier transform can be performed exactly for all $k$, leading to
\be
h(k) = \pi g \, e^{-a_0 |k|} \sgn(k) \quad, \quad  ({\rm for} s=0) \;.
\ee
The short-range case $s>1$ is discussed below.
More generally, for other values of $s$, the introduction of the parameter $a_0$ introduces an exponential cutoff for $k\sim a_0^{-1}$.

Inserting \eqref{fourierHgeneral} into \eqref{dean_fourierx_step1}, we arrive at the following equation, written in matrix form
\be
\partial_t \begin{bmatrix} \delta \hat\rho_s(k,t) \\ \delta \hat\rho_d(k,t)
\end{bmatrix}
= - \mathbb{M}_k \begin{bmatrix} \delta \hat\rho_s(k,t) \\ \delta \hat\rho_d(k,t)
\end{bmatrix}
+ \mathbb{R}(k,t) \;,
\ee
with
\be
\mathbb{M}_k = \begin{bmatrix} \rho k h(k) & iv_0 k \\ iv_0 k & 2\gamma
\end{bmatrix}
\quad , \quad \mathbb{R}(k,t) = \begin{bmatrix} 0 \\ 2 \sqrt{\gamma \rho} \, \hat \zeta_K(k,t)
\end{bmatrix} \;.
\ee 
Introducing
\bea
&& \gamma_\pm(k)=\gamma \pm \frac{1}{2} \rho k h(k) \quad , \quad \Delta(k) = \gamma_-(k)^2 - v_0^2 k^2 \;, \\ \label{gamma_delta}
&&\lambda_\pm(k) = \gamma_+(k) \pm \sqrt{\Delta(k)} \quad , \quad \mu_\pm(k)
= \gamma_-(k) \pm \sqrt{\Delta(k)} \;,  \label{lambda_mu}
\eea
the matrix $\mathbb{M}_k$ can be diagonalized as
\be
\mathbb{M}_k = \mathbb{P}_k \begin{bmatrix}
    \lambda_+(k) & 0 \\ 0 & \lambda_-(k)
\end{bmatrix} \mathbb{P}_k^{-1} \quad {\rm with} \quad \mathbb{P}_k =
\begin{bmatrix}
    1 & 1 \\ \frac{\mu_+(k)}{iv_0k} & \frac{\mu_-(k)}{iv_0k}
\end{bmatrix} \quad , \quad \ \mathbb{P}_k^{-1} = \frac{1}{2\sqrt{\Delta(k)}}
\begin{bmatrix}
    -\mu_-(k) & iv_0k \\ \mu_+(k) & -iv_0k
\end{bmatrix} \;.
\ee
Here we have assumed that $k\neq0$ and $\Delta(k)\neq0$; the zero mode is fixed by particle-number conservation, while the case $\Delta(k)=0$ is obtained by continuity.

\subsubsection{Relaxation properties}

From now on, we will assume that (i) $kh(k) \to 0$ as $k\to 0$ and (ii) $h(k)$ is an odd function of $k$ such that $kh(k)>0$ for all $k\neq 0$. These properties are satisfied in particular for interactions of the form \eqref{Riesz_regularized}. More generally, (ii) implies that we focus on repulsive interactions. 

Under the assumption (i), one has $\Delta(k)>0$ for small $k$, i.e. the eigenvalues $\lambda_\pm(k)$ are real. Nevertheless, the function $\Delta(k)$ may change sign if the equation
\be 
|\gamma - \frac{\rho}{2} k h(k) | = v_0 k \;,
\ee 
has real roots, which generally happens for $k$ large enough if assumption (ii) is satisfied. In this case there is an overdamped/underdamped transition in the relaxation. Note that we have the relation
\be 
\lambda_+(k) \lambda_-(k) = [\gamma + \frac{\rho}{2} kh(k)]^2 - \Delta(k) = 2 \gamma \rho k h(k) + v_0^2 k^2 > 0 \;.
\ee 
Hence under the assumption (ii), and in the regime where $\Delta(k)>0$, we have
\be 
\lambda_+(k) > \lambda_-(k) >0 \;,
\ee 
which ensures the stability of the system. In the regime where $\Delta(k)<0$, the two eigenvalues are complex conjugate of each other and their real part is again positive.

In the limit $k\to 0$, the fast relaxation rate becomes $\lambda_+\to 2\gamma$, while for the slow relaxation rate behaves as
\be 
\lambda_-(k) = \rho k h(k) + \frac{v_0^2}{2 \gamma} k^2 + o(k^2)  \;.
\ee 

We call the interaction short-ranged when the first moment of the force
$F(x)=-W'(x)$ is finite. It is then also equal to the total integral, i.e. the zero-th moment $W_0$, 
of the interaction potential. In this case
\be
h(k)=kW_0+o(k),
\qquad
W_0
=
2\int_0^\infty dx\,xF(x)
=
\int_{-\infty}^{+\infty}dx\,W(x),
\ee
where in the last equality we assumed that $W(x)$ is even and
$xW(x)\to0$ as $x\to\infty$. A special case are interactions of the form \eqref{Riesz_regularized}
with $s > 1$ for which one has
\be
W_0=2g\int_0^{+\infty} dx\,
\frac{x^2}{x^{2+s}+a_0^{2+s}}
=
\frac{2\pi g}{s+2}\,a_0^{1-s}
\csc\left(\frac{3\pi}{s+2}\right),
\qquad s>1 \;.
\ee
Note that the value of $W_0$ is non-universal and depends on the short distance cutoff.

This leads to the long-wavelength diffusive behavior 
\be \label{lambdaminusdiffusion}
\lambda_-(k) =  \left( \frac{v_0^2}{2 \gamma} + \rho W_0 \right) k^2 + o(k^2) \;,
\ee 
with an interacting effective diffusion coefficient  
\be \label{defDeff}
D_{\rm eff}= D_a +\rho W_0 \quad , \quad D_a= \frac{v_0^2}{2\gamma} \;,
\ee 
where $D_a$ is the non-interacting effective ``active" diffusion coefficient.

For the Riesz gas with $0<s<1$, $W_0$ diverges and the expression \eqref{fourierH} gives
\be 
\lambda_-(k) = C_s g \rho |k|^{1+s} + \frac{v_0^2}{2\gamma}k^2+o(k^2)  \;,
\ee 
i.e. the relaxation is superdiffusive.

\subsubsection{Two-point correlation functions}

In the stationary state, one has
\be
\begin{bmatrix} \delta \hat\rho_s(k,t) \\ \delta \hat\rho_d(k,t)
\end{bmatrix}
= \int_{-\infty}^t ds \, e^{-\mathbb{M}_k (t-s)} \mathbb{R}(k,s) \;,
\ee
from which we can compute the correlation matrix, using the noise correlations \eqref{Fourier_etacorr} to evaluate the integrals. In the regime where $\Delta(k)>0$, we find
\bea \label{corr_fourier1} 
\!\!\!\!\!\!\!\! \langle \delta \hat\rho_s(k,t) \delta \hat\rho_s^*(k',t') \rangle &=& 2\pi \delta(k-k') C_{ss}(k,t-t') \;, \\
\!\!\!\!\!\!\!\! \langle \delta \hat\rho_d(k,t) \delta \hat\rho_d^*(k',t') \rangle &=& 2\pi \delta(k-k') C_{dd}(k,t-t') \nn \;, \\
\!\!\!\!\!\!\!\! \langle \delta \hat\rho_s(k,t) \delta \hat\rho_d^*(k',t') \rangle &=& 2\pi \delta(k-k') C_{sd}(k,t-t') \;, \nn
\eea
with
\bea \label{corr_fourier2} 
\!\!\!\!\!\!\!\! C_{ss}(k,\tau) &=& \frac{\gamma \rho v_0^2 k^2}{v_0^2 k^2 + 2\gamma \rho k h(k)} e^{-\gamma_+(k)|\tau|} \Bigg[ \frac{\cosh(\sqrt{\Delta(k)} |\tau|)}{\gamma_+(k)} + \frac{\sinh(\sqrt{\Delta(k)} |\tau|)}{\sqrt{\Delta(k)}} \Bigg] , \\
\!\!\!\!\!\!\!\! C_{dd}(k,\tau) &=& \frac{\gamma \rho}{v_0^2 k^2 + 2\gamma \rho k h(k)} e^{-\gamma_+(k)|\tau|} \Bigg[ \frac{v_0^2k^2+2\rho kh(k)\gamma_+(k)}{\gamma_+(k)} \cosh(\sqrt{\Delta(k)} |\tau|) \nn \\
&& \hspace{4.2cm}  - \frac{v_0^2k^2 + 2\rho kh(k) \gamma_-(k)}{\sqrt{\Delta(k)}} \sinh(\sqrt{\Delta(k)} |\tau|) \Bigg] \nn , \\
\!\!\!\!\!\!\!\! C_{sd}(k,\tau) &=& -i \gamma \rho v_0 k e^{-\gamma_+(k)|\tau|} \Bigg[ \frac{\rho k h(k)}{v_0^2 k^2 + 2\gamma \rho k h(k)} \bigg( \frac{\cosh({\sqrt{\Delta(k)} |\tau|})}{\gamma_+(k)} + \frac{\sinh(\sqrt{\Delta(k)} |\tau|)}{\sqrt{\Delta(k)}} \bigg) + \frac{\sinh\big(\sqrt{\Delta(k)}(\tau)\big)}{\gamma_+(k)\sqrt{\Delta(k)}}\Bigg] . \nn 
\eea
The expressions in the regime $\Delta(k)<0$ can be directly obtained by analytic continuation with $\sqrt{\Delta(k)}=i\sqrt{|\Delta(k)|}$.

\subsection{Application:  Time integrated current fluctuations for interacting RTPs} \label{sec:typical_current}

\subsubsection{General expression of the second moment}

We now apply the results above to study the fluctuations of the time integrated current flowing through the origin,
\be 
Q(t_f) = \int_0^{+\infty} dx [\rho_s(x,t_f) - \rho_s(x,0) ] \;,
\ee 
Note that we are now considering particles on the real line. Nevertheless, the expressions above obtained in the thermodynamic limit still apply in this setting. We consider an annealed initial condition, where at $t=0$ the particles are in the stationary state, with a uniform average density $\rho$. The first moment of the integrated current then vanishes, $\langle Q(t_f) \rangle=0$. Consider now its
second moment. Using the time translation symmetry, we can then write
\be 
\langle Q(t_f)^2 \rangle = 2 \int_0^{+\infty} dx \int_0^{+\infty} dx' \left[ \langle \rho_s(x,0) \rho_s(x',0) \rangle - \langle \rho_s(x,0) \rho_s(x',t_f) \rangle \right] .
\ee 
In the stationary state, $\rho_s(x,t)=\rho+\delta \rho_s(x,t)$. The mean stationary density cancels out, and we obtain
\be
\langle Q(t_f)^2 \rangle = 2 \int_{0}^{+\infty} dx \int_{0}^{+\infty} dx' \int_{-\infty}^{+\infty} \frac{dk}{2\pi} \int_{-\infty}^{+\infty} \frac{dk'}{2\pi} e^{ikx} e^{-ik'x'} \left[ \langle \delta \hat \rho_s(k,0) \delta \hat \rho_s^*(k',0) \rangle - \langle \delta \hat \rho_s(k,0) \delta \hat \rho_s^*(k',t_f) \rangle \right] ,
\ee
where we can use the expression given in \eqref{corr_fourier1} for the correlations of the density in Fourier space. This gives
\be
\left\langle Q(t_f)^2\right\rangle
=
2\int_0^\infty dx\int_0^\infty dx'
\int_{-\infty}^{+\infty}\frac{dk}{2\pi}
e^{ik(x-x')}
\left[
C_{ss}(k,0)-C_{ss}(k,t_f)
\right].
\ee
To ensure convergence, we regularize the half-line integrals by inserting $e^{-\epsilon(x+x')}$ before exchanging the integrals. Using the identity
\be
\int_0^\infty dx\int_0^\infty dx'\, e^{ik(x-x')-\epsilon(x+x')} = \frac{1}{k^2+\epsilon^2} \;,
\ee
we then obtain the relation
\be
\left\langle Q(t_f)^2\right\rangle = 2\lim_{\epsilon\to 0^+} \int_{-\infty}^{+\infty}\frac{dk}{2\pi} \frac{C_{ss}(k,0)-C_{ss}(k,t_f)}{k^2+\epsilon^2} \;.
\ee
Assuming that in the limit $k\to 0$ the numerator vanishes fast enough to avoid convergence issues, we arrive at the general formula
\be \label{Qvar_Css_formula}
\left\langle Q(t_f)^2\right\rangle
= \int_{-\infty}^{+\infty}\frac{dk}{\pi} \frac{C_{ss}(k,0)-C_{ss}(k,t_f)}{k^2} \;.
\ee
Using the stationary correlator expressed in \eqref{corr_fourier2}, one obtains explicitly
\be \label{Qvar_general}
\left\langle Q(t_f)^2\right\rangle = \frac{2\rho}{\pi}
\int_{0}^{+\infty} dk\, \frac{\gamma}{\gamma_+(k)}
\frac{v_0^2}{v_0^2 k^2+2\gamma\rho k h(k)}
\Bigg[ 1-e^{-\gamma_+(k)t_f}
\left( \cosh(\sqrt{\Delta(k)}t_f) + \frac{\gamma_+(k)}{\sqrt{\Delta(k)}}\sinh(\sqrt{\Delta(k)}t_f) \right) \Bigg],
\ee
where we have used the fact that the integrand is an even function of $k$, and we recall that
\be \label{gammadelta}
\gamma_+(k)=\gamma + \frac{1}{2} \rho k h(k) \quad , \quad  \Delta(k) = [\gamma - \frac{1}{2} \rho k h(k)]^2 - v_0^2 k^2 \;.
\ee  
When evaluating \eqref{Qvar_general} using \eqref{gammadelta}, one should remember that $\Delta(k)$ can become negative depending on the value of $k$.

\subsubsection{Convergence of the integral, short and long-time behavior}

We now analyze the integral in \eqref{Qvar_general}. Let us first check that it is convergent. Expanding the term in square brackets in \eqref{Qvar_general} at small $k$, retaining terms linear in $k h(k)$ and in $k^2$, the result is  
exactly proportional to the denominator $v_0^2 k^2+2\gamma\rho k h(k)$ in \eqref{Qvar_general}.
Writing $\left\langle Q(t_f)^2\right\rangle = \int_{0}^{+\infty} dk f(k)$ one has 
\be 
f(k) \underset{k\to 0}{\simeq} \frac{2\rho}{\pi} D_a \left(t_f-\frac{e^{-\gamma t_f}}{\gamma}\sinh(\gamma t_f)\right)
\ee
as $k\to0$, which ensures the convergence of the integral at small $k$.
Similarly, assuming that $k h(k)\to 0$ at large $k$, the integrand behaves as
\be \label{UVapprox}
f(k) \underset{k\to\infty}{\sim} \frac{2\rho}{\pi k^2} \left[ 1 - e^{-\gamma t_f} \left( \cos(v_0 |k| t_f) + \frac{\gamma}{v_0 |k|} \sin (v_0 |k| t_f) \right) \right] \;,
\ee
which ensures the convergence of the integral at large $k$.

In the short-time limit, the integral is dominated by large $k$. For the
regularized interaction \eqref{Riesz_regularized}, one has $k h(k)\to0$ as
$k\to\infty$, for any $s\geq0$. Using \eqref{UVapprox} and performing the
change of variable $q=v_0 k t_f$, we find
\be \label{ballisitic_limit}
\left\langle Q(t_f)^2\right\rangle
\simeq \frac{2\rho v_0 t_f}{\pi}
\int_0^{+\infty} dq\,\frac{1-\cos q}{q^2}
= \rho v_0 t_f \;,
\ee
i.e. a ballistic short-time behavior for any $s\geq0$.

The large-time limit of the integral \eqref{Qvar_general} is dominated by small $k$. 
Let us start with the short-range case $s>1$. We perform the change of variable 
$k=p/\sqrt{t_f}$ in \eqref{Qvar_general}. In the limit of large $t_f$ we find, 
using $h(k)=W_0 k+o(k)$ and $D_{\rm eff}=D_a+\rho W_0$,
\be \label{Qvar_largetimeSR}
\langle Q(t_f)^2 \rangle \simeq \frac{2 D_a}{\pi D_{\rm eff}} \sqrt{t_f}\,\rho 
\int_0^{+\infty} dp \frac{1-e^{-D_{\rm eff}p^2}}{p^2}
= \frac{2\rho D_a}{\sqrt{\pi D_{\rm eff}}}\sqrt{t_f} \;.
\ee

Let us now consider the long-range case $0<s<1$. We perform the change of variable
$k=p/t_f^{1/(1+s)}$ in \eqref{Qvar_general}. In the limit of large $t_f$ we find,
using $h(k)\simeq C_s g\,{\rm sgn}(k)|k|^s$,
\be \label{Qvar_largetimeLR}
\langle Q(t_f)^2 \rangle \simeq
\frac{2\rho D_a}{\pi}\,t_f^{\frac{s}{1+s}}
\int_0^{+\infty} dp\,
\frac{1-e^{-C_s\rho g\,p^{1+s}}}{C_s\rho g\,p^{1+s}}
=
\frac{2\Gamma(\frac{1}{1+s})\rho D_a}
{\pi s(C_s\rho g)^{\frac{1}{1+s}}}
\,t_f^{\frac{s}{1+s}} \;.
\ee

In the case of the log-gas $s=0$, this integral diverges at large $p$ and for convergence we should in principle keep the next term in the small $k$ expansion. The leading behavior in $t_f$ is however correctly recovered replacing 
$t_f^{\frac{s}{1+s}} \to t_f^{\frac{s}{1+s}}-t_0^{\frac{s}{1+s}}$ in \eqref{Qvar_largetimeLR} where $t_0$ is a microscopic time, and taking the limit $s \to 0$. 
The result is for $s=0$ 
\be \label{Qvar_largetime_log_gas}
\left\langle Q(t_f)^2\right\rangle = \frac{2 D_a}{\pi^2 g} \log ( t_f/t_0)  + O(1) \;.
\ee

Finally, in the marginal case $s=1$, the first moment of the force diverges logarithmically and one has
$h(k)\simeq 2g\,k\log(1/(a_0|k|))$ at small $k$. As a consequence, the slow relaxation rate behaves as
$\lambda_-(k)\simeq 2\rho g\,k^2\log(1/|k|)$, leading at large times to
\be
\left\langle Q(t_f)^2\right\rangle
\simeq
2D_a\sqrt{\frac{\rho}{\pi g}}\,
\sqrt{\frac{t_f}{\log(t_f/t_0)}} \;,
\ee
where $t_0$ is a microscopic time scale.

\subsubsection{Non-interacting case}

In the non-interacting case $h(k)=0$, one has $\Delta(k)=\gamma^2-v_0^2k^2$, and the equation \eqref{Qvar_general} can be written under the form (writing $u=v_0 k/\gamma$)
\be \label{Qvar_free_u}
\left\langle Q(t_f)^2\right\rangle = \frac{\rho v_0}{\gamma} \mathcal{I}(\gamma t_f) \;, \quad \text{with } \mathcal{I}(\tau)
= \frac{2}{\pi} \int_{0}^{+\infty} \frac{du}{u^2} \left[1-e^{-\tau} \left( \cosh\left(\tau\sqrt{1-u^2}\right) + \frac{\sinh\left(\tau\sqrt{1-u^2}\right)}{\sqrt{1-u^2}} \right) \right]  .
\ee
We first notice that
\be
\mathcal{I}'(\tau) = e^{-\tau} J(\tau) \quad , \quad J(\tau) = \frac{2}{\pi} \int_{0}^{+\infty} du \frac{\sinh\left(\tau\sqrt{1-u^2}\right)}{\sqrt{1-u^2}} \;.
\ee
Taking the Laplace transform $\mathcal{L}\{J\}(p)= \int_0^{+\infty} d\tau\, e^{-p\tau}J(\tau)$ and exchanging the integrals gives for $p>1$,
\be
\mathcal{L}\{J\}(p) = \frac{2}{\pi} \int_0^{+\infty} \frac{du}{u^2+p^2-1} = \frac{1}{\sqrt{p^2-1}} \;.
\ee
We recognize the Laplace transform of the modified Bessel function, i.e.
\be
J(\tau) = I_0(\tau) \;.
\ee
Using that $\mathcal{I}(0)=0$, we obtain
\be
\mathcal{I}(\tau) =  \int_0^{\tau} d\tau' e^{-\tau'} I_0(\tau') = \tau e^{-\tau} \left[ I_0(\tau)+I_1(\tau) \right] \;.
\ee
Thus, in the non-interacting case, we arrive at the following explicit expression
\be \label{Qvar_free_Bessel}
\left\langle Q(t_f)^2\right\rangle
= \rho v_0 t_f e^{-\gamma t_f} \left[ I_0(\gamma t_f)+I_1(\gamma t_f) \right].
\ee
The limiting behaviors ($\sim \sqrt{t_f}$ at large time and ballistic at small time) agree with the equations \eqref{Qvar_largetimeSR} and \eqref{ballisitic_limit} above. 

This result can be directly compared with Refs.~\cite{FreeRTPs2020,JoseRossoRamola2023,JoseRossoRamola2024},
where the current fluctuations of non-interacting RTPs were studied for an annealed step
initial condition with density $\rho$ on the negative half-line and no particles on the
positive half-line. In that case the current is Poisson distributed and its variance is
\[
\left\langle Q_{\rm step}(t_f)^2\right\rangle_c
=
\frac{\rho v_0 t_f}{2}e^{-\gamma t_f}
\left[I_0(\gamma t_f)+I_1(\gamma t_f)\right].
\]
In the homogeneous stationary state considered here, the net current is the difference
of two independent currents flowing from left to right and from right to left. Its mean
therefore vanishes and its variance is twice the step result, recovering exactly
\eqref{Qvar_free_Bessel}.

\subsubsection{Comparison with previous works}

Let us compare these results with previous studies of dynamical fluctuations in Riesz gases. 
In Ref.~\cite{RieszGas2023}, Dandekar, Krapivsky and Mallick considered passive Brownian particles interacting through the singular Riesz potential. Since the interaction diverges at contact, the particles preserve their ordering and the dynamics is single-file, in contrast to the regularized interaction considered here for which particles can cross.
Their starting point is a fluctuating hydrodynamic equation for the coarse-grained density, 
with a non-local Riesz contribution to the current for $0<s<1$ (they note that this current 
can equivalently be obtained from the Kawasaki--Dean approach). They then use a 
Martin--Siggia--Rose path-integral representation and a saddle-point construction analogous 
to MFT, followed by a perturbative expansion around the homogeneous state. In this way they 
compute directly the variance of the integrated current $Q$. Since their dynamics is 
single-file, the displacement $X$ of a tagged particle can then be related to the current: 
to the order relevant for the variance one has 
$\langle Q^2\rangle=\rho^2\langle X^2\rangle$ (this relation does not extend to higher cumulants). For $0<s<1$, their annealed result for $\langle Q^2\rangle$ agrees 
with \eqref{Qvar_largetimeLR}, including the prefactor, upon replacing the Brownian diffusion 
coefficient $D$ by $D_a=v_0^2/(2\gamma)$. They also obtain the same $\sqrt{t_f/\log t_f}$ 
scaling in the marginal case $s=1$. The logarithmic case $s=0$ was subsequently studied 
directly in Ref.~\cite{DysonCurrent2024}, using a non-local MFT derived from the 
Dean--Kawasaki fluctuating hydrodynamics; the resulting annealed current variance agrees 
with \eqref{Qvar_largetime_log_gas}, again with $D$ replaced by $D_a$.

A complementary approach was developed by us in Ref.~\cite{RieszFluct}. There the
microscopic dynamics is linearized in the weak-noise regime around the equally spaced
crystal configuration, and the small particle displacements are obtained by diagonalizing
the corresponding Hessian. This gives directly the space-time correlations of the particle
displacements, and in particular their mean-square displacement, rather than the integrated
current. In the passive case, the resulting long-time amplitudes were shown in
Ref.~\cite{RieszFluct} to coincide, including their numerical prefactors, with the results
of Ref.~\cite{RieszGas2023}, obtained from fluctuating hydrodynamics. Since the singular
Riesz gas is single-file, the current variance can be inferred in the hydrodynamic regime
from the tagged-particle variance using $\langle Q^2\rangle\simeq
\rho^2\langle X^2\rangle$. For $0<s<1$, this gives precisely
\eqref{Qvar_largetimeLR}. For RTPs the result of Ref.~\cite{RieszFluct} reduces at times
larger than the persistence and interaction time scales to the same expression with the
Brownian diffusion coefficient replaced by $D_a=v_0^2/(2\gamma)$. In fact the comparison
can be extended beyond the long-time asymptotics: expanding \eqref{Qvar_general} in the
weak-noise regime to leading order in $v_0^2$, its full time-dependent kernel coincides
with the one obtained from the phonon modes in Ref.~\cite{RieszFluct}, upon identifying
the hydrodynamic relaxation rate $\rho k h(k)$ with the long-wavelength phonon dispersion.

This agreement has a simple interpretation. At long wavelengths, the smooth part of the
density of a weakly deformed crystal is related to its displacement field $u(x,t)$ by
$\delta\rho(x,t)\simeq-\rho\,\partial_x u(x,t)$, and the continuity equation similarly
gives $j(x,t)\simeq\rho\,\partial_tu(x,t)$. Thus the linearized Dean--Kawasaki theory and
the phonon calculation describe the same long-wavelength mode, respectively in Eulerian
and Lagrangian variables. The agreement is particularly robust for $0<s<1$, since the
nonanalytic dispersion $\sim |k|^{1+s}$, including its coefficient, is fixed by the
long-range tail of the interaction. For $s>1$, on the other hand, the $k^2$ coefficient
depends on short-distance physics: for the singular Riesz crystal of
Refs.~\cite{RieszGas2023,RieszFluct} it is
$(s+1)\zeta(s)g\rho^s$, whereas for the regularized interaction considered here it is
$\rho W_0$ (and $D_{\rm eff}=D_a+\rho W_0$). Hence the $\sqrt{t_f}$ scaling and the
general structure of the amplitude agree, but the numerical SR amplitudes need not
coincide between the singular and regularized models. The crystalline harmonics omitted
from the smooth density description also imply that differences can arise at microscopic
length and time scales.

Finally, for short-range interactions, a general MFT description of interacting Brownian 
particles was recently given in Ref.~\cite{Grabsch2025,Grabsch2026}. In terms of the collective 
diffusion coefficient $D(\rho)$ and mobility $\sigma(\rho)$, it yields for an equilibrium 
one-dimensional system
$\langle Q(t)^2\rangle \simeq \sigma(\rho)\sqrt{t}/\sqrt{\pi D(\rho)}$. 
For Brownian particles $\sigma(\rho)=2D\rho$, whereas $D(\rho)$ contains the effect of the 
interactions through the equilibrium equation of state. This has precisely the same 
structure as \eqref{Qvar_largetimeSR}, with $D\to D_a$ and 
$D(\rho)\to D_{\rm eff}$. Notice that, unlike the tracer result, this current formula does 
not require a single-file constraint and applies also to soft, crossing particles, making 
it particularly close in spirit to the present regularized model.

We stress that these previous works mainly concern singular Riesz interactions for which particles cannot cross, whereas here the interaction is regularized and crossings are allowed. Note also that \eqref{Qvar_general} also describes the finite-time crossover at finite tumbling rate.

\section{Macroscopic fluctuation theory: General method} \label{sec:MFT_gen}

In the remaining part of this paper, we go beyond the second moment and we use MFT to compute the moment generating function of the current and density fluctuations, from which one can deduce their large deviations. In the present section we present the general method. In the next section \ref{sec:MFT_step} we apply it to the case of free RTPs on a line. In particular, we show how this approach recovers the results of \cite{FreeRTPs2020,JoseRossoRamola2023,JoseRossoRamola2024} for the fluctuations of the current through the origin for a step initial condition. For free RTPs on a finite interval coupled to reservoirs at the boundaries, we study in Section \ref{sec:MFT_reservoirs}
the large deviations of the space-integrated current in the stationary state. We then extend some of these results to the case where interactions are present, to first order in a perturbation theory in the strength of the interaction.

\subsection{Large deviation form for the densities} \label{sec:large_dev_dens}

As in the previous section, we use the coarse-grained density $\rho_\sigma(x,t)$ defined
in \eqref{coarse_graining}. The coarse-graining scale $\ell$ should be large enough to
contain many particles, while remaining small compared to the relevant spatial scales of
the interaction and of the observable considered. 
We start from the exact version of the
DK dynamics underlying Eq.~\eqref{dean1}, i.e. before making the approximation
\eqref{PoissonDK}. Indeed, for the large-deviation analysis, it is important to retain
the Poissonian nature of the tumbling events, rather than the Gaussian approximation used
in \eqref{dean1} to describe typical fluctuations. The dynamics can then be written as

\bea \label{deanLDpm} 
\partial_t \rho_+(x,t) &=& -\partial_x J_+(x,t) - K_+(x,t) + K_-(x,t) \;, \\
\partial_t \rho_-(x,t) &=& -\partial_x J_-(x,t) - K_-(x,t) + K_+(x,t) \;, \nn
\eea
where $J_+(x,t)$ and $J_-(x,t)$ are Gaussian variables such that
\be \label{LJpm1}
P(J_+,J_-) \propto \exp(-\mathcal{L}_J dxdt) \quad , \quad
\mathcal{L}_J =
\frac{1}{2}
\begin{bmatrix}
J_+ - \bar{J}_+ \\
J_- - \bar{J}_-
\end{bmatrix}^T
\mathbf{C}^{-1}
\begin{bmatrix}
J_+ - \bar{J}_+ \\
J_- - \bar{J}_-
\end{bmatrix}
\quad , \quad
\mathbf{C} = 2D
\begin{bmatrix}
\rho_+ & 0 \\
0 & \rho_-
\end{bmatrix} \;,
\ee
with
\bea \label{defJbar}
&&\bar{J}_+ = v_0 \rho_+ - V'(x) \rho_+ + \rho_+ \mathcal{H}[\rho_s]
- D \partial_x \rho_+ \;, \\
&&\bar{J}_- = -v_0 \rho_- - V'(x) \rho_-
 + \rho_- \mathcal{H}[\rho_s]
- D \partial_x \rho_- \;, \nonumber
\eea
where we recall that
$\mathcal{H}[f](x) = -\int dy \, W'(x-y) f(y)$.
For $D=0$, the Gaussian current weight in \eqref{LJpm1} is understood as the limit
$D\to0$, in which it imposes $J_\pm=\bar J_\pm$ deterministically.
The  terms $K_\pm$ in \eqref{deanLDpm} account for the tumbling events: the numbers $K_\pm dxdt$ of tumbling events in a space-time cell are Poisson
variables with mean $\gamma\rho_\pm dxdt$.
The main difference compared to Section~\ref{sec:typical} is that we no longer treat
the Poissonian noise due to the tumbling events at the Gaussian level.
Following \cite{Agranov2022}, we write their large deviation function as
\be
-\ln P(K_\pm) \simeq \mathcal{L}_{K_\pm} dxdt
\quad , \quad
\mathcal{L}_{K_\pm}
= K_\pm \ln \frac{K_\pm}{\gamma\rho_\pm} - K_\pm + \gamma\rho_\pm \;.
\ee
This is the standard local large-deviation cost associated with a Poisson process
with mean number of events $\gamma\rho_\pm dxdt$ in the space-time cell.

Introducing the notation $\vec{\rho}=(\rho_+,\rho_-)$, one obtains the large deviation
form for the densities
\bea \label{LDpm1}
P(\vec{\rho}(x,t_f)|\vec{\rho}(x,0))
&\simeq&
{\cal N}
\int \mathcal{D}\rho_+ \mathcal{D}\rho_-
\mathcal{D}J_+ \mathcal{D}J_-
\mathcal{D}K_+ \mathcal{D}K_-
\, e^{-\int_0^{t_f} dt \int dx
(\mathcal{L}_J+\mathcal{L}_{K_+}+\mathcal{L}_{K_-})}
\nonumber \\
&& \hspace{1cm}
\delta(\partial_t \rho_+ + \partial_x J_+ + K_+ -K_-)
\delta(\partial_t \rho_- + \partial_x J_- + K_- -K_+)
\nonumber \\
&\simeq&
{\cal N}
\int \mathcal{D}\rho_+ \mathcal{D}\rho_-
\mathcal{D}J_+ \mathcal{D}J_-
\mathcal{D}K_+ \mathcal{D}K_-
\mathcal{D}p_+ \mathcal{D}p_-
\, e^{-\mathcal{\hat S}} \;,
\eea
with
\be \label{LDSpm1}
\mathcal{\hat S}
=
\int_0^{t_f} dt \int dx
\left[
\mathcal{L}_J+\mathcal{L}_{K_+}+\mathcal{L}_{K_-}
+p_+(\partial_t\rho_+ + \partial_xJ_+ +K_+-K_-)
+p_-(\partial_t\rho_- + \partial_xJ_- +K_- -K_+)
\right] \;.
\ee
In the first step we have written the transition probability as a path integral over
density, current and tumbling histories, with the evolution equations enforced by delta
functionals. In the second step we have represented these delta functionals by introducing
the auxiliary response fields $p_\pm(x,t)$, after the usual rotation of the response-field
contours, leading to the MSR representation used in MFT.
The normalization ${\cal N} $ includes factors such as $\prod_{xt} (\det \mathbf{C}(x,t))^{-1/2}$; these cancel when the Gaussian currents are integrated
out and do not contribute to the leading MFT action.
{At this level of the calculation, we keep the large-deviation speed implicit, i.e., the
large parameter controlling the saddle-point approximation below. It will depend on the
setting and on the observable considered. In the infinite-line problem of
Section~\ref{sec:MFT_step}, this large parameter arises from the long-time limit
$t_f\to+\infty$, with a scaling that will be specified below. In the boundary-driven
problem of Section~\ref{sec:MFT_reservoirs}, the MFT description is understood in the
hydrodynamic large-system-size limit, while the stationary cumulant generating function
is obtained from the additional long-time limit $t_f\to+\infty$.
}

Note that the equations \eqref{deanLDpm} can also be written as a function of
$\rho_s=\rho_++\rho_-$ and $\rho_d=\rho_+-\rho_-$.
Introducing $J_s=J_++J_-$ and $J_d=J_+-J_-$, we have
\bea \label{deanLDrhos}
\partial_t \rho_s(x,t) &=& -\partial_x J_s(x,t) \;, \\
\partial_t \rho_d(x,t) &=& -\partial_x J_d(x,t)
-2[K_+(x,t)-K_-(x,t)] \;. \label{deanLDrhod}
\eea
where
\be \label{LJsd}
P(J_s,J_d)
\propto \exp(-\mathcal{\tilde L}_J dxdt)
\quad , \quad
\mathcal{\tilde L}_J
=
\frac{1}{2}
\begin{bmatrix}
J_s-\bar J_s\\
J_d-\bar J_d
\end{bmatrix}^T
 \mathbf{\tilde C}^{-1}
\begin{bmatrix}
J_s-\bar J_s\\
J_d-\bar J_d
\end{bmatrix}
\quad , \quad
\mathbf{\tilde C}
=
2D
\begin{bmatrix}
\rho_s & \rho_d\\
\rho_d & \rho_s
\end{bmatrix} \;,
\ee
with
\bea \label{defJsbar}
&&\bar{J}_s
= v_0\rho_d -V'(x)\rho_s
+\rho_s\mathcal H[\rho_s]
-D\partial_x\rho_s \;,\\
&&\bar{J}_d
= v_0\rho_s -V'(x)\rho_d
 +\rho_d\mathcal H[\rho_s]
-D\partial_x\rho_d \;. \label{defJdbar}
\eea
This leads to the large deviation form for
$\vec{\tilde\rho}=(\rho_s,\rho_d)$
\bea \label{LDsd}
P(\vec{\tilde\rho}(x,t_f)|\vec{\tilde\rho}(x,0))
&\simeq&
{\cal N}
\int \mathcal{D}\rho_s\mathcal{D}\rho_d
\mathcal{D}J_s\mathcal{D}J_d
\mathcal{D}K_+\mathcal{D}K_-
\,e^{-\int_0^{t_f}dt\int dx
(\mathcal{\tilde L}_J+\mathcal L_{K_+}+\mathcal L_{K_-})}
\\
&&\hspace{1cm}
\delta(\partial_t\rho_s+\partial_xJ_s)
\delta\left(\partial_t\rho_d+\partial_xJ_d
+2(K_+-K_-)\right)
\nonumber\\
&\simeq&
{\cal N}
\int \mathcal{D}\rho_s\mathcal{D}\rho_d
\mathcal{D}J_s\mathcal{D}J_d
\mathcal{D}K_+\mathcal{D}K_-
\mathcal{D}p_s\mathcal{D}p_d
\,e^{-\mathcal{\hat S}}\;,\nonumber
\eea
with
\be \label{LDSsd}
\mathcal{\hat S}
=
\int_0^{t_f}dt\int dx
\left[
\mathcal{\tilde L}_J+\mathcal L_{K_+}+\mathcal L_{K_-}
+p_s( \partial_t\rho_s+\partial_xJ_s)
+p_d\left(\partial_t\rho_d+\partial_xJ_d +2(K_+-K_-)\right)
\right]\;.
\ee
The equations \eqref{LDsd}-\eqref{LDSsd} can be obtained from
\eqref{LDpm1}-\eqref{LDSpm1} through the change of variables
\be \label{pm_to_sd}
\rho_{s/d}=\rho_+\pm\rho_-
\quad,\quad
J_{s/d}=J_+\pm J_-
\quad,\quad
p_{s/d}=\frac{p_+\pm p_-}{2}
\;.
\ee

In \eqref{LDsd}-\eqref{LDSsd}, the densities $\rho_\pm$ entering
$\mathcal L_{K_\pm}$ are of course understood as
$\rho_\pm=(\rho_s\pm\rho_d)/2$.
Following \cite{Agranov2022}, one could simplify equation \eqref{LDsd} by performing
the change of variables
$(K_+,K_-)\to(K_+,K=K_+-K_-)$ and performing a saddle-point integration over $K_+$.
Instead, here we will derive the saddle-point equations from
\eqref{LDpm1}-\eqref{LDSpm1} and use the change of variables \eqref{pm_to_sd}
whenever it is useful to solve these equations.

Our goal for the remainder of this paper is to compute the moment generating function
(and then the large-deviation rate function via a Legendre transform)
of some observable
$\mathcal O(\{\rho_\pm,J_\pm\})$, which is a functional of the densities and currents,
\be \label{LDgeneral}
\langle e^{\lambda\mathcal O}\rangle
\simeq
{\cal N}
\int
\mathcal D\rho_+\mathcal D\rho_-
\mathcal DJ_+\mathcal DJ_-
\mathcal DK_+\mathcal DK_-
\mathcal Dp_+\mathcal Dp_-
\,e^{\lambda\mathcal O-\mathcal{\hat S}}
P(\vec\rho(x,0))\;.
\ee

For this, we should first derive the saddle-point equations extremizing the argument of the exponential, which we do in
Section~\ref{sec:SaddlePoint_general} for a generic observable.
The last two parts of the paper will focus on two different observables.
In Section~\ref{sec:MFT_step}, we will first  consider particles on the real line with a step
initial condition, and study the same observable as in
Section~\ref{sec:typical_current}, namely the time-integrated current through the origin,
which reads
\be \label{current_through_zero_def}
Q(t_f)
=
\int_0^{t_f}dt\,J_s(0,t)
=
\int_0^{+\infty}dx\,
[\rho_s(x,t_f)-\rho_s(x,0)]\;,
\ee
where the second equality is obtained using equation \eqref{deanLDrhos}
with the boundary condition $J_s(x\to+\infty,t)=0$.
In Section~\ref{sec:MFT_reservoirs}, we will then consider particles on an interval $[0,L]$,
with densities fixed by reservoirs at $x=0$ and $L$, and study instead the
space-integrated current in the stationary state,
\be \label{integrated_current_def}
\mathcal Q_{t_f}(L)
=
\frac{1}{L}\int_0^{t_f}dt\int_0^Ldx\,J_s(x,t)\;.
\ee
These two observables will lead to slightly different sets of saddle-point equations,
with different boundary conditions. For each observable, we first focus on the case of
non-interacting RTPs, before adding interactions. In each case, we start by
(re)deriving the results for Brownian particles for the sake of comparison.

\subsection{Saddle-point equations} \label{sec:SaddlePoint_general}

We now eliminate $K_\pm$ and $J_\pm$ from \eqref{LDgeneral}. Since the exponent is
$\lambda\mathcal O-\mathcal{\hat S}$, the relevant saddle point extremizes
$\mathcal{\hat S}-\lambda\mathcal O$, rather than $\mathcal{\hat S}$ alone.
Since by assumption $\mathcal O$ does not depend explicitly on $K_\pm$, the corresponding saddle-point equations give
\be \label{saddle_point_K}
\frac{\delta (\hat{\cal S}-\lambda {\cal O})}{\delta K_+} = 0 \ \Rightarrow \ K_+
=
\gamma\rho_+e^{-(p_+-p_-)}
\quad,\quad
\frac{\delta (\hat{\cal S}-\lambda {\cal O})}{\delta K_-} = 0 \ \Rightarrow \ K_-
=
\gamma\rho_-e^{p_+-p_-}\;.
\ee
Note that these equations do not have any boundary terms, and that they do not depend
on the choice of observable $\mathcal O$ since we have assumed that $\mathcal O$ only
depends on the densities $\rho_\pm$ and currents $J_\pm$.
These two observations do not extend to the currents $J_+$ and $J_-$, for which the
saddle-point values read
\be \label{saddle_point_J}
\begin{bmatrix}
J_+\\
J_-
\end{bmatrix}
=
\begin{bmatrix}
\bar J_+\\
\bar J_-
\end{bmatrix}
+
\mathbf C
\left(
\begin{bmatrix}
\partial_xp_+\\
\partial_xp_-
\end{bmatrix}
+
\lambda\frac{\delta\mathcal O}{\delta\vec J}
\right)
\equiv
\begin{bmatrix}
J_+^{\rm sp}\\
J_-^{\rm sp}
\end{bmatrix}
\;,
\ee
where $\vec J=[J_+\ J_-]^T$ and where $\bar J_\pm$ are given in
\eqref{defJbar}. In addition, if the particles are confined to an interval $[0,L]$,
with the particle density fixed at $x=0$ and $x=L$, some of the $p_\pm$ fields will be constrained to vanish at the boundary, see discussion in Section~\ref{sec:MFT_reservoirs}.
When the term proportional to $\lambda$ in \eqref{saddle_point_J} is nonzero,
it is convenient to define
\be \label{defhatp}
\hat p_\pm(x,t)
=
p_\pm(x,t)
+
\lambda\int_0^xdy\,
\frac{\delta\mathcal O}{\delta J_\pm(y,t)}\;.
\ee

Inserting the solutions \eqref{saddle_point_K} and \eqref{saddle_point_J} into
\eqref{LDgeneral}, with the action $\mathcal{\hat S}$ given in \eqref{LDSpm1},
we obtain
\be \label{largedeviationform_general}
\langle e^{\lambda\mathcal O}\rangle
\simeq
{{\cal N}}
\int\mathcal D\rho_+\mathcal D\rho_-
\mathcal Dp_+\mathcal Dp_-\,
e^{\lambda\mathcal O-\mathcal S}
P(\vec\rho(x,0))\;,
\ee
where, from this point on, $\mathcal O$ is understood to be evaluated at the
saddle-point currents $J_\pm=J_\pm^{\rm sp}$ given by \eqref{saddle_point_J}, with

\bea
\mathcal S
&=&
\int_0^{t_f}dt\int dx\,
\Bigg[
-
\begin{bmatrix}
\partial_xp_+-\frac{1}{2}\partial_x\hat p_+\\
\partial_xp_- -\frac{1}{2}\partial_x\hat p_-
\end{bmatrix}^{T}
\mathbf C
\begin{bmatrix}
\partial_x\hat p_+\\
\partial_x\hat p_-
\end{bmatrix}
\nn\\
&&\hspace{1.5cm}
-\gamma\rho_+\left(e^{-(p_+-p_-)}-1\right)
-\gamma\rho_-\left(e^{p_+-p_-}-1\right)
+p_+\partial_t\rho_+
+p_-\partial_t\rho_-
-\bar J_+\partial_xp_+
-\bar J_-\partial_xp_-
\Bigg]
\nn\\
&&
 +
\int_0^{t_f}dt\,
\left[
p_+J_+^{\rm sp}
+
p_-J_-^{\rm sp}
\right]_{L_1}^{L_2} 
\;.
\label{action_after_JK_general}
\eea
The last term is a spatial boundary contribution generated when
$p_\pm\partial_xJ_\pm$ is integrated by parts in deriving
\eqref{saddle_point_J}: it is written in the case of an interval $[L_1,L_2]$. On the
infinite line $L_1 \to -\infty$, $L_2 \to +\infty$, this spatial boundary term vanishes under the appropriate
decay conditions. On a finite interval it must in general be retained and
treated according to the boundary conditions. This will be discussed in
Section~\ref{sec:MFT_reservoirs} and, for RTPs, in
Appendix~\ref{app:RTP_boundary_conditions}.
Also, note that the action ${\cal S}$ in \eqref{action_after_JK_general} 
depends implicitly on $\lambda$ since the saddle-point currents
\eqref{saddle_point_J} have already been substituted into it. This
dependence is encoded in the shifted fields $\hat p_\pm$ through
\eqref{defhatp}. For observables linear in the currents, it is often more
convenient to combine these two contributions and work directly with the
tilted action $\widetilde{\cal S}={\cal S}-\lambda{\cal O}$, as we will do
in Section~\ref{sec:MFT_reservoirs}.

The bulk part of the action can be rewritten in canonical form as
\be \label{action_general}
\mathcal S
=
\int_0^{t_f}dt
\int dx\,
\left(
p_+\partial_t\rho_+
+
p_-\partial_t\rho_-
\right)
-
H
+
\int_0^{t_f}dt\,
\left[
p_+J_+^{\rm sp}
+
p_-J_-^{\rm sp}
\right]_{L_1}^{L_2}
\;,
\qquad
H=\int_0^{t_f}dt\,{\sf H} \;,
\ee
where ${\sf H}$ is the instantaneous bulk Hamiltonian,
\bea \label{H_general}
{\sf H}
&=&
\int dx\,
\Bigg[
2D\rho_+\partial_x\hat p_+
\left(
\partial_xp_+
-\frac{1}{2}\partial_x\hat p_+
\right)
+
2D\rho_-\partial_x\hat p_-
\left(
\partial_xp_-
-\frac{1}{2}\partial_x\hat p_-
\right)
\nn\\
&&\hspace{1.5cm}
+
\bar J_+\partial_xp_+
+
\bar J_-\partial_xp_-
+
\gamma\rho_+
\left(
e^{-(p_+-p_-)}-1
\right)
+
\gamma\rho_-
\left(
e^{p_+-p_-}-1
\right)
\Bigg] \;.
\eea

We want to calculate the cumulant generating function
\be \label{CGFdef}
\Sigma(\lambda)=\ln\langle e^{\lambda\mathcal O}\rangle\;.
\ee
In the corresponding large-deviation regime, the functional integral can be evaluated
by a saddle-point approximation\footnote{Here and below we use $\sup$ although, strictly speaking, it is an extremization over $p_\pm$.}
\be \label{CGF_SP}
\Sigma(\lambda)
\simeq
\sup_{\rho_\pm,p_\pm}
\left[
\lambda\mathcal O-\mathcal S
+\ln P( \vec\rho(x,0))
\right]\;.
\ee

The optimal trajectories which maximize the argument of the exponential are obtained
by solving the saddle-point equations 
\be \label{SPeq_general0}
\partial_t\rho_\pm
=
\frac{\delta H}{\delta p_\pm(x,t)}
+\lambda\frac{\delta\mathcal O}{\delta p_\pm(x,t)}
\quad,\quad
\partial_t p_\pm
=
-\frac{\delta H}{\delta\rho_\pm(x,t)}
-\lambda\frac{\delta\mathcal O}{\delta\rho_\pm(x,t)}\;,
\ee
valid for $0<t<t_f$. 
Here $H=\int_0^{t_f}dt\,{\sf H}$ denotes the
time-integrated bulk Hamiltonian functional defined in \eqref{action_general}-\eqref{H_general}. The spatial surface term in
\eqref{action_general} does not modify these bulk equations, but must be
retained when deriving the spatial boundary conditions.
The initial and final time conditions, arising from the temporal boundary terms, are
\bea \label{BCgeneral}
&&p_\pm(x,0)
=
-\lambda\frac{\delta\mathcal O}{\delta\rho_\pm(x,0)}
-\frac{\delta\ln P(\vec\rho(x,0))}
{\delta\rho_\pm(x,0)}
\;,\\
&&p_\pm(x,t_f)
=
\lambda\frac{\delta\mathcal O}{\delta\rho_\pm(x,t_f)}
\;. \label{FCgeneral}
\eea

The initial condition \eqref{BCgeneral} is only valid for an {\it annealed} initial
condition, i.e. when the initial densities $\rho_\pm(x,0)$ are indeed drawn from some
probability density $P(\vec\rho(x,0))$. Another possibility is to consider a
{\it quenched} initial condition, by fixing the value of the initial densities.
In this case the initial condition simply reads
\be
\rho_\pm(x,0)=\rho_\pm^0(x)\;.
\ee
One may also consider a mixed initial condition, where the total density
$\rho_s(x,0)$ is fixed but the initial orientation of the particles is random.

Once the optimal trajectories have been obtained, the cumulant generating function
can be calculated from the relation
\be
\Sigma'(\lambda)=\mathcal O^*
\quad,\quad
\Sigma(0)=0\;, \label{SigmaPrime}
\ee
where $\mathcal O^*$ denotes the value of $\mathcal O$ at the optimum in
\eqref{CGF_SP}. 
The first relation follows by differentiating \eqref{CGF_SP}. The
variations of the optimal fields do not contribute because of the
saddle-point equations. Moreover, although both $\mathcal S$ and
$\mathcal O$ depend implicitly on $\lambda$ through the saddle-point
currents $J_\pm^{\rm sp}$, the corresponding terms proportional to
$\partial_\lambda J_\pm^{\rm sp}$ cancel by virtue of the current
saddle-point equation \eqref{saddle_point_J}.
A Legendre--Fenchel transform then gives, at the same leading
exponential order, the large-deviation rate function,
\be \label{LegendreTransform}
P(\mathcal O)\, \asymp\,e^{-I(\mathcal O)}
\quad,\quad
I(\mathcal O)
=
\sup_\lambda
\left[
\lambda\mathcal O-\Sigma(\lambda)
\right]\;.
\ee

In the two remaining sections, we will use these saddle-point equations in two
different settings to analyze the large deviations of the observables
\eqref{current_through_zero_def} and \eqref{integrated_current_def}, respectively,
first without and then with interactions.

\section{MFT for the step initial condition on the full real line} \label{sec:MFT_step}

\subsection{General setting and saddle-point equations}

In this section, we consider particles on the real line. We consider an initial condition where the mean density is a step function,
\be \label{meanIC}
\langle \rho_s(x,t=0) \rangle_{\rm IC} = r(x) \quad , \quad
r(x) = \rho_0 \theta(-x) + \rho_1 \theta(x) \;,
\ee
where $\langle \cdot \rangle_{\rm IC}$ denotes an average over the initial condition.
The case $\rho_1=0$ is the one considered in \cite{FreeRTPs2020} for free RTPs.
There are several ways to implement this initial condition: one can allow some fluctuations around this mean (annealed case) or enforce a deterministic initial condition (quenched case). For RTPs, one can also choose whether the initial orientations are random or fixed. We will discuss several of these possible choices.

We want to study the fluctuations of the observable
\be \label{current_through_zero_def2}
Q(t_f) = \int_0^{t_f} dt\, J_s(0,t) = \int_0^{+\infty} dx \, [\rho_s(x,t_f)-\rho_s(x,0)] \;,
\ee
namely of the integrated current flowing through the origin between $t=0$ and $t=t_f$. For free RTPs, the cumulants of this quantity grow as $\sqrt{t_f}$ at large times, as found in \cite{FreeRTPs2020}. More generally, however, interactions can modify this scaling, as already seen for the second cumulant in Section~\ref{sec:typical_current}. We therefore keep the large-deviation speed implicit in the general discussion below.

Using the second representation of $Q(t_f)$ in \eqref{current_through_zero_def2}, the observable $\mathcal{O}=Q(t_f)$ depends only on the densities at $t=0$ and $t=t_f$, and has no explicit dependence on the currents $J_\pm$. Hence, \eqref{defhatp} gives $\hat p_\pm=p_\pm$, and the saddle-point equations \eqref{SPeq_general0} read
\bea \label{SPeq_general1}
\partial_t \rho_+ &=& -\partial_x [ 2D \rho_+ \partial_x p_+ + \bar J_+ ] -\gamma \rho_+ e^{-(p_+-p_-)} +\gamma \rho_- e^{p_+-p_-} \;, \\
\partial_t \rho_- &=& -\partial_x [ 2D \rho_- \partial_x p_- + \bar J_- ] -\gamma \rho_- e^{p_+-p_-} + \gamma \rho_+ e^{-(p_+-p_-)} \;, \\
\partial_t p_+ &=& -D (\partial_x p_+)^2 + [-v_0 + V'(x) - \mathcal{H}[\rho_s](x)] \partial_x p_+ -D\partial_x^2 p_+ - \gamma (e^{-(p_+-p_-)}-1) \\
&& \hspace{0.5cm} + \int dy \, W'(y-x) [\rho_+(y) \partial_y p_+(y) + \rho_-(y) \partial_y p_-(y)] \;, \nn \\
\partial_t p_- &=& -D (\partial_x p_-)^2 + [v_0 + V'(x) - \mathcal{H}[\rho_s](x) ] \partial_x p_- -D\partial_x^2 p_- - \gamma (e^{p_+-p_-}-1) \label{SPeq_general4} \\
&& \hspace{0.5cm} + \int dy \, W'(y-x) [\rho_+(y) \partial_y p_+(y) + \rho_-(y) \partial_y p_-(y)] \;, \nn 
\eea
where $\bar J_\pm$ are defined in \eqref{defJbar}. Note that if instead one uses the first current representation of $Q(t_f)$ in \eqref{current_through_zero_def2}, the observable depends explicitly on $J_s$ and $\hat p_\pm$ are not equal to $p_\pm$. The two representations are of course equivalent, but here we consistently use the density representation.

In addition, the final time condition \eqref{FCgeneral} becomes
\be \label{final_time_step}
p_\pm(x,t_f) = \lambda \theta(x) \;.
\ee
The initial condition will be discussed below.

Written in terms of $\rho_{s/d}$ and $p_{s/d}$, the saddle-point equations (\ref{SPeq_general1})-(\ref{SPeq_general4}) become (using the relations \eqref{pm_to_sd})
\bea
\partial_t \rho_s &=&  -\partial_x [ 2D \rho_s \partial_x p_s + 2D \rho_d \partial_x p_d + \bar J_s ] \;,\\
\partial_t \rho_d &=& -\partial_x [ 2D \rho_s \partial_x p_d + 2D \rho_d \partial_x p_s + \bar J_d ] + 2\gamma \rho_s \sinh(2p_d) - 2\gamma \rho_d \cosh(2p_d) \;, \\
\partial_t p_s &=&  - D (\partial_x p_s)^2 - D (\partial_x p_d)^2 -v_0\partial_xp_d +
[V'(x)-\mathcal{H}[\rho_s](x)]\partial_xp_s
-D\partial_x^2p_s-2\gamma\sinh^2(p_d)\nonumber\\
&&+\int dy\,W'(y-x)\left[\rho_s(y)\partial_y p_s(y)+\rho_d(y)\partial_y p_d(y)\right] \;, \\
\partial_t p_d &=& -2D (\partial_x p_s) (\partial_x p_d) - v_0 \partial_x p_s + [V'(x) -\mathcal{H}[\rho_s](x)] \partial_x p_d - D\partial_x^2 p_d + \gamma \sinh(2p_d) \;,
\eea
with $\bar J_s$ and $\bar J_d$ defined in \eqref{defJsbar}-\eqref{defJdbar}, and with the final time condition
\be
p_s(x,t_f) = \lambda \theta(x) \quad , \quad p_d(x,t_f) = 0 \;.
\ee

In the rest of this section, we solve these saddle-point equations both for Brownian and run-and-tumble particles, for different types of initial conditions. We start with the non-interacting case, for which a Cole--Hopf transformation allows  us to obtain a simple set of forward and backward Fokker--Planck equations. We then add interactions using a perturbative approach. Once the solutions are obtained, the cumulant generating function can be deduced in each case by calculating the integrated current at the saddle point $Q^*(t_f)$ and using the relation \eqref{SigmaPrime}. Before that, let us discuss the initial conditions.

\subsection{Annealed versus quenched initial condition}

We now introduce the different types of initial conditions that we will be considering in this section.
\\

\noindent {\bf Annealed case.} We start with the annealed case, where at $t=0$ the particles are independently and uniformly distributed on the interval $[-L,0]$ with density $\rho_0$, and on the interval $[0,L]$ with density $\rho_1$, and we take the limit $L\to+\infty$ with $\rho_0$ and $\rho_1$ fixed. In addition, each particle is in the $+$ or $-$ state with equal probability $1/2$. In the thermodynamic limit, this corresponds locally to two independent Poisson point processes of intensities $r(x)/2$ for the two orientations.
In the non-interacting case, this can be interpreted as placing a reflecting boundary at $x=0$, letting the system reach the stationary state, and then suddenly removing the wall at $t=0$. This interpretation does not hold in the presence of interactions as the stationary state will generally differ from a uniform distribution. In this case, a different form of annealed initial condition can be considered, as we will discuss in Section~\ref{sec:step_interaction}. For now, we focus on the uniform case.

Since the number of particles of orientation $\sigma$ inside a small interval $[x,x+dx]$ at time $t=0$ follows a Poisson distribution with parameter $r(x)dx/2$, where $r(x)$ is the step density defined in \eqref{meanIC}, we have in the large deviation limit
\be
-\ln P(\vec{\rho}(x,0)) \simeq \frac{1}{2} \sum_{\sigma=\pm1} \int_{-\infty}^\infty dx \left(2\rho_\sigma(x,0) \ln \frac{2\rho_\sigma(x,0)}{r(x)} -2\rho_\sigma(x,0) + r(x) \right) \;.
\ee
This is the sum of the local relative entropies of two independent Poisson fields of intensity $r(x)/2$.
From \eqref{BCgeneral}, we then have the initial condition
\be \label{BCannealed1}
p_\pm(x,0) = \lambda \theta(x) + \ln \left( \frac{2\rho_\pm(x,0)}{r(x)} \right) \;.
\ee
\\

\noindent {\bf Quenched positions and annealed orientations.}
In addition to the annealed case, the authors of \cite{FreeRTPs2020} consider a ``quenched'' initial condition, where the particles are equally spaced at $t=0$, i.e. they enforce the deterministic condition
\be \label{IC_quenched_rhos}
\rho_s(x,0) = r(x) \;.
\ee
However, in their setting the orientations of the particles are left to fluctuate as in the annealed case, i.e. each particle starts in the $+$ or $-$ state with equal probability. This distinction between quenched positions (or density) and quenched orientations was further emphasized in Refs.~\cite{JoseRossoRamola2023,JoseRossoRamola2024}.
The large deviations of $\rho_d(x,0)$ are thus given by the conditional entropy corresponding to a binomial distribution,
\bea
-\ln P(\rho_d(x,0)|\rho_s(x,0)) &=& \int_{-\infty}^\infty dx \left(\rho_+ \ln \frac{2\rho_+}{\rho_s} + \rho_- \ln \frac{2\rho_-}{\rho_s} \right) \\
&=& \int_{-\infty}^\infty dx \left[ \frac{\rho_s+\rho_d}{2} \ln \left(1+\frac{\rho_d}{\rho_s} \right) + \frac{\rho_s-\rho_d}{2} \ln \left(1-\frac{\rho_d}{\rho_s} \right) \right] \;.
\eea
This leads to
\be \label{IC_quenched_pd}
p_d(x,0) = -\lambda \frac{\delta Q(t_f)}{\delta\rho_d(x,0)} - \frac{\delta\ln  P(\rho_d(x,0)|\rho_s(x,0))}{\delta \rho_d(x,0)}
= \frac{1}{2} \ln \left( \frac{\rho_s(x,0)+\rho_d(x,0)}{\rho_s(x,0)-\rho_d(x,0)} \right)
= \frac{1}{2} \ln \left( \frac{\rho_+(x,0)}{\rho_-(x,0)} \right) \;.
\ee
Here $\delta Q/\delta\rho_d(x,0)=0$, since $Q$ depends only on the total density $\rho_s$.
In terms of $\rho_\pm$ and $p_\pm$, the initial condition can thus be summarized as
\bea \label{IC_quenched_rhopm}
&&\rho_\pm(x,0)= \frac{\rho_s(x,0) \pm \rho_d(x,0)}{2} = \frac{r(x) \pm \rho_d(x,0)}{2} \;, \\
&&p_\pm(x,0) = p_s(x,0) \pm p_d(x,0) = p_s(x,0) \pm \frac{1}{2} \ln \left( \frac{\rho_+(x,0)}{\rho_-(x,0)} \right) \;. \label{IC_quenched_ppm}
\eea
The fields $\rho_d(x,0)$ and $p_s(x,0)$ are unknown, but we have enough equations to determine them. In particular, no
initial condition is imposed on $p_s(x,0)$ since its conjugate field $\rho_s(x,0)$ is fixed.
\\

\noindent {\bf Quenched positions and orientations.}
Instead of letting the orientations of the particles fluctuate, we now prepare them in a specific state, which amounts to fixing the densities $\rho_\pm(x,0)$. This corresponds to a fully quenched initial condition for both the density and orientation fields. In this case, we can simply replace the initial condition on $p_\pm$ by an initial condition on $\rho_\pm$,
\be \label{fullquenched}
\rho_\pm(x,0) = r_\pm(x) \quad , \quad
r_\pm(x) = \rho_0^\pm \theta(-x) + \rho_1^\pm \theta(x) \;,
\ee
with $\rho_0^++\rho_0^-=\rho_0$ and $\rho_1^++\rho_1^-=\rho_1$.
\\

In the rest of this section, we will denote respectively by $\Sigma_{\rm an}$, $\Sigma_{\rm qu}$ and $\Sigma_{\rm or}$ 
the cumulant generating functions corresponding to these different initial conditions, and similarly for the large deviation rate functions (where the subscripts stand for ``annealed'', ``quenched'' and ``orientation'' respectively).
\\

\subsection{Non-interacting case} \label{sec:MFT_step_noninteracting}

We now solve the saddle-point equations \eqref{SPeq_general1}-\eqref{SPeq_general4} for the different initial conditions defined above. Before introducing interactions, we consider the case of free, non-interacting particles, i.e. $V'(x)=0$ and $W'(x)=0$.
Before going to the RTP case, it is useful to first review the calculation for Brownian particles, $v_0=0$, using the method of Derrida and Gerschenfeld~\cite{Derrida2009}, based on a Cole-Hopf transformation. This method is then generalized to the RTP case in the following sections.

\subsubsection{Non-interacting Brownian particles}\label{sec:Brownian}

Before going to the RTP case, we consider Brownian particles, $v_0=0$. In this case one can consistently set $\rho_d=0$ and $p_d=0$. Denoting $\rho=\rho_s$ and $p=p_s$, 
the saddle-point equations then simplify to (see also Appendix
\ref{app:Brownian_boundary_conditions} for the direct derivation from the Brownian MFT action)
\bea \label{eqopt_brownian1}
\partial_t \rho &=& \partial_x [ D\partial_x\rho  - 2D \rho \partial_x p ] \;, \\
\partial_t p &=& -D (\partial_x p)^2 -D\partial_x^2 p \;,
\label{eqopt_brownian2}
\eea
with the final time condition
\be \label{final_time_br}
p(x,t_f) = \lambda \theta(x) \;.
\ee
Of course, in this case the initial condition for the orientation is irrelevant and we will only consider two initial conditions: annealed (uniform) 
\be \label{BCannealedBr1}
p(x,0) = \lambda \theta(x) + \ln \left( \frac{\rho(x,0)}{r(x)} \right) \;,
\ee
and quenched
\be \label{BCquenchedBr1}
\rho(x,0) = r(x) = \rho_0 \theta(-x) + \rho_1 \theta(x) \;.
\ee
The results will serve as a benchmark for the RTP case. This calculation should allow us to recover the results of Derrida and Gerschenfeld~\cite{Derrida2009} for non-interacting diffusive particles with a step initial condition: a difference of two independent Poisson processes in the annealed case (reducing to a Poisson distribution when $\rho_1=0$), and a distinct non-trivial cumulant generating function in the quenched case.
\\

\noindent {\bf Cole--Hopf transformation.} The equations \eqref{eqopt_brownian1}-\eqref{eqopt_brownian2} can be solved by performing a standard Cole--Hopf transformation \cite{KrapivskyVoid,Elgart2004,Derrida2009},
\be
F=\rho e^{-p} \quad , \quad G=e^p \;,
\ee
which simplifies the problem into a pair of diffusion equations (forward and backward)
\bea
\partial_t F &=& D \partial_x^2 F \;, \\
\partial_t G &=& - D \partial_x^2 G \;.
\eea
The final time condition \eqref{final_time_br} implies
\bea \label{final_time_brownian1}
F(x,t_f)&=&\rho(x,t_f) e^{-\lambda\theta(x)}
= \rho(x,t_f) [\theta(-x)+e^{-\lambda}\theta(x)] \;, \\
G(x,t_f)&=&e^{\lambda\theta(x)}
= \theta(-x)+e^{\lambda}\theta(x) \;.
\label{final_time_brownian2}
\eea
Note that the final time condition \eqref{final_time_br} fixes the backward field $G$,
whereas the expression for $F(x,t_f)$ is only an identity and not an independent boundary
condition. The field $F$ is instead fixed at the initial time in the annealed case, whereas
in the quenched case its initial value is determined in terms of $G(x,0)$ through the
prescribed density, as discussed below.
\\

\noindent {\bf Annealed initial condition.} Written in terms of the variables $F$ and $G$, the annealed initial condition \eqref{BCannealedBr1} directly fixes $F(x,0)$ and implies
\bea
F(x,0)&=& \rho_0\theta(-x) +\rho_1 e^{-\lambda} \theta(x) \;, \\ 
G(x,0)&=& \frac{\rho(x,0)}{\rho_0}\theta(-x) + \frac{\rho(x,0)}{\rho_1} e^{\lambda} \theta(x) \;.
\eea
The second relation is only formal when one of the densities $\rho_0$ or $\rho_1$ vanishes. In what follows only the initial condition for $F$ and the final condition for $G$ are needed to solve the diffusion equations.

Solving the diffusion equation using the initial condition for $F(x,t)$ and the final time condition for $G(x,t)$ leads to
\bea \label{F_Brownian_annealed}
F(x,t)&=&
\frac{\rho_1 e^{-\lambda} + \rho_0}{2} + \frac{\rho_1 e^{-\lambda} - \rho_0}{2}{\rm erf}\Big(\frac{x}{\sqrt{4Dt}}\Big) \;, \\
G(x,t)&=&\frac{e^\lambda -1}{2} {\rm erf}\Big(\frac{x}{\sqrt{4D(t_f-t)}}\Big) + \frac{e^\lambda +1}{2} \;. \label{G_Brownian_annealed}
\eea
Using the inverse transformation $\rho=FG$ and $p=\ln G$ and inserting into \eqref{current_through_zero_def2}, we obtain the saddle-point value
\be
Q^*(t_f)= \frac{\rho_0 e^{\lambda}-\rho_1 e^{-\lambda}}{2} \int_0^{+\infty} dx \, {\rm erfc}\Big(\frac{x}{\sqrt{4Dt_f}}\Big)
= (\rho_0 e^\lambda - \rho_1 e^{-\lambda}) \sqrt{\frac{Dt_f}{\pi}}\;.
\ee
Using the relation \eqref{SigmaPrime} (which here reads $\Sigma_{\rm an}'(\lambda)=Q^*(t_f)$) and integrating over $\lambda$ (with $\Sigma_{\rm an}(0)=0$) finally gives us the expression of the cumulant generating function
\be \label{sigmafinal_brownian_annealed}
\Sigma_{\rm an}(\lambda) \equiv \ln \, \langle e^{\lambda Q(t_f)} \rangle \simeq \big[\rho_0(e^\lambda -1) + \rho_1 (e^{-\lambda} -1) \big] \sqrt{\frac{Dt_f}{\pi}}\;.
\ee
This is precisely the annealed Brownian result of Derrida and Gerschenfeld~\cite{Derrida2009}, i.e. the cumulant generating function of the difference of two independent Poisson variables, with respective means $\rho_0\sqrt{Dt_f/\pi}$ and $\rho_1\sqrt{Dt_f/\pi}$. Notice that, since there are no interactions, the particles initially on the two sides contribute independently to opposite current directions, which will also be the case in the next few examples.

For $\rho_0,\rho_1>0$, the expression \eqref{sigmafinal_brownian_annealed} also satisfies the fluctuation symmetry emphasized in Ref.~\cite{Derrida2009},
\be
\Sigma_{\rm an}(\lambda)
=
\Sigma_{\rm an}\left(\ln\frac{\rho_1}{\rho_0}-\lambda\right) 
\quad , \quad 
I_{\rm an}(Q)-I_{\rm an}(-Q)
=
Q\ln\frac{\rho_1}{\rho_0} \;.
\ee

Performing a Legendre transform as explained in \eqref{LegendreTransform}, we obtain the large deviation form $P_{\rm an}(Q(t_t) \simeq Q)\,{\asymp}\, e^{-I_{\rm an}(Q)}$ with
\be
I_{\rm an}(Q) = \rho_0 \sqrt{\frac{Dt_f}{\pi}} \Psi_{\rm an} \left( \sqrt{\frac{\pi}{Dt_f}} \frac{Q}{\rho_0} \right) \quad , \quad \Psi_{\rm an}(q) = \lambda_q q - \tilde \Sigma_{\rm an}(\lambda_q) \quad , \quad \tilde \Sigma_{\rm an}(\lambda) = (e^\lambda -1) + \frac{\rho_1}{\rho_0} (e^{-\lambda} -1) \;,
\ee
where
\be
\lambda_q = \ln \left( \frac{q}{2} + \sqrt{\frac{q^2}{4} + \frac{\rho_1}{\rho_0}} \, \right) \;. 
\ee
Equivalently, writing $b=\rho_1/\rho_0$, the rate function can be written explicitly as
\be \label{psi_annealed_br_step}
\Psi_{\rm an}(q)
=
q\ln\left(\frac{q+\sqrt{q^2+4b}}{2}\right)
-\sqrt{q^2+4b}+1+b \;.
\ee
In the case where $\rho_1=0$, the function $\Psi_{\rm an}(q)$ simplifies to 
\be
\Psi_{\rm an}(q) = q \ln q - q + 1 \;, \qquad q\geq0 \;,
\ee
which is simply the rate function of a Poisson process. In this case $Q\geq0$ and the rate function is infinite for $q<0$.
\\

\noindent {\bf Quenched initial condition.} We now consider the quenched initial condition \eqref{BCquenchedBr1}, which in terms of the variables $F$ and $G$ reads
\bea
F(x,0)&=& [\rho_0\theta(-x)+\rho_1 \theta(x)] e^{-p(x,0)} = \frac{\rho_0\theta(-x)+\rho_1 \theta(x)}{G(x,0)} \;, \\
G(x,0)&=& e^{p(x,0)} \;, 
\eea
Since the final time condition is the same as in the annealed case, $G(x,t)$ is still given by \eqref{G_Brownian_annealed}, while $F(x,t)$ now reads
\be \label{F_Brownian_quenched}
F(x,t) = \int_{-\infty}^{+\infty} dy \, \frac{e^{-\frac{(x-y)^2}{4Dt}}}{\sqrt{4\pi Dt}} \frac{r(y)}{G(y,0)} \;.
\ee 
Inserting into the expression for the integrated current \eqref{current_through_zero_def2} with $\rho=FG$, we get at the saddle point
\bea
Q^*(t_f)
&=& \int_0^{+\infty} dx \, [e^{\lambda} F(x,t_f) - r(x)] \nn \\
&=& \int_{-\infty}^{+\infty} dy \, r(y) \int_0^{+\infty} dx
\left[
\frac{e^{-\frac{(x-y)^2}{4Dt_f}}}{\sqrt{4\pi Dt_f}}
\frac{e^\lambda}{G(y,0)}-\delta(x-y)
\right] \nn \\
&=& \rho_0 \int_{-\infty}^{0} dy \,
\frac{\frac{e^\lambda}{2}
\left(1+{\rm erf}\big(\frac{y}{\sqrt{4Dt_f}}\big)\right)}
{\frac{e^\lambda +1}{2}
+\frac{e^\lambda -1}{2}
{\rm erf}\big(\frac{y}{\sqrt{4Dt_f}}\big)}
+\rho_1 \int_0^{+\infty} dy \,
\frac{\frac{e^{-\lambda}}{2}
\left({\rm erf}\big(\frac{y}{\sqrt{4Dt_f}}\big)-1\right)}
{\frac{1+e^{-\lambda}}{2}
+\frac{1-e^{-\lambda}}{2}
{\rm erf}\big(\frac{y}{\sqrt{4Dt_f}}\big)}
\;.
\eea
Using again the relation \eqref{SigmaPrime} and integrating over $\lambda$ gives us the cumulant generating function (after a change of variables $u=y/\sqrt{Dt_f}$)
\be \label{sigma_free_brownian_quenched}
\!\!\!\Sigma_{\rm qu}(\lambda) = \rho_0 \sqrt{Dt_f} \int_{-\infty}^{0} du \, \ln \left(\frac{e^\lambda +1}{2} +\frac{e^\lambda -1}{2} \, {\rm erf}\Big(\frac{u}{2}\Big) \right) + \rho_1 \sqrt{Dt_f} \int_0^{+\infty} du \, \ln \left(\frac{1+e^{-\lambda}}{2} +\frac{1-e^{-\lambda}}{2} \, {\rm erf}\Big(\frac{u}{2}\Big)\right) .
\ee
This result coincides with equation~(51) of Derrida--Gerschenfeld~\cite{Derrida2009} after restoring the diffusion coefficient $D$, which is set to unity there.
Applying a Legendre transform gives us the large deviation function
\be \label{LD_free_Br_quenched}
I_{\rm qu}(Q) = \rho_0 \sqrt{\frac{Dt_f}{\pi}} \Psi_{\rm qu} \left( \sqrt{\frac{\pi}{Dt_f}} \frac{Q}{\rho_0} \right) \quad , \quad \Psi_{\rm qu}(q) = \sup_\lambda \left[ \lambda q - \tilde \Sigma_{\rm qu}(\lambda) \right] \;,
\ee 
where
\be
\tilde \Sigma_{\rm qu}(\lambda) = \sqrt{\pi} \int_0^{+\infty}du\, \left[ \ln\left( 1+\frac{e^\lambda-1}{2}{\rm erfc}\left(\frac{u}{2}\right) \right) + \frac{\rho_1}{\rho_0} \ln\left(
1+\frac{e^{-\lambda}-1}{2}{\rm erfc}\left(\frac{u}{2}\right) \right) \right] .
\ee

It is useful to compare the typical fluctuations in the two ensembles. Expanding the two cumulant generating functions around $\lambda=0$ gives
$\langle Q(t_f)\rangle_{\rm an}
=
\langle Q(t_f)\rangle_{\rm qu}
=
(\rho_0-\rho_1)\sqrt{\frac{Dt_f}{\pi}}$,
while the variances are
\be
\langle Q(t_f)^2\rangle_{\rm c,an}
=
(\rho_0+\rho_1)\sqrt{\frac{Dt_f}{\pi}}
\quad,\quad
\langle Q(t_f)^2\rangle_{\rm c,qu}
=
(\rho_0+\rho_1)\sqrt{\frac{Dt_f}{2\pi}} \;.
\ee
Thus the annealed and quenched ensembles have the same mean current but the annealed
variance is larger by a factor $\sqrt{2}$, due to additional fluctuations in the 
initial density profile in that case.

The distinction becomes even stronger in the far tails. For simplicity, consider
$\rho_1=0$. In the annealed case one has the Poisson asymptotics
$\Psi_{\rm an}(q)\sim q\ln q$ for $q\to+\infty$. By contrast, from
\eqref{sigma_free_brownian_quenched} one finds
$\tilde\Sigma_{\rm qu}(\lambda)
\simeq \frac{4\sqrt{\pi}}{3}\lambda^{3/2}$
as $\lambda\to+\infty$, 
which yields
\be \label{psi_qu_diff}
\Psi_{\rm qu}(q)
\simeq \frac{q^3}{12\pi}
{\quad} {\rm as} \qquad q\to+\infty \quad , \quad {\rm implying} \qquad  
P_{\rm qu}(Q(t_f) \simeq Q)
\asymp
\exp\left[-\frac{Q^3}{12\rho_0^2Dt_f}\right]
\ee
in the extreme positive-current regime. This cubic tail is the independent-particle
example of the super-Gaussian extreme-current behavior discussed by Derrida and
Gerschenfeld~\cite{Derrida2009}. In the annealed problem an atypically favorable initial configuration can help produce a large current, whereas in the quenched problem the particles must produce the same fluctuation dynamically from a fixed initial density profile.

It is also useful to contrast this result with the interacting SSEP studied exactly by
Derrida and Gerschenfeld in Ref.~\cite{DerridaGerschenfeldSSEP2009}. There, all current
cumulants again grow as $\sqrt{t_f}$, but the cumulant generating function is no longer
that of independent Poisson processes. In the following RTP calculation, activity introduces yet another
distinction: although the long-time typical fluctuations become diffusive, the finite
propagation speed remains visible in sufficiently large current fluctuations.

The above exact solution also allows to compute the {\it optimal space-time histories} of the density field $\rho(x,t)$ for a fixed value $Q$ of the current. The calculation is performed 
in Appendix \ref{app:density_br}. The results have been discussed in great detail in the introduction section \ref{sec:intro_results}.

\subsubsection{Non-interacting RTPs}

We now consider the same problem for free RTPs. For simplicity we do not include any additional thermal noise, i.e. we set $D=0$. The saddle-point equations \eqref{SPeq_general1}-\eqref{SPeq_general4} now read
\bea \label{SPeq_RTPstep1}
\partial_t \rho_+ &=& -v_0 \partial_x \rho_+ -\gamma \rho_+ e^{-(p_+-p_-)} +\gamma \rho_- e^{p_+-p_-} \;, \\
\partial_t \rho_- &=& v_0 \partial_x \rho_- -\gamma \rho_- e^{p_+-p_-} + \gamma \rho_+ e^{-(p_+-p_-)} \;, \\
\partial_t p_+ &=& -v_0 \partial_x p_+ - \gamma (e^{-(p_+-p_-)}-1) \;,\\
\partial_t p_- &=& v_0 \partial_x p_- - \gamma (e^{p_+-p_-}-1) \;, \label{SPeq_RTPstep4}
\eea
with the final time condition \eqref{final_time_step}. As in the Brownian case, our goal is to obtain the cumulant generating function and large deviation rate function of the integrated current through the origin, defined in \eqref{current_through_zero_def2}, for a step initial condition \eqref{meanIC}. In the limit $\rho_1\to0$, we should be able to recover the results from \cite{FreeRTPs2020}, both for an annealed initial condition and in the quenched case with annealed orientations. As discussed above, since the particles have a persistent motion, there is a third type of initial condition not considered in \cite{FreeRTPs2020}, namely the case of quenched positions and quenched orientations, which we will also consider. This case, together with other types of annealed/quenched choices for the initial orientation field, was subsequently studied directly at the microscopic level in \cite{JoseRossoRamola2023,JoseRossoRamola2024}.
As we will see, all these cases lead to different forms of the large deviations.
\\

\noindent {\bf Cole--Hopf transformation.} We now use a straightforward generalization of the Cole--Hopf transformation used above for Brownian particles, by introducing the variables $F_\pm=\rho_\pm e^{-p_\pm}$ and $G_\pm=e^{p_\pm}$. The saddle-point equations \eqref{SPeq_RTPstep1}-\eqref{SPeq_RTPstep4} become
\bea \label{eqFp_freeRTP}
\partial_t F_+ &=& -v_0 \partial_x F_+ -\gamma F_+ +\gamma F_- \;, \\
\partial_t F_- &=& v_0 \partial_x F_- -\gamma F_- +\gamma F_+ \;, \label{eqFm_freeRTP} \\
-\partial_t G_+ &=& v_0 \partial_x G_+ - \gamma G_+ + \gamma G_- \;, \label{eqGp_freeRTP} \\
-\partial_t G_- &=& -v_0 \partial_x G_- - \gamma G_- + \gamma G_+ \;. \label{eqGm_freeRTP}
\eea
These are simply the equations obeyed by the propagator of a free RTP, forward in time for $F_\pm$, and backward in time for $G_\pm$. These equations can be rewritten by introducing $F_s=F_++F_-$ and $F_d=F_+-F_-$ and similarly for $G_s$ and $G_d$,
\bea \label{eqFs_freeRTP}
\partial_t F_s &=& -v_0 \partial_x F_d \;, \\
\partial_t F_d &=& -v_0 \partial_x F_s -2\gamma F_d \;, \label{eqFd_freeRTP}\\
-\partial_t G_s &=& v_0 \partial_x G_d \;, \label{eqGs_freeRTP}\\
-\partial_t G_d &=& v_0 \partial_x G_s - 2\gamma G_d \;. \label{eqGd_freeRTP}
\eea
The final time condition \eqref{final_time_step} implies
\bea \label{FGpm_final}
F_\pm(x,t_f)&=&\rho_\pm(x,t_f) e^{-\lambda\theta(x)}
= \rho_\pm(x,t_f) [\theta(-x)+e^{-\lambda}\theta(x)] \;, \\
G_\pm(x,t_f)&=&e^{\lambda\theta(x)}
= \theta(-x)+e^{\lambda}\theta(x) \;,
\eea
or in terms of $F_{s/d}$ and $G_{s/d}$
\bea \label{Fsd_final}
F_s(x,t_f)&=& \rho_s(x,t_f) [\theta(-x)+e^{-\lambda}\theta(x)]
\quad , \quad
F_d(x,t_f)=\rho_d(x,t_f) [\theta(-x)+e^{-\lambda}\theta(x)] \;,\\
G_s(x,t_f)&=& 2[\theta(-x)+e^{\lambda}\theta(x)]
\quad , \quad
G_d(x,t_f)=0 \;.
\label{Gsd_final}
\eea
Note that, as in the Brownian case, the final time condition fixes the backward fields $G_\pm$, whereas the expressions for $F_\pm(x,t_f)$ are identities rather than independent boundary conditions. The forward fields $F_\pm$ will instead be fixed by the initial condition.

We will now solve these equations for the different types of initial conditions introduced above. The original variables can then be recovered as $\rho_\pm = F_\pm G_\pm$ and $p_\pm=\ln G_\pm$, which allows the calculation of the cumulant generating function through the relation \eqref{SigmaPrime}, as we have done in the Brownian case.
\\

\noindent {\bf Annealed case.} We start with the fully annealed case, where at time $t=0$ the particles again obey a Poisson distribution with average density $r(x)=\rho_0\theta(-x) + \rho_1 \theta(x)$, and each particle is a $+$ or a $-$ particle with equal probability. The corresponding initial condition for $p_\pm(x,0)$ is given in \eqref{BCannealed1}. In terms of $F_\pm$ and $G_\pm$, this condition reads
\bea
F_\pm(x,0)&=& \frac{\rho_0}{2} \theta(-x) + \frac{\rho_1}{2} e^{-\lambda} \theta(x) \;, \\
G_\pm(x,0)&=& \frac{2\rho_\pm(x,0)}{\rho_0}\theta(-x)
+ \frac{2\rho_\pm(x,0)}{\rho_1} e^{\lambda} \theta(x) \;,
\eea
or in terms of $F_{s/d}$ and $G_{s/d}$,
\bea \label{IC_step_annealed_Fs}
F_s(x,0)&=& \rho_0\theta(-x) + \rho_1 e^{-\lambda} \theta(x)
\quad , \quad F_d(x,0) = 0 \;, \\
G_s(x,0)&=& \rho_s(x,0)
\left[\frac{2}{\rho_0}\theta(-x)
+ \frac{2}{\rho_1} e^{\lambda} \theta(x) \right]
\quad , \quad
G_d(x,0) = \rho_d(x,0)
\left[\frac{2}{\rho_0}\theta(-x)
+ \frac{2}{\rho_1} e^{\lambda} \theta(x) \right] \;.
\label{IC_step_annealed_Gs}
\eea
As above, the relations for $G_{s/d}(x,0)$ are not independent boundary conditions and are only formal if one of the densities $\rho_0,\rho_1$ vanishes. The solution below requires only the initial conditions for $F_{s/d}$ and the final conditions for $G_{s/d}$.

Combining the equations \eqref{eqFs_freeRTP} and \eqref{eqFd_freeRTP} gives the telegrapher's equation
\be
\partial_t^2 F_s +2\gamma \partial_t F_s = v_0^2 \partial_x^2 F_s \;.
\ee
Solving this equation for the initial condition $P_s(x,t=0)=\delta(x-x_0)$ and $\partial_t P_s(x,t=0)=-v_0\partial_x P_d(x,t=0)=0$ gives the free propagator for the total density of an RTP, averaged over the initial orientations. 
In Fourier space and Laplace in time, denoting respectively the Fourier transform in $x$
by a hat and the Laplace transform in $t$ by a tilde, i.e. $\hat{\tilde P}_s(k,{\sf s})=\int_{-\infty}^{+\infty} dx e^{-ikx} \int_0^{+\infty} dt e^{-{\sf s}t} P_s(x,t)$,
it reads \cite{Masoliver1993}
\be \label{FreePropagator_FourierLaplace}
\hat{\tilde P}_s(k,{\sf s}|x_0) = \frac{{\sf s}+2\gamma}{{\sf s}^2+2\gamma {\sf s} + v_0^2 k^2} e^{-i k x_0} \;.
\ee
Inverting first the Fourier transform in space leads to
\be
\tilde P_s(x,{\sf s}|x_0) = \frac{1}{2v_0} \sqrt{\frac{{\sf s}+2\gamma}{{\sf s}}} e^{-\sqrt{{\sf s}({\sf s}+2\gamma)}\frac{|x-x_0|}{v_0}} \;.
\ee
Introducing ${\sf \tilde s}={\sf s}/\gamma$ and $\tilde t=\gamma t$ and using the Laplace inversion formula (C6) in \cite{LMS2019}, we obtain the propagator in real space and real time \cite{Masoliver1993,LMS2019}
\bea
P_s(x,t|x_0) &=& \frac{e^{-\gamma t}}{2} \Bigg[
\delta(x-x_0-v_0 t) + \delta (x-x_0+v_0 t)
+\theta(v_0 t - |x-x_0|)
\Bigg(
\frac{\gamma}{v_0}
I_0\Big(\sqrt{\gamma^2t^2 -
\Big(\frac{\gamma (x-x_0)}{v_0}\Big)^2}\Big)
\nn\\
&& \hspace{1cm}
+\frac{\gamma t}{\sqrt{v_0^2t^2 -(x-x_0)^2}}
I_1\Big(\sqrt{\gamma^2t^2 -
\Big(\frac{\gamma (x-x_0)}{v_0}\Big)^2}\Big)
\Bigg) \Bigg] \;.
\label{freeRTP_propagator}
\eea

Here, we have the initial condition \eqref{IC_step_annealed_Fs}, whose Fourier transform reads
\be \label{FsFourierLaplaceIC}
\hat F_s(k,0)
=(\rho_0+\rho_1 e^{-\lambda})\pi \delta(k)
+ \frac{i}{k}(\rho_0-\rho_1 e^{-\lambda}) \;.
\ee
The Fourier transform of the step function should be understood distributionally, with the usual principal value prescription for $1/k$. We can then use the propagator \eqref{FreePropagator_FourierLaplace} to write
\be \label{FsFourierLaplace}
\hat{\tilde F}_s(k,{\sf s}) =
\frac{{\sf s}+2\gamma}{{\sf s}^2+2\gamma {\sf s} + v_0^2 k^2}
\hat F_s(k,0) \;.
\ee
Inverting the Fourier transform we obtain
\be
\tilde F_s(x,{\sf s}) =
\frac{\rho_0 \theta(-x) + \rho_1 e^{-\lambda} \theta(x)}{{\sf s}}
+ \frac{1}{2} (\rho_0 - \rho_1 e^{-\lambda})
\, \sgn(x) \,
\frac{e^{-\sqrt{{\sf s}({\sf s}+2\gamma)} \frac{|x|}{v_0}}}{{\sf s}}  \;.
\ee
Inverting the Laplace transform then leads to
\be \label{Fs_real0}
F_s(x,t) =
\rho_0 \theta(-x) + \rho_1 e^{-\lambda} \theta(x)
+ (\rho_0 - \rho_1 e^{-\lambda})
\, \sgn(x) \, U(|x|,t) \;.
\ee
where
\be \label{Udef}
U(x,t) = \int_0^{+\infty} dz\, P_s(z,t|-x) \;.
\ee
The quantity $U(x,t)$ was introduced in \cite{FreeRTPs2020}. It is the probability that a particle starting at $X(0)=-x$ at time zero will be at position $X(t)>0$ at time $t$, i.e if $x>0$ it is a crossing probability. 
It is most easily defined through its Laplace transform in time
\be \label{UdefLaplace}
\tilde U(x,{\sf s}) =
\int_0^{+\infty} dt \, e^{-{\sf s}t} U(x,t)
=
\frac{e^{-\sqrt{{\sf s}({\sf s}+2\gamma)} \frac{x}{v_0}}}{2{\sf s}} \;.
\ee
For completeness, $U(x,t)$ can also be written explicitly in real time as
\be \label{Urealtime}
U(x,t)
=
\frac{e^{-\gamma t}}{2}\theta(v_0t-x)
\Bigg[
1+\gamma t
\int_{x/(v_0t)}^1dq\,
\Bigg\{
I_0\left(\gamma t\sqrt{1-q^2}\right) \, +
\frac{
I_1\left(\gamma t\sqrt{1-q^2}\right)
}{\sqrt{1-q^2}}
\Bigg\}
\Bigg],
\qquad x\geq0.
\ee

Using the symmetry property
\be \label{Usym}
U(x,t)=1-\int_{-\infty}^0 dz P_s(z,t|-x)=1-\int_{-\infty}^0 dz P_s(-z,t|x) = 1-U(-x,t) \;, 
\ee
the expression \eqref{Fs_real0} can be written in a simpler form
\bea \label{Fs_real}
F_s(x,t) &=&
\rho_0  + (\rho_1 e^{-\lambda} - \rho_0) U(-x,t) \\
&=& \rho_1 e^{-\lambda} + (\rho_0 - \rho_1 e^{-\lambda}) U(x,t) \;. \nn
\eea

We can then calculate $F_d(x,t)$ using equation \eqref{eqFs_freeRTP}, which in Fourier space and Laplace time reads
\be
\hat{\tilde F}_d(k,{\sf s})
=
\frac{i}{v_0 k}
({\sf s} \hat{\tilde F}_s(k,{\sf s}) - \hat F_s(k,0))
=
\frac{-i v_0 k}{{\sf s}^2+2\gamma {\sf s} + v_0^2 k^2}
\hat F_s(k,0) \;,
\ee
where $\hat F_s(k,0)$ is given in \eqref{FsFourierLaplaceIC}. This leads to
\be
\tilde F_d(x,{\sf s}) =
(\rho_0-\rho_1 e^{-\lambda})
\frac{e^{-\sqrt{{\sf s}({\sf s}+2\gamma)} \frac{|x|}{v_0}}}
{2\sqrt{{\sf s}({\sf s}+2\gamma)}} \;.
\ee
This can be inverted to give
\be \label{Fd_real}
F_d(x,t) =
\frac{e^{-\gamma t}}{2}
(\rho_0-\rho_1 e^{-\lambda})
I_0\Big(\sqrt{\gamma^2 t^2 -
\Big(\frac{\gamma x}{v_0}\Big)^2}\Big)
\theta (v_0 t - |x|) \;.
\ee
From \eqref{Fs_real} and \eqref{Fd_real}, we can finally obtain
$F_\pm = \frac{1}{2} (F_s \pm F_d)$.

We now recall that $G_s(x,t)$ and $G_d(x,t)$ follow the backward version of the equations for the free RTP propagator. Introducing the backward time $\tau=t_f-t$, these equations are mapped onto the forward equations for $F_{s/d}$ together with the exchange $+\leftrightarrow-$. The initial condition \eqref{IC_step_annealed_Fs} for $F_{s/d}$ and the final condition \eqref{Gsd_final} for $G_{s/d}$ have the same step form, up to their prefactors. We can therefore deduce the solutions for $G_s(x,t)$ and $G_d(x,t)$ directly from those for $F_s(x,t)$ and $F_d(x,t)$. This gives
\bea \label{Gs_sol_freeRTP}
G_s(x,t) &=&  2 + 2(e^{\lambda} - 1) U(-x,t_f-t) \\
&=& 2e^{\lambda} + 2(1-e^{\lambda}) U(x,t_f-t) \;, \nn \\
G_d(x,t) &=&
e^{-\gamma (t_f-t)} (e^{\lambda}-1)
I_0\Big(\sqrt{\gamma^2 (t_f-t)^2 -
\Big(\frac{\gamma x}{v_0}\Big)^2}\Big)
\theta (v_0 (t_f-t) - |x|) \;.
\label{Gd_sol_freeRTP}
\eea
We then have $G_\pm = \frac{1}{2} (G_s \pm G_d)$.

To evaluate the integrated current \eqref{current_through_zero_def2} at the saddle point, we use the relation
\be \label{rhos_from_FG}
\rho_s = F_+G_+ + F_-G_-
= \frac{1}{2} (F_sG_s+F_dG_d) \;.
\ee
This gives (note that $F_d G_d = 0$ both at $t=0$ and at $t=t_f$, so that we could have computed only $F_s$ and $G_s$)
\bea
&&\rho_s(x,0)
= \rho_0 \theta(-x) + \rho_1 \theta(x)
+ (1-e^\lambda)
[-\rho_0\theta(-x) + \rho_1 e^{-\lambda} \theta(x)]
\, U(|x|,t_f) \;, \\
&&\rho_s(x,t_f)
= \rho_0 \theta(-x) + \rho_1 \theta(x)
+ (\rho_0 - \rho_1 e^{-\lambda})
[-\theta(-x)+e^{\lambda}\theta(x)]
\, U(|x|,t_f) \;.
\eea
We thus obtain
\bea \label{QRTP_final}
Q^*(t_f)
=
\left(e^{\lambda}
- \frac{\rho_1}{\rho_0} e^{-\lambda}\right)
\mu(t_f)
\quad , \quad
\mu(t) = \rho_0 \int_0^{+\infty} dx \, U(x,t) \;.
\eea
The integral $\mu(t)$ was evaluated in \cite{FreeRTPs2020}. Its Laplace transform,
\be
\tilde \mu({\sf s})
=
\frac{\rho_0 v_0}{2{\sf s}\sqrt{{\sf s}({\sf s}+2\gamma)}} \;,
\ee
can be inverted explicitly, leading to
\bea \label{muexpr}
\mu(t_f)
=
\frac{1}{2} \rho_0 v_0 t_f e^{-\gamma t_f}
[I_0(\gamma t_f)+I_1(\gamma t_f)] \;.
\eea
Using the relation \eqref{SigmaPrime}
($\Sigma_{\rm an}'(\lambda) = Q^*(t_f)$)
and integrating over $\lambda$, we finally obtain the cumulant generating function
\be \label{sigmafinal_RTP_annealed}
\Sigma_{\rm an}(\lambda)
=
\left[
(e^{\lambda}-1)
+
\frac{\rho_1}{\rho_0}
(e^{-\lambda}-1)
\right]
\mu(t_f) \;.
\ee
For $\rho_1=0$, this coincides (upon identifying the counting variables) with equation~(42) of \cite{FreeRTPs2020}, namely the cumulant generating function of a Poisson distribution with parameter $\mu(t_f)$. As in the Brownian case, when $\rho_1>0$ the particles initially on the two sides contribute independently to opposite current directions. The resulting net-current distribution is then a Skellam distribution, i.e. the difference of two independent Poisson variables with means $\mu(t_f)$ and $(\rho_1/\rho_0)\mu(t_f)$, respectively.

Taking the Legendre transform of \eqref{sigmafinal_RTP_annealed} gives us the large deviation form $P_{\rm an}(Q(t_f)\simeq Q) \, {\asymp} \, e^{-I_{\rm an}(Q)}$ with
\be\label{LDP_annealed_RTP_step}
I_{\rm an}(Q)
=
\mu(t_f)
\Psi_{\rm an}
\left( \frac{Q}{\mu(t_f)} \right)
\quad , \quad
\Psi_{\rm an}(q)
=
\lambda_q q
-\tilde \Sigma_{\rm an}(\lambda_q)
\quad , \quad
\tilde \Sigma_{\rm an}(\lambda)
=
(e^\lambda -1)
+
\frac{\rho_1}{\rho_0}(e^{-\lambda}-1) \;,
\ee
where
\be
\lambda_q
=
\ln \left(
\frac{q}{2}
+
\sqrt{\frac{q^2}{4}+\frac{\rho_1}{\rho_0}}
\right) \;.
\ee
In the case where $\rho_1=0$, the function $\Psi_{\rm an}(q)$ is simply the rate function of a Poisson process,
\be
\Psi_{\rm an}(q) = q \ln q - q + 1 \;,
\qquad q\geq0 \;.
\ee
This corresponds to the result in equations~(12)--(13) of \cite{FreeRTPs2020}. Near $q=1$, it gives Gaussian typical fluctuations with variance equal to the mean $\mu(t_f)$. For $q\to0$ it gives the probability of the rare event where all the particles are back on the left at time $t_f$,
$P_{\rm an}(Q(t_f)\simeq 0)\sim e^{-\mu(t_f)}$, while for $q\to+\infty$ it gives the Poisson right tail
$\Psi_{\rm an}(q)\sim q\ln q$.

We thus see that in the annealed setting the full current statistics has the same universal form as for non-interacting Brownian particles: for $\rho_1=0$ the current is Poisson distributed, while for $\rho_1>0$ it follows a Skellam distribution.
The details of the single-particle dynamics enter only through the function $\mu(t_f)$
, which in this case plays the role of the large deviation parameter. Looking at the asymptotics of the expression \eqref{muexpr} (given in \cite{FreeRTPs2020}), we see that the Brownian result \eqref{sigmafinal_brownian_annealed} is recovered at large times, with $\mu(t_f) \simeq \rho_0\sqrt{D_a t_f/\pi}$ (where $D_a=\frac{v_0^2}{2\gamma}$ is the effective diffusion coefficient), while at short times there is a ballistic regime $\mu(t_f)\simeq \rho_0v_0t_f/2$.

Before moving to the quenched case, let us use the cumulant generating function \eqref{sigmafinal_RTP_annealed} to obtain the first two cumulants of the integrated current,
\be
\langle Q(t_f) \rangle
=
\left(1-\frac{\rho_1}{\rho_0}\right)\mu(t_f)
\quad , \quad
\langle Q^2(t_f) \rangle
-\langle Q(t_f) \rangle^2
=
\left(1+\frac{\rho_1}{\rho_0}\right)\mu(t_f) \;.
\ee
The expression for the variance coincides with the result \eqref{Qvar_free_Bessel} in Section~\ref{sec:typical} in the case of a uniform initial density $\rho_0=\rho_1=\rho$, as it should.

It is worth stressing a qualitative difference with the quenched problem studied below. In the annealed ensemble, arbitrarily large currents can be generated partly by an atypically large number of particles initially located in the region from which the origin can be reached during the observation time. Consequently the finite propagation speed $v_0$ does not produce a hard upper bound on $Q$ in the annealed ensemble, and activity enters the full distribution only through $\mu(t_f)$. When the initial positions are quenched, these initial number fluctuations are absent; the finite velocity then produces a distinct far-right large-deviation regime with $Q=O(t_f)$, which retains a strong signature of the active dynamics, as found in \cite{FreeRTPs2020}.
\\

\noindent {\bf Quenched positions and annealed orientations.} Let us now consider the case where the initial positions of the particles are quenched, but where the particle orientations are still drawn uniformly between $+$ and $-$. This corresponds to the ``quenched'' case of \cite{FreeRTPs2020}. Instead of using the initial condition \eqref{BCannealed1}, we now directly impose the total density at time $t=0$, $\rho_s(x,0) = r(x)$. However, since the particle orientations are still fluctuating in the initial state, we have the additional condition \eqref{IC_quenched_pd} for $p_d(x,0)$. 

Importantly, since $G_s$ and $G_d$ are completely determined by the backward equations \eqref{eqGs_freeRTP}-\eqref{eqGd_freeRTP} together with the final time conditions \eqref{Gsd_final}, they are independent of the choice of initial condition, which therefore only affects $F_s$ and $F_d$. Thus, they are still given by equations \eqref{Gs_sol_freeRTP} and \eqref{Gd_sol_freeRTP}.

Using the initial condition $\rho_s(x,0) = r(x)$ and the final time condition \eqref{Gsd_final}, together with the relation $\rho_s = (F_s G_s + F_d G_d)/2$, the integrated current \eqref{current_through_zero_def2} evaluated at the saddle point reads
\be \label{Qstar_quenched_start}
Q^*(t_f) = \int_0^{+\infty} dx \, [e^\lambda F_s(x,t_f)-r(x)] \;.
\ee
We thus only need to calculate $F_s(x,t)$. The initial conditions \eqref{IC_quenched_rhopm}-\eqref{IC_quenched_ppm} give
\be
F_\pm (x,0) = \rho_\pm(x,0)e^{-p_\pm(x,0)} = e^{-p_s(x,0)} \sqrt{\rho_+(x,0)\rho_-(x,0)} = \frac{e^{-p_s(x,0)}}{2} \sqrt{r(x)^2-\rho_d^2(x,0)} \;,
\ee
and thus
\be
F_s(x,0) = e^{-p_s(x,0)} \sqrt{r(x)^2-\rho_d^2(x,0)} \quad , \quad F_d(x,0) = 0 \;.
\ee
The relation $\rho_s = (F_s G_s + F_d G_d)/2$ finally implies
\be \label{Fs_x0}
F_s(x,0) = \frac{2r(x)}{G_s(x,0)} \;,
\ee
with $G_s(x,0)$ obtained from \eqref{Gs_sol_freeRTP} for $t=0$.
The solution for $F_s(x,t)$ can be formally expressed using this initial condition and the free RTP propagator \eqref{freeRTP_propagator},
\be \label{Fs_quenched}
F_s(x,t)
= \int_{-\infty}^{+\infty} dy  \, \frac{2 r(y)}{G_s(y,0)} P_s(x,t|y) \;.
\ee
Inserting into \eqref{Qstar_quenched_start} and exchanging the integrals, we obtain, using the function $U(x,t)$ defined in \eqref{Udef},
\be \label{Q_free_rtp_qa_step1}
Q^*(t_f) = \int_{-\infty}^{+\infty} dy \, r(y) \left[ 2 e^\lambda \frac{U(-y,t_f)}{G_s(y,0)} - \theta(y) \right] \;.
\ee
{Using the expression \eqref{Gs_sol_freeRTP} for $G_s$ finally gives}
\be \label{Q_free_rtp_qa_step2}
Q^*(t_f) = \rho_0 \int_{0}^{+\infty} dy \, \frac{e^\lambda U(y,t_f)}{1 + (e^\lambda-1) U(y,t_f)} - \rho_1 \int_{0}^{+\infty} dy \, \frac{e^{-\lambda} U(y,t_f)}{1 + (e^{-\lambda}-1) U(y,t_f)} \;,
\ee
which integrates to (using $\Sigma_{\rm qu}'(\lambda)=Q^*(t_f)$)
\be \label{Sigma_qa}
\Sigma_{\rm qu}(\lambda) = \rho_0 \int_{0}^{+\infty} dy \, \ln\left[1 + (e^\lambda-1) U(y,t_f)\right] + \rho_1 \int_{0}^{+\infty} dy \, \ln \left[1 + (e^{-\lambda}-1) U(y,t_f)\right] \;.
\ee
For $\rho_1=0$, this coincides with equation~(16) of \cite{FreeRTPs2020}.
We now focus on extreme positive-current fluctuations. Although the integrated current $Q$ can take either sign when both initial densities are nonzero, the results for negative currents $Q<0$ follow directly by exchanging $\rho_0$ and $\rho_1$ and replacing $Q$ with $|Q|$ in the expressions below, where we consider $Q>0$. Contrary to the annealed case, it is not possible to obtain a simple expression for the Legendre transform of this function $\Sigma_{\rm qu}(\lambda)$ in (\ref{Sigma_qa}). At large times, two distinct fluctuation regimes emerge~\cite{FreeRTPs2020}. On the diffusive scale $Q=O(\rho_0\sqrt{D_a t_f})$, the result reduces to the quenched Brownian problem \eqref{LD_free_Br_quenched}, with $D$ replaced by $D_a=v_0^2/(2\gamma)$. In contrast, the extreme right tail occurs on the ballistic scale $Q=O(\rho_0v_0t_f)$ and retains a genuinely active large-deviation form even at asymptotically long times. To analyse this regime, we show, following \cite{FreeRTPs2020}, that $\Sigma_{\rm qu}(\lambda)$ in Eq. (\ref{Sigma_qa}) takes the following scaling form in the limit $\lambda \to \infty$, $\gamma t_f \to \infty$, with $\lambda/(\gamma t_f)$ held fixed:
\bea \label{sigma_tf_sq}
\Sigma_{\rm qu}(\lambda) \simeq \rho_0 v_0 \gamma t_f^2 \mathcal{W}\left( \frac{\lambda}{\gamma t_f}\right) \;,
\eea
where
\be \label{def_W}
\mathcal{W}(x)=
\begin{cases}
\displaystyle \int_0^x dy\,\sqrt{y(2-y)}, & 0<x<1,\\[2mm]
& \\
\displaystyle \frac{\pi}{4}+x-1, & \hspace*{0.7cm} x>1.
\end{cases}
\ee

To establish Eqs.~(\ref{sigma_tf_sq})--(\ref{def_W}), we first note that the term proportional to $\rho_0$ in the expression for $\Sigma_{\rm qu}(\lambda)$ in Eq. (\ref{Sigma_qa}) is exactly the same as the one analysed in Ref.~\cite{FreeRTPs2020}, where it was shown that
\bea \label{ldf_rho0}
\rho_0 \int_{0}^{+\infty} dy \, \ln\left[1 + (e^\lambda-1) U(y,t_f)\right] \simeq \rho_0 v_0 \gamma t_f^2 \mathcal{W}\left( \frac{\lambda}{\gamma t_f}\right) \;,
\eea
with $W(x)$ given in (\ref{def_W}). Let us now examine the term proportional to $\rho_1$ in Eq. (\ref{Sigma_qa}) in the same scaling limit. Since $\lambda \to \infty$, we have
\bea \label{ldf_rho1.1}
\rho_1 \int_{0}^{+\infty} dy \, \ln \left[1 + (e^{-\lambda}-1) U(y,t_f)\right] \underset{\lambda \to \infty} {\longrightarrow} \rho_1 \int_{0}^{+\infty} dy \, \ln \left[1 -U(y,t_f)\right] \;.
\eea
Furthermore, for large $\gamma t_f$, the ballistic large-deviation form of $U(y,t_f)$ given in Ref.~\cite{FreeRTPs2020} reads
\be \label{ldf_rho1.2}
U(y,t_f)
{\asymp}
\exp\left[-\gamma t_f
\left(1-\sqrt{1-\left( \frac{y}{v_0 t_f}\right)^2}\right)\right] \Theta(v_0 t_f - y) \;.
\ee
The exponential rate in this expression indicates that the integral over $y$ in Eq. (\ref{ldf_rho1.1}) is concentrated on the diffusive scale $y=O(\sqrt{t_f})$. Using the diffusive scaling form of $U(y,t_f)$ on this scale {(obtained by Taylor-expanding the square root in \eqref{ldf_rho1.2} for $y\ll v_0 t_f$)} and changing variables to $y=u\sqrt{t_f}$ shows that the $\rho_1$ contribution is of order $O(\sqrt{t_f})$. It is therefore subleading compared with the $O(t_f^2)$ contribution proportional to $\rho_0$ in Eq. (\ref{ldf_rho0}), yielding Eq. (\ref{sigma_tf_sq}). This neglect of the $\rho_1$ term applies only to the extreme positive-current regime considered here. On the diffusive scale, both densities generally contribute to the current fluctuations (see Ref.~\cite{Derrida2009} and the discussion above in Section~\ref{sec:Brownian}).
Therefore, Eq. (\ref{sigma_tf_sq}) gives
\be \label{Pqu_ldf}
P_{\rm qu}(Q(t_f) \simeq Q)\asymp\exp[-\rho_0v_0\gamma t_f^2\Psi_{\rm RTP}(Q/(\rho_0v_0t_f))]
\ee
with $0<Q\leq\rho_0v_0t_f$ and the same rate function as in the case $\rho_1=0$~\cite{FreeRTPs2020}
\be
\Psi_{\rm RTP}(q)
=
\sup_{x\geq0}\left[qx-\mathcal{W}(x)\right],
\qquad
0\le q\le 1 \;.
\ee
We refer the reader to Ref.~\cite{FreeRTPs2020}, in particular Secs.~II and V B, for a detailed derivation. While the rate function $\Psi_{\rm RTP}(q)$ can be computed explicitly, here we give only its asymptotic behaviors, namely
\bea
\Psi_{\rm RTP}(q)\simeq
\begin{cases}
&\dfrac{q^3}{6}
\quad, \quad \quad \hspace*{0.1cm}q\to 0\;, \\
& \\
&1-\dfrac{\pi}{4} \quad, \quad q \to 1 \;.
\end{cases}
\eea
As shown in Ref.~\cite{FreeRTPs2020}, the cubic behavior at small $q$, once inserted in Eq. (\ref{Pqu_ldf}), matches exactly the large-$q$ behavior of the large-deviation rate function obtained on the diffusive scale for $Q = O(\rho_0 \sqrt{D_a t_f})$, which coincides with the one given for Brownian particles in Eq. (\ref{psi_qu_diff}) with the substitution $D \to D_a=v_0^2/(2 \gamma)$.

The difference between the quenched and annealed cases has a simple interpretation.
In the annealed case there is the additional freedom to chose an optimal initial conditions where the particles accumulate close to the origin. This leads to 
the Poisson statistics found above.
By contrast, for quenched initial conditions, each particle contributes a Bernoulli variable with crossing probability $U(y,t_f)$, which explains the logarithm in \eqref{Sigma_qa}.
At long times and fixed $\lambda$, the relevant distances are $y=O(\sqrt{D_a t_f})$, and the RTP therefore becomes effectively Brownian. This limit is, however, nonuniform in $\lambda$: probing the far-right sector $Q=O(\rho_0v_0t_f)$ requires $\lambda=O(\gamma t_f)$ and leads to the ballistic large-deviation form derived above. Thus, while moderate long-time fluctuations are effectively Brownian, extreme current fluctuations retain the signature of the underlying active dynamics.
\\

The above exact solution also allows to compute the {\it optimal space-time histories} of the density field $\rho_s(x,t)$ for a fixed value $Q$ of the current. The calculation is performed in Appendix \ref{app:density_RTP}. The results have been discussed in great detail in the introduction section \ref{sec:intro_results}.

\vspace*{0.5cm}

\noindent {\bf Quenched positions and orientations.} We now consider a third type of initial condition, which was not studied in \cite{FreeRTPs2020}. This is the case where both the particle positions and the particle orientations are fixed. This fully quenched setting was subsequently considered in \cite{JoseRossoRamola2023,JoseRossoRamola2024}. This condition is the easiest to write down, since we simply fix the densities $\rho_\pm(x,0)$ in the initial state. We write
\be \label{fullquenched2}
\rho_\pm(x,0) = r_\pm(x) \quad , \quad r_\pm(x) = \rho_0^\pm \theta(-x) + \rho_1^\pm \theta(x) \;,
\ee
with $\rho_0^++\rho_0^-=\rho_0$ and $\rho_1^++\rho_1^-=\rho_1$. The fraction of $+$ particles on the left-hand and right-hand sides are thus given by $\rho_0^+/\rho_0$ and $\rho_1^+/\rho_1$, respectively, and can be different from $1/2$.

In this case, $G_s$ and $G_d$ are still given by \eqref{Gs_sol_freeRTP}-\eqref{Gd_sol_freeRTP}, and the integrated current at the saddle point takes the same form \eqref{Qstar_quenched_start} as in the previous case. The initial condition for $F_\pm$ reads
\be
F_\pm(x,0) = \frac{r_\pm(x)}{G_\pm(x,0)} \;,
\ee
with
\be \label{Gpm_zero}
G_\pm(x,0) = \theta(-x) + e^{\lambda} \theta(x)
+ (1-e^\lambda) \sgn(x) U(|x|,t_f)
\pm \frac{e^{-\gamma t_f}}{2} (e^{\lambda}-1)
I_0\Big(\sqrt{\gamma^2 t_f^2 - \Big(\frac{\gamma x}{v_0}\Big)^2}\Big)
\theta (v_0 t_f - |x|) \;.
\ee

The difference with the previous case, where the orientations in the initial state were drawn randomly with uniform probability, is that we can no longer use the orientation-averaged propagator \eqref{freeRTP_propagator} to obtain $F_s(x,t)$, since each particle now starts with a fixed orientation. We thus introduce $P_s^{\pm}(x,t|x_0)$, the position propagator of a free RTP starting at $x_0$ with initial orientation $\pm$, summed over its orientation at time $t$  (such that $P_s=(P_s^++P_s^-)/2$). Its Fourier--Laplace transform reads \cite{Masoliver1993,JoseRossoRamola2023,JoseRossoRamola2024}
\be \label{FreePropagator_FourierLaplace_or}
\hat{\tilde P}_s^{\pm}(k,{\sf s}|x_0) = \frac{{\sf s}+2\gamma \mp i v_0 k}{{\sf s}^2+2\gamma {\sf s} + v_0^2 k^2} e^{-i k x_0} \;.
\ee
In real space and real time this gives
\bea
P_s^{\pm}(x,t|x_0) &=& e^{-\gamma t} \Bigg[ \delta(x-x_0 \mp v_0 t)
+ \frac{\gamma}{2v_0} \theta(v_0 t - |x-x_0|)
\Bigg( I_0\Big(\sqrt{\gamma^2t^2 - \Big(\frac{\gamma (x-x_0)}{v_0}\Big)^2}\Big) \nn \\
&& \hspace{1cm}
+ \frac{v_0 t \pm (x-x_0)}
{\sqrt{v_0^2t^2 -(x-x_0)^2}}
I_1\Big(\sqrt{\gamma^2t^2 - \Big(\frac{\gamma (x-x_0)}{v_0}\Big)^2}\Big)
\Bigg) \Bigg] \;.
\label{freeRTP_propagator_or}
\eea

Since $P_s^{\pm}$ is the total position propagator conditioned on the initial orientation $\sigma$, it is $F_s$, rather than $F_\sigma$, which is obtained by summing the two propagated initial contributions:
\be
F_s(x,t) =
\sum_{\sigma=\pm}
\int_{-\infty}^{+\infty} dy\,
\frac{r_\sigma(y)}{G_\sigma(y,0)}
P_s^{\sigma}(x,t|y) \;.
\ee
Performing the same manipulations as in the previous case, we arrive at
\be \label{Q_free_rtp_qa_step1_or}
Q^*(t_f) = \sum_{\sigma=\pm} \int_{-\infty}^{+\infty} dy \, r_\sigma(y)
\left[
e^\lambda \frac{U_\sigma(-y,t_f)}{G_\sigma(y,0)}
-\theta(y)
\right] \;,
\ee
where we introduced
\be \label{Upmdef}
U_\pm(x,t) = \int_0^{+\infty} dz\, P_s^{\pm}(z,t|-x) \;,
\ee
the probability that a particle starting at $X(0)=-x$ at time zero 
with orientation $\pm$ will be at position $X(t)>0$ at time $t$, i.e.
for $x>0$ they are orientation-resolved crossing probabilities.
Note that the function $U(x,t)$ defined in \eqref{Udef} is recovered by averaging over the initial state, $U=(U_++U_-)/2$. The relation \eqref{Usym} becomes
\be\label{Upmsym}
U_\sigma(x,t)=1-U_{-\sigma}(-x,t) \;.
\ee

These orientation-resolved crossing probabilities
$U_\pm(x,t)$ can also be written explicitly. Their expression in Laplace time is given in \cite{JoseRossoRamola2023,JoseRossoRamola2024}
\be \label{UdefLaplace}
\tilde U_\pm(x,{\sf s}) =
\int_0^{+\infty} dt \, e^{-{\sf s}t} U_\pm(x,t)
=
\frac{e^{-\sqrt{{\sf s}({\sf s}+2\gamma)} \frac{x}{v_0}}}{2{\sf s}} \left( 1 \pm \sqrt{\frac{\sf s}{{\sf s}+2\gamma}} \right) \;, \quad x\geq 0 \;.
\ee

Using
Eq.~\eqref{freeRTP_propagator_or} in the definition \eqref{Upmdef}, one finds the relation
\be
U_\pm(x,t)
=
U(x,t)
\pm
\frac{e^{-\gamma t}}{2}
I_0\left(
\gamma t\sqrt{1-\frac{x^2}{v_0^2t^2}}
\right)
\theta(v_0t-x),
\qquad x\geq0,
\label{Upm_realtime}
\ee
where $U(x,t)$ is given in Eq.~\eqref{Urealtime}.

Let us now note that, using a similar method as the one used to derive \eqref{Gs_sol_freeRTP}, but solving directly for $G_\sigma(x,t)$ instead of $G_s(x,t)$, we obtain an expression of $G_\sigma(x,t)$ in terms of $U_\sigma(x,t)$, which is equivalent to \eqref{Gpm_zero} (in particular, the contribution proportional to $G_d$ is fully contained in the orientation dependence of $U_\sigma$),  
\bea \label{Gpm_quenched_or}
G_\sigma(x,t) &=& 1 + (e^{\lambda} - 1) U_\sigma(-x,t_f-t) \\
&=& e^{\lambda} + (1-e^{\lambda}) U_{-\sigma}(x,t_f-t) \;. \nn
\eea
Alternatively, this expression can be obtained by noticing that, since $G_\sigma(x,t)$ satisfies the backward Fokker-Planck equation for a free RTP \eqref{eqGp_freeRTP}-\eqref{eqGm_freeRTP} with the final time condition $G_\sigma(x,t_f)=e^{\lambda\theta(x)}$, it can be written as
\be \label{G_identity}
G_\sigma(x,t)
= \int dy \, P_s^{\sigma}(y,t_f-t|x) e^{\lambda\theta(y)} \equiv
{\mathbb E}_{x,\sigma}
\left[
e^{\lambda\theta(X_{t})}
\right].
\ee
Indeed, $P_s^{\sigma}(y,t_f-t|x)=\sum_{\sigma_f} p_t(y,\sigma_f,t_f|x,\sigma,t)$ also satisfies the backward Fokker-Planck equation for a free RTP with final time condition $P_s^{\sigma}(y,0|x)=\delta(x)$, thus both sides of \eqref{G_identity} satisfy the same equation with the same final time condition.
Hence we can write
\be
G_\sigma(y,t) = 1+(e^\lambda-1)
{\mathbb P}_{y,\sigma}\left(X_{t_f}>0\right),
\ee
with
\be
{\mathbb P}_{x,\sigma}\left(X_{t_f}>0\right)
=
\int_0^{+\infty}dz\,
P_s^\sigma(z,t_f-t|x).
\ee
Using the definition \eqref{Upmdef},
we immediately obtain the first equality of \eqref{Gpm_quenched_or}, while the symmetry relation \eqref{Upmsym} gives the second equality.
Summing the expression \eqref{Gpm_quenched_or} over $\sigma=\pm 1$ with $U=(U_+ + U_-)/2$ correctly recovers the result \eqref{Gs_sol_freeRTP} for $G_s$. 

Inserting into \eqref{Q_free_rtp_qa_step1_or} yields
\be \label{Q_free_rtp_qq_step2}
Q^*(t_f) = \sum_{\sigma=\pm}\int_{0}^{+\infty} dy \left[ \rho_0^\sigma \frac{e^\lambda U_\sigma(y,t_f)}{1 + (e^\lambda-1) U_\sigma(y,t_f)} - \rho_1^\sigma \frac{e^{-\lambda} U_{-\sigma}(y,t_f)}{1 + (e^{-\lambda}-1) U_{-\sigma}(y,t_f)} \right] \;,
\ee
which integrates to
\be \label{Sigma_or}
\Sigma_{\rm or}(\lambda) = \sum_{\sigma=\pm} \int_{0}^{+\infty} dy \, \Big\{ \rho_0^\sigma \ln\left[1 + (e^\lambda-1) U_\sigma(y,t_f)\right] + \rho_1^\sigma \ln \left[1 + (e^{-\lambda}-1) U_{-\sigma}(y,t_f)\right] \Big\} \;.
\ee
This result reduces to the one of Ref. \cite{JoseRossoRamola2023,JoseRossoRamola2024} in the case
$\rho_1^\pm=0$ and $\rho_0^+=\rho_0^-=\rho/2$ obtained by a different method. This case illustrates 
the difference with quenched positions but annealed orientations.
Indeed, the fully quenched result contains the average of the two logarithms associated with $U_+$ and $U_-$, whereas \eqref{Sigma_qa} contains the logarithm of their average $U=(U_++U_-)/2$. In particular, \cite{JoseRossoRamola2023,JoseRossoRamola2024} finds in this symmetric case $\langle Q^2\rangle_c\simeq \rho_0v_0\gamma t_f^2/2$ at short times, instead of the linear behavior obtained when the orientations are annealed, while at long times the diffusive behavior $\langle Q^2\rangle_c\simeq\rho_0\sqrt{D_at_f/(2\pi)}$ is recovered. Thus the memory of the initial orientations is especially pronounced in the short-time ballistic regime.
\\

Let us now analyze Eq.~\eqref{Sigma_or} in the long-time limit. In the diffusive regime, $\gamma t_f\gg 1$, the relevant initial distances are $y=O(\sqrt{D_a t_f})$, with $D_a=v_0^2/(2\gamma)$. On this scale, the memory of the initial orientation is lost and
\be
U_\pm(y,t_f)
\simeq
U(y,t_f)
\simeq
\frac{1}{2}\,{\rm erfc}
\left(\frac{y}{\sqrt{4D_a t_f}}\right).
\ee
Using $\sum_\sigma\rho_0^\sigma=\rho_0$ and $\sum_\sigma\rho_1^\sigma=\rho_1$, Eq.~\eqref{Sigma_or} therefore reduces to Eq.~\eqref{Sigma_qa}. Hence the fully quenched problem has, at leading order in the diffusive regime, the same large-deviation function as the quenched Brownian problem discussed above, with $D$ replaced by $D_a$.

We now turn to the extreme positive-current regime, $Q=O(\rho_0v_0t_f)$, for which $\lambda=O(\gamma t_f)$. For $y=v_0t_f z$, with fixed $0<z<1$, the orientation-resolved crossing probabilities have the same leading exponential behavior,
\be
U_\pm(v_0t_f z,t_f)
\asymp
\exp\left[
-\gamma t_f\left(1-\sqrt{1-z^2}\right)
\right],
\ee
the dependence on the initial orientation entering only through subexponential factors. Therefore, in Eq.~\eqref{Sigma_or}, the contribution proportional to $\rho_1^\sigma$ is subleading for the same reason as in Eqs.~\eqref{ldf_rho1.1}--\eqref{ldf_rho1.2}, while the two contributions proportional to $\rho_0^\sigma$ have the same leading exponential form. Since $\rho_0^++\rho_0^-=\rho_0$, we obtain
\be
\Sigma_{\rm or}(\lambda)
\simeq
\rho_0v_0\gamma t_f^2
\mathcal{W}\left(\frac{\lambda}{\gamma t_f}\right),
\ee
with the same function $\mathcal{W}$ as in Eq.~\eqref{def_W}. Consequently, the fully quenched problem has the same leading ballistic large-deviation form as Eq.~\eqref{Pqu_ldf},
\be
P_{\rm or}(Q(t_f) \simeq Q)
\asymp
\exp\left[
-\rho_0v_0\gamma t_f^2
\Psi_{\rm RTP}
\left(\frac{Q}{\rho_0v_0t_f}\right)
\right],
\qquad
0<Q\leq \rho_0v_0t_f,
\ee
with the same rate function $\Psi_{\rm RTP}$. Thus the fixed initial orientation fractions affect subleading terms, but not the leading diffusive or ballistic large-deviation functions.

\subsection{Weakly interacting particles} 
\label{sec:step_interaction}

In this subsection we consider weakly interacting particles. We use the MFT method together
with a perturbative approach to include the interaction. We compute the large deviations
of the integrated current through the origin to first order in perturbation theory, for
particles on the infinite line with annealed initial conditions. We start with Brownian
particles and then treat RTPs.

\subsubsection{Current fluctuations for weakly interacting Brownian particles}

We start with interacting Brownian particles described by the equation of motion
\be
\dot x_i(t)
=
-\varepsilon\sum_{j\neq i}W'(x_i(t)-x_j(t))
+
\sqrt{2D}\,\xi_i(t) \;,
\ee
where the parameter $\varepsilon$ controls the interaction strength. Introducing the
density $\rho(x,t)$, we consider the same integrated current observable as in the free
Brownian case,
$Q(t_f)
=
\int_0^{+\infty}dx\,
[
\rho(x,t_f)-\rho(x,0)
]$.
The MFT action reads
\bea
{\cal S}_\varepsilon[\rho,p]
&=&
\int_0^{t_f}dt\int_{-\infty}^{+\infty}dx
\left[
p\partial_t\rho
-
D\rho(\partial_xp)^2
+
D\partial_x\rho\,\partial_xp
-
\varepsilon \rho\,\mathcal H[\rho]\,\partial_xp
\right]
\nn\\
&=&
{\cal S}_0[\rho,p]
+
\varepsilon {\cal S}_1[\rho,p] \;,
\eea
with
\be \label{S1_expr_br}
{\cal S}_1[\rho,p]
=
\int_0^{t_f}dt
\int_{-\infty}^{+\infty}dx
\int_{-\infty}^{+\infty}dy\,
\rho(x,t)\rho(y,t)W'(x-y)\partial_xp(x,t).
\ee

{\bf Initial condition}. There are two natural meanings of an annealed initial condition.
The first possibility is to average with respect to the non-interacting initial measure.
In that case the initial density is sampled from a local Poisson field with intensity
$r(x)=\rho_0\theta(-x)+\rho_1\theta(x)$. The initial density distribution is then {the same as in Sec.~\ref{sec:MFT_step_noninteracting},}
\be \label{initannealed}
P_{\rm an}^{(0)}[\rho(x,0)]
\asymp
\exp\left\{
-\int_{-\infty}^{+\infty}dx
\left[
\rho(x,0)\ln\frac{\rho(x,0)}{r(x)}
-\rho(x,0)
+r(x)
\right]
\right\}.
\ee
This preparation corresponds to switching on the interaction only during the observation
time interval $[0,t_f]$, while keeping the same initial density fluctuations as in the
free Brownian problem.

The second possibility is to prepare the system with the interaction already present
before the measurement starts. For a homogeneous system, $\rho_0=\rho_1$, the initial
density is then sampled from the interacting equilibrium measure. More generally, for
the step profile $\rho_0\neq\rho_1$, this preparation can be realized by equilibrating
the two half-lines separately, at bulk densities $\rho_0$ and $\rho_1$, in the presence
of a separating wall, and removing the wall at $t=0$. We denote the corresponding
initial large-deviation functional by
\be \label{init_noninteracting_br}
P_{\rm an}^{(\varepsilon)}[\rho(x,0)]
\asymp
\exp\left[
-{\cal V}_\varepsilon[\rho(x,0)]
\right],
\ee
where
\be
{\cal V}_\varepsilon[\rho]
=
{\cal V}_0[\rho]
+
\varepsilon{\cal V}_1[\rho]
+
O(\varepsilon^2),
\ee
where ${\cal V}_0[\rho]$ is the term already present in \eqref{initannealed}, while the expression of ${\cal V}_1[\rho]$ will be derived below.
In this preparation, the current fluctuations receive both a dynamical correction during
the interval $[0,t_f]$ and a static correction coming from the interacting initial
correlations.

We now treat these two initial preparations of the system separately.

\subsubsection{Annealed initial condition from the non-interacting measure}

For the non-interacting annealed preparation, the cumulant generating function is
\be
\Sigma_\varepsilon^{\rm an,0}(\lambda)
=
\ln \left\langle e^{\lambda Q(t_f)}\right\rangle
\simeq
\underset{\rho,p}{\sup}
\left[
\lambda Q(t_f)
-
{\cal S}_\varepsilon[\rho,p]
+
\ln P_{\rm an}^{(0)}[\rho(x,0)]
\right].
\ee
where $\underset{\rho,p}{\sup}$ means evaluation at the saddle-point.
We expand
\be
\Sigma_\varepsilon^{\rm an,0}(\lambda)
=
\Sigma_0^{\rm an}(\lambda)
+
\varepsilon
\Sigma_1^{\rm an,0}(\lambda)
+
O(\varepsilon^2).
\ee
Since the free optimal path, denoted here $(\rho^{(0)},p^{(0)})$ (this is to avoid confusion with the initial density variable $\rho_0$), extremizes the tilted free variational
problem, the first-order correction to the path itself does not contribute at order~$\varepsilon$. Therefore
\be
\Sigma_1^{\rm an,0}(\lambda)
=
-{\cal S}_1[\rho^{(0)},p^{(0)}].
\ee
Using the expression for ${\cal S}_1$ {given in \eqref{S1_expr_br}}, this gives
\bea
\Sigma_1^{\rm an,0}(\lambda)
&=&
-\int_0^{t_f}dt
\int_{-\infty}^{+\infty}dx
\int_{-\infty}^{+\infty}dy\,
\rho^{(0)}(x,t)\rho^{(0)}(y,t)
W'(x-y)\partial_x p^{(0)}(x,t)
\\
&=&
-\int_0^{t_f}dt
\int_{-\infty}^{+\infty}dx
\int_{-\infty}^{+\infty}dy\,
F_0(x,t)\partial_xG_0(x,t)
\,F_0(y,t)G_0(y,t) W'(x-y) \;,
\eea
where in the second line we have substituted the Cole--Hopf variables
\be
\rho^{(0)}(x,t)=F_0(x,t)G_0(x,t),
\qquad
p^{(0)}(x,t)=\ln G_0(x,t).
\ee
This is an explicit result since, for the annealed Brownian problem,
$F_0$ and $G_0$ are the error-function profiles obtained previously (see \eqref{F_Brownian_annealed}-\eqref{G_Brownian_annealed}),
\bea \label{FG0_Brownian_ann}
F_0(x,t)
&=&
\frac{\rho_1 e^{-\lambda}+\rho_0}{2}
+
\frac{\rho_1 e^{-\lambda}-\rho_0}{2}
{\rm erf}\left(\frac{x}{\sqrt{4Dt}}\right),
\nn\\
G_0(x,t)
&=&
\frac{e^\lambda-1}{2}
{\rm erf}\left(\frac{x}{\sqrt{4D(t_f-t)}}\right)
+
\frac{e^\lambda+1}{2}.
\eea
In particular,
\be
\partial_xG_0(x,t)
=
\frac{e^\lambda-1}{\sqrt{4\pi D(t_f-t)}}
\exp\left[-\frac{x^2}{4D(t_f-t)}\right].
\ee
We have thus obtained, to first order in the interaction,
\be
\Sigma_\varepsilon^{\rm an,0}(\lambda)
=
\left[
\rho_0(e^\lambda-1)
+
\rho_1(e^{-\lambda}-1)
\right]
\sqrt{\frac{Dt_f}{\pi}}
+
\varepsilon
\Sigma_1^{\rm an,0}(\lambda)
+
O(\varepsilon^2),
\ee
where the first term is the free Brownian annealed result, and
$\Sigma_1^{\rm an,0}$ is given by the explicit space-time integral above.
The zeroth-order term is exact for arbitrary $t_f$ for the Poisson initial
condition. The interaction correction, on the other hand, has been obtained from the
MFT saddle-point approximation and should be understood in the corresponding
large-deviation regime. For the infinite-line problem considered here, we will now
analyze its long-time behavior.

Let us consider an even, short-ranged interaction potential $W(x)$ which decays
sufficiently fast at infinity. We denote by $a$ its characteristic range 
(denoted $a_0$ in the case of the Riesz interaction \eqref{Riesz_regularized}) 
and assume
that $a$ remains fixed in the long-time limit. In particular, we assume that the moments
\be
W_0=\int_{-\infty}^{+\infty}dr\,W(r)
\quad,\quad
W_2=\int_{-\infty}^{+\infty}dr\,r^2W(r)
\ee
are finite. Since $W$ is even, $W'$ is odd and
$\int_{-\infty}^{+\infty}dr\,W'(r)=0$.

At long times, in the bulk of the time interval, i.e. for $t=\tau t_f$ with
$0<\tau<1$ fixed, the free optimal fields vary on diffusive length scales
$O(\sqrt{Dt_f})$ much larger than $a$. We can therefore perform a gradient expansion
of the interaction term. Writing $r=x-y$, we obtain
\be
\int_{-\infty}^{+\infty}dy\,
W'(x-y)\rho^{(0)}(y,t)
=
\int_{-\infty}^{+\infty}dr\,
W'(r)\rho^{(0)}(x-r,t)
=
W_0\,\partial_x\rho^{(0)}(x,t)
+
\frac{W_2}{2}\,\partial_x^3\rho^{(0)}(x,t)
+\cdots,
\ee
where we used integration by parts $\int_{-\infty}^{+\infty}dr\,rW'(r)=-W_0$, 
$\int_{-\infty}^{+\infty}dr\,r^3W'(r)=-3W_2$. 
Thus, to leading order in the ratio between the interaction range and the diffusive
length scale,
\bea \label{Sigma1_Brownian_local}
\Sigma_1^{\rm an,0}(\lambda)
&\simeq&
-W_0
\int_0^{t_f}dt
\int_{-\infty}^{+\infty}dx\,
F_0(x,t)\partial_xG_0(x,t)
\partial_x\left[F_0(x,t)G_0(x,t)\right].
\eea
For $t=O(t_f)$, and assuming $W_2/W_0=O(a^2)$, the next term in the
gradient expansion is smaller than the leading one by a factor
$O(a^2/(Dt_f))$. The gradient expansion is not uniform within temporal boundary
layers of width $O(a^2/D)$ near $t=0$ and $t=t_f$. These boundary layers can modify
subleading terms, but not the leading $O(\sqrt{t_f})$ contribution derived below.

This expression also makes the long-time scaling explicit. The free saddle-point fields
satisfy $F_0=O(\rho)$ and $G_0=O(1)$, where $\rho$ denotes a typical density scale.
In the bulk of the time interval, their gradients vary on the diffusive scale
$\sqrt{Dt_f}$, so that
\be
\partial_xG_0
=
O\left((Dt_f)^{-1/2}\right),
\qquad
\partial_x(F_0G_0)
=
O\left(\rho(Dt_f)^{-1/2}\right).
\ee
The relevant space-time region has spatial extent $O(\sqrt{Dt_f})$ and temporal
extent $O(t_f)$. Hence
\be
\Sigma_1^{\rm an,0}(\lambda)
=
O\left(
\rho^2\frac{W_0}{D}\sqrt{Dt_f}
\right),
\qquad
W_0=\int_{-\infty}^{+\infty}dr\,W(r).
\ee
Thus the first-order correction scales with the same diffusive length
$\sqrt{Dt_f}$ as the free contribution. The corresponding dimensionless perturbative
parameter is of order
\be
 \varepsilon\,\rho\,\frac{W_0}{D} \;,
\ee
and the validity of the perturbation theory requires that its absolute value is small compared to unity. 
The temporal boundary layers discussed above can contribute subleading terms, but do
not modify this leading large-$t_f$ scaling.

It is possible to evaluate the remaining dimensionless space-time integral explicitly,
which is done in Appendix \ref{app:Brownian_first_order}. We obtain

\be \label{Sigma_ann0_SR}
\Sigma_\varepsilon^{\rm an,0}(\lambda)
=
\sqrt{\frac{Dt_f}{\pi}}
\left\{
\rho_0(e^\lambda-1)
+
\rho_1(e^{-\lambda}-1)
-
\varepsilon
\frac{W_0}{D}
\Phi(\lambda;\rho_0,\rho_1)
+
O(\varepsilon^2)
\right\}
+
o(\sqrt{t_f}),
\ee
where
\bea \label{defPhi} 
\Phi(\lambda;\rho_0,\rho_1)
=
\frac14
\Bigg\{
&&
\rho_0^2
\left[
(\sqrt2-1)e^{2\lambda}
+
(1-2\sqrt2)e^\lambda
+
\sqrt2
\right]
+
\rho_0\rho_1
\left[
e^\lambda+e^{-\lambda}-2
\right]
\nn\\
&&
+
\rho_1^2
\left[
\sqrt2
+
(1-2\sqrt2)e^{-\lambda}
+
(\sqrt2-1)e^{-2\lambda}
\right]
\Bigg\}.
\eea
As consistency checks,
\be
\Phi(0;\rho_0,\rho_1)=0,
\qquad
\Phi(\lambda;\rho_0,\rho_1)
=
\Phi(-\lambda;\rho_1,\rho_0),
\ee
so that $\Sigma_\varepsilon^{\rm an,0}(0)=0$ and the result has the expected
left-right symmetry.

The first three cumulants follow by differentiating the cumulant generating function at
$\lambda=0$. To first order in $\varepsilon$, and in the large $t_f$ limit, the mean current is
\be
\langle Q(t_f) \rangle 
\simeq 
\left[
\rho_0-\rho_1
+
\varepsilon
\frac{W_0}{4D}
(\rho_0^2-\rho_1^2) +
O(\varepsilon^2)
\right] 
\sqrt{\frac{Dt_f}{\pi}} .
\ee
The variance is
\be
\langle Q(t_f)^2 \rangle_c
\simeq 
\left[
\rho_0+\rho_1
+
\varepsilon
\frac{W_0}{4D}
\left\{
(3-2\sqrt2)(\rho_0^2+\rho_1^2)
-
2\rho_0\rho_1
\right\} +
O(\varepsilon^2)
\right]
\sqrt{\frac{Dt_f}{\pi}},
\ee
and the third cumulant is
\be
\langle Q(t_f)^3 \rangle_c
\simeq 
\left[
\rho_0-\rho_1
-
\varepsilon
\frac{W_0}{4D}
(6\sqrt2-7)(\rho_0^2-\rho_1^2) +
O(\varepsilon^2)
\right]
\sqrt{\frac{Dt_f}{\pi}}.
\ee

A few physical remarks can be made about these results. First, to leading order at long
times, the interaction enters only through the zeroth moment
$W_0=\int dr\,W(r)$. Thus, all sufficiently short-ranged even interactions with the
same value of $W_0$ give the same leading correction to the current statistics. The
details of the shape of the interaction potential, through moments such as $W_2$, only
enter subleading terms in the long-time expansion. This can also be understood directly
at the hydrodynamic level. Indeed, the gradient expansion of the interaction force gives
\be
{\cal H}[\rho]
\simeq
-W_0\,\partial_x\rho,
\ee
so that the deterministic current becomes, to first order in $\varepsilon$,
\be
\bar J
\simeq
-\left(D+\varepsilon W_0\rho\right)\partial_x\rho.
\ee
Thus, for a repulsive interaction with $W_0>0$, the collective diffusion is enhanced.
In particular, for $\rho_0>\rho_1$, this is consistent with the positive interaction
correction to the mean integrated current,
\be
\langle Q(t_f)\rangle
=
\left[
\rho_0-\rho_1
+
\varepsilon\frac{W_0}{4D}
(\rho_0^2-\rho_1^2)
+O(\varepsilon^2)
\right]
\sqrt{\frac{Dt_f}{\pi}}.
\ee

The effect on the fluctuations is different. For a homogeneous initial state,
$\rho_0=\rho_1=\rho$, left-right symmetry implies that all odd cumulants vanish, while
the variance becomes
\be \label{var2}
\langle Q(t_f)^2\rangle_c
=
2\rho
\left[
1+
\varepsilon\frac{\rho W_0}{2D}(1-\sqrt2)
+O(\varepsilon^2)
\right]
\sqrt{\frac{Dt_f}{\pi}}.
\ee
Since $1-\sqrt2<0$, a repulsive interaction suppresses the current fluctuations even
though it enhances the collective diffusion. Physically, the repulsion tends to smooth
density fluctuations and therefore reduces the fluctuations of the number of particles
crossing the origin.

Finally, it is important to recall that the results obtained in this subsection correspond
to a Poisson initial condition which is independent of the interaction. They therefore
describe the dynamical effect of switching on the interaction at $t=0$, without the
static correlations that would be present in an interacting equilibrium initial state.
In particular, for $\rho_0\neq\rho_1$, the fluctuation symmetry of the free problem,
which involves the bare density ratio $\rho_1/\rho_0$, is in general not preserved by
the dynamical correction alone. The contribution of the interaction to the initial
large-deviation functional must \st{also} be included when the system is prepared with the interaction already present, as we discuss next.

For power-law long-range interactions, $W(r)\sim g/|r|^s$ with $0<s<1$, the first-order
interaction correction scales instead as
\be
\Sigma_1^{\rm an,0}(\lambda)
=
O\left(
\frac{\rho^2 g}{D}
(Dt_f)^{1-s/2}
\right),
\ee
which grows faster than the free contribution $O(\rho\sqrt{Dt_f})$. Therefore, for any
fixed $\varepsilon>0$, the naive first-order perturbative expansion becomes nonuniform
at long times and cannot be used to determine the asymptotic current fluctuations in
this regime.

\subsubsection{Annealed initial condition from the interacting measure}

We now consider the second annealed preparation, in which the interaction is already
present before the measurement starts. As discussed above, for the step profile this
corresponds to preparing two interacting equilibrium half-systems at bulk densities
$\rho_0$ and $\rho_1$ and joining them at $t=0$. The initial density is therefore
sampled from
\be
P_{\rm an}^{(\varepsilon)}[\rho(x,0)]
\asymp
e^{-{\cal V}_\varepsilon[\rho(x,0)]},
\qquad
{\cal V}_\varepsilon
=
{\cal V}_0+\varepsilon{\cal V}_1+O(\varepsilon^2).
\ee
The cumulant generating function is
\be
\Sigma_\varepsilon^{\rm an,\varepsilon}(\lambda)
\simeq
\sup_{\rho,p}
\left[
\lambda Q(t_f)
-
{\cal S}_\varepsilon[\rho,p]
-
{\cal V}_\varepsilon[\rho(x,0)]
\right].
\ee
Since the zeroth-order tilted problem is identical to the one considered above, its
optimal path is again the free saddle $(\rho^{(0)},p^{(0)})$. Therefore the first-order
correction is
\be
\Sigma_1^{\rm an,\varepsilon}(\lambda)
=
-{\cal S}_1[\rho^{(0)},p^{(0)}]
-
{\cal V}_1[\rho^{(0)}(x,0)].
\ee
The first term is precisely the dynamical correction calculated in the previous
paragraph and in Appendix~\ref{app:Brownian_first_order}. Thus the only new contribution
comes from the interacting initial measure.

To first order in the interaction, the equilibrium density large-deviation functional
contains the interaction contribution
\be
{\cal V}_1[\rho]
=
\frac{1}{2D}
\int dx\,dy\,
[\rho(x)-r(x)]W(x-y)[\rho(y)-r(y)],
\ee
up to terms linear in $\rho-r$ which fix the bulk chemical potentials on the two sides,
and microscopic contributions associated with the separating wall. For a short-ranged
interaction, the leading hydrodynamic contribution is therefore
\be \label{V1_local_Brownian}
{\cal V}_1[\rho]
\simeq
\frac{W_0}{2D}
\int_{-\infty}^{+\infty}dx\,
[\rho(x)-r(x)]^2,
\qquad
W_0=\int_{-\infty}^{+\infty}dr\,W(r).
\ee
The omitted wall and gradient corrections are subleading compared with the
$O(\sqrt{t_f})$ term considered below.

Using the free saddle \eqref{FG0_Brownian_ann}, its initial density satisfies
\be
\rho^{(0)}(x,0)-r(x)
=
\begin{cases}
\displaystyle
\frac{\rho_0}{2}(e^\lambda-1)
\,{\rm erfc}\left(\frac{-x}{\sqrt{4Dt_f}}\right),
& x<0,
\\[3mm]
\displaystyle
-\frac{\rho_1}{2}(1-e^{-\lambda})
\,{\rm erfc}\left(\frac{x}{\sqrt{4Dt_f}}\right),
& x>0.
\end{cases}
\ee
Substituting this expression into \eqref{V1_local_Brownian}, and using
$\int_0^{+\infty}du\,{\rm erfc}^2(u)
=
\frac{2-\sqrt2}{\sqrt{\pi}}$,
gives
\be
{\cal V}_1[\rho^{(0)}(x,0)]
=
\frac{W_0}{D}
\sqrt{\frac{Dt_f}{\pi}}\,
\Phi_{\rm init}(\lambda;\rho_0,\rho_1)
+
o(\sqrt{t_f}),
\ee
with
\be
\Phi_{\rm init}(\lambda;\rho_0,\rho_1)
=
\frac{2-\sqrt2}{4}
\left[
\rho_0^2(e^\lambda-1)^2
+
\rho_1^2(e^{-\lambda}-1)^2
\right].
\ee

Combining this static contribution with the dynamical correction
\eqref{Sigma_ann0_SR}, we obtain
\be \label{Sigma2eme}
\Sigma_\varepsilon^{\rm an,\varepsilon}(\lambda)
=
\sqrt{\frac{Dt_f}{\pi}}
\left\{
\rho_0(e^\lambda-1)
+
\rho_1(e^{-\lambda}-1)
-
\varepsilon\frac{W_0}{D}
\Phi_{\rm tot}(\lambda;\rho_0,\rho_1)
+
O(\varepsilon^2)
\right\}
+
o(\sqrt{t_f}),
\ee
where
\be
\Phi_{\rm tot}(\lambda;\rho_0,\rho_1)
=
\Phi(\lambda;\rho_0,\rho_1)
+
\Phi_{\rm init}(\lambda;\rho_0,\rho_1).
\ee
Using \eqref{defPhi}, the factors involving $\sqrt2$ cancel and the result simplifies to
\bea
\Phi_{\rm tot}(\lambda;\rho_0,\rho_1)
=
\frac14
\Bigg\{
&&
\rho_0^2
\left[
e^{2\lambda}
-
3e^\lambda
+
2
\right]
+
\rho_0\rho_1
\left[
e^\lambda+e^{-\lambda}-2
\right]
+
\rho_1^2
\left[
e^{-2\lambda}
-
3e^{-\lambda}
+
2
\right]
\Bigg\}.
\eea
This leads to the following expressions for the first three cumulants
at large $t_f$ 
\bea \label{cum123}
\langle Q(t_f)\rangle_c
&\simeq &
\left[
\rho_0-\rho_1
+
\varepsilon\frac{W_0}{4D}
(\rho_0^2-\rho_1^2)
+
O(\varepsilon^2)
\right]
\sqrt{\frac{Dt_f}{\pi}}
,
\nn\\
\langle Q(t_f)^2\rangle_c
&\simeq&
\left[
\rho_0+\rho_1
-
\varepsilon\frac{W_0}{4D}
(\rho_0+\rho_1)^2
+
O(\varepsilon^2)
\right]
\sqrt{\frac{Dt_f}{\pi}}
,
\nn\\
\langle Q(t_f)^3\rangle_c
&\simeq&
\left[
\rho_0-\rho_1
-
5\varepsilon\frac{W_0}{4D}
(\rho_0^2-\rho_1^2)
+
O(\varepsilon^2)
\right]
\sqrt{\frac{Dt_f}{\pi}}.
\eea

It is useful to compare this result with the direct calculation of typical current
fluctuations in the stationary state. For Brownian particles, linearizing the
Dean--Kawasaki equation around the homogeneous stationary state of density $\rho$, as we did in Section \ref{sec:linearized_RTP} for the RTPs,
gives in Fourier space,
\be
\partial_t\delta\hat\rho(k,t)
=
-\Omega(k)\delta\hat\rho(k,t)
-ik\sqrt{2D\rho}\,\hat\xi(k,t),
\qquad
\Omega(k)=Dk^2+\varepsilon\rho k h(k).
\ee
The stationary density correlation function is therefore
\be
C(k,t)
=
\frac{D\rho k^2}{\Omega(k)}
e^{-\Omega(k)|t|}.
\ee
Using the general relation \eqref{Qvar_Css_formula}, we obtain
\be
\langle Q(t_f)^2\rangle
=
\frac{2D\rho}{\pi}
\int_0^\infty dk\,
\frac{1-e^{-\Omega(k)t_f}}
{Dk^2+\varepsilon\rho k h(k)}.
\ee
For a short-range interaction, $h(k)\simeq W_0k$ at small $k$. Using the effective
diffusion coefficient introduced above, $D_{\rm eff}=D+\varepsilon\rho W_0$,
the long-time behavior becomes
\bea
\langle Q(t_f)^2\rangle
&\simeq&
\frac{2D\rho}{\pi D_{\rm eff}}
\int_0^\infty dk\,
\frac{1-e^{-D_{\rm eff}k^2t_f}}{k^2}
=
2\rho\sqrt{\frac{Dt_f}{\pi}}
\sqrt{\frac{D}{D_{\rm eff}}},
\eea
Expanding to first order in $\varepsilon$ gives
\be
\langle Q(t_f)^2\rangle
\simeq
2\rho\sqrt{\frac{Dt_f}{\pi}}
\left[
1-\varepsilon\frac{\rho W_0}{2D}
+O(\varepsilon^2)
\right].
\ee
This agrees exactly with the variance obtained in \eqref{cum123} from the MFT for
$\rho_0=\rho_1=\rho$ when the initial condition is sampled from the interacting
equilibrium measure. In contrast, it differs from the result \eqref{var2} obtained for a Poisson
initial condition, as expected since the latter does not contain the stationary
interaction-induced density correlations.
\\

\subsubsection{Weakly interacting RTPs on the line: Annealed initial condition from the non-interacting measure}

We now turn to RTPs and consider the same weak pair interaction as above, with no
additional thermal noise, i.e. $D=0$,
\be
\dot x_i(t)
=
v_0\sigma_i(t)
-
\varepsilon\sum_{j\neq i}W'(x_i(t)-x_j(t)),
\qquad
\sigma_i=\pm1.
\ee
For simplicity, we consider the annealed initial condition drawn from the
non-interacting RTP measure, already studied in the free case. This preparation means
that the weak interaction acts only during the observation time. Thus the two initial
densities $\rho_\pm(x,0)$ are independent Poisson fields, and at $\varepsilon=0$
the tilted optimal fields are precisely the free RTP saddle $F_\pm,G_\pm$ obtained above.

The interaction contribution to the MFT action is
\be
{\cal S}_1
=
-\int_0^{t_f}dt\int dx\,
{\cal H}[\rho_s](x,t)
\left[
\rho_+(x,t)\partial_xp_+(x,t)
+
\rho_-(x,t)\partial_xp_-(x,t)
\right].
\ee
The spatial surface term generated by the integration by parts in the general MFT
action vanishes on the full line. 

As in the Brownian case, the first-order correction to the optimal path itself does not
contribute. Writing 
\be
\Sigma_\varepsilon^{\rm an,0}(\lambda)
=
\Sigma_0^{\rm an}(\lambda)
+
\varepsilon\Sigma_1^{\rm an,0}(\lambda)
+
O(\varepsilon^2),
\ee
we therefore have
$\Sigma_1^{\rm an,0}(\lambda)
=-{\cal S}_1[\rho_\pm^{(0)},p_\pm^{(0)}]$.
In terms of the Cole--Hopf variables introduced in the free RTP calculation,
\be \label{Sigma1_RTP_exact}
\Sigma_1^{\rm an,0}(\lambda)
=
-\int_0^{t_f}dt\int dx \int dy\,
W'(x-y)\,
\rho_s^{(0)}(y,t)\,
B^{(0)}(x,t),
\ee
where
\bea
\rho_s^{(0)}
&=&
F_+G_+ + F_-G_-
=
\frac12\left(F_sG_s+F_dG_d\right),
\nn\\
B^{(0)}
&=&
F_+\partial_xG_+ + F_-\partial_xG_-
=
\frac12\left(F_s\partial_xG_s+F_d\partial_xG_d\right).
\eea
All the fields entering \eqref{Sigma1_RTP_exact} are thus known explicitly from
\eqref{Urealtime}, \eqref{Fs_real}, \eqref{Fd_real}, \eqref{Gs_sol_freeRTP} and
\eqref{Gd_sol_freeRTP}.

For a short-range even interaction, the same gradient expansion as in the Brownian
case can be used wherever the free saddle varies smoothly on scales large compared
with the interaction range $a$. 
To leading order,
$\int dy\,W'(x-y)\rho_s^{(0)}(y,t)
\simeq W_0\partial_x\rho_s^{(0)}(x,t)$, with
$W_0=\int dr\,W(r)$, and hence
\be \label{Sigma1_RTP_local}
\Sigma_1^{\rm an,0}(\lambda)
\simeq
-W_0
\int_0^{t_f}dt\int dx\,
B^{(0)}(x,t)\,
\partial_x\rho_s^{(0)}(x,t).
\ee
Equivalently,
\be \label{SigmaannRTP}
\Sigma_1^{\rm an,0}(\lambda)
\simeq
-\frac{W_0}{4}
\int_0^{t_f}dt\int dx\,
\left[
F_s\partial_xG_s+F_d\partial_xG_d
\right]
\partial_x\left[
F_sG_s+F_dG_d
\right].
\ee
This local form will apply in the diffusive regime below. It is, however, not
uniformly valid in the persistent regime, where the free RTP saddle contains sharp
ballistic fronts.
\\

\noindent{\bf Diffusive regime.} We first consider $\gamma t_f\gg1$. On scales
$x=O(\sqrt{D_at_f})$, with $D_a=v_0^2/(2\gamma)$, the orientation degree of
freedom has locally relaxed and the free RTP saddle reduces to the Brownian one:
$F_+\simeq F_-\simeq F_0/2$ and $G_+\simeq G_-\simeq G_0$, where $F_0,G_0$
are the Brownian saddle-point fields \eqref{FG0_Brownian_ann} with $D$ replaced by $D_a$. Consequently
$\rho_s^{(0)}\simeq F_0G_0$ and $B^{(0)}\simeq F_0\partial_xG_0$, so that
\eqref{Sigma1_RTP_local} reduces precisely to the Brownian expression
\eqref{Sigma1_Brownian_local}, with $D\to D_a$. Using the result obtained above,
\be \label{SigmaannRTPexpl}
\Sigma_\varepsilon^{\rm an,0}(\lambda)
=
\sqrt{\frac{D_at_f}{\pi}}
\left\{
\rho_0(e^\lambda-1)
+
\rho_1(e^{-\lambda}-1)
-
\varepsilon\frac{W_0}{D_a}
\Phi(\lambda;\rho_0,\rho_1)
+
O(\varepsilon^2)
\right\}
+
o(\sqrt{t_f}),
\ee
where $\Phi$ is the same function as in \eqref{defPhi}. Thus, at long times, the
leading weak-interaction correction for RTPs coincides with the Brownian result upon
replacing $D$ by $D_a=v_0^2/(2\gamma)$.
\\

\noindent{\bf Persistent ballistic regime.} Let us first recall 
the result for the annealed SCGF without interactions, which reads,
from \eqref{muexpr} and \eqref{sigmafinal_RTP_annealed}
\be
\Sigma_0^{\rm an}(\lambda)
=
\frac{v_0t_f}{2}  \chi(\tau_f)
\left[
\rho_0(e^\lambda-1)
+
\rho_1(e^{-\lambda}-1)
\right]\quad , \quad  \chi(\tau_f) =
e^{-\tau_f}
\left[
I_0(\tau_f)+I_1(\tau_f)
\right], \quad , \quad 
\tau_f=\gamma t_f , 
\ee
which is exact for arbitrary $t_f$. 
Hence all cumulants $\langle Q(t_f)^n \rangle_c$, with integer $n \geq 1$, once divided by $t_f$, are functions of the parameter 
$\tau_f=\gamma t_f$, more precisely they are, for $n \geq 1$
\be
\left\langle Q(t_f)^n\right\rangle_c
=
\frac{v_0t_f}{2}\,
\chi(\tau_f)
\left[
\rho_0+(-1)^n\rho_1
\right].
\ee
In the presence of a sufficiently short-range interaction with a characteristic decay length $a$ (e.g. $a=a_0$ for the regularized Riesz gas \eqref{Riesz_regularized}), we consider 
the following regime which we denote the persistent ballistic regime
\be \label{ballistic_scaling_RTP}
\rho v_0t_f\gg1,
\qquad
\frac{v_0t_f}{a}\gg1,
\qquad
\tau_f=\gamma t_f=O(1),
\ee
where $\rho$ denotes a typical density scale. 
The first condition ensures a large
number of particles in the ballistic region contributing to the current, the second
makes the interaction short-ranged on the scale $v_0t_f$, and the third keeps
persistence effects finite. This can be viewed as the joint limit
$t_f\to\infty$, $\gamma\to0$ with $\tau_f$ fixed.

In this regime it is important not to apply the gradient expansion
\eqref{SigmaannRTP} uniformly, since the free RTP saddle contains sharp
ballistic fronts. We instead start from the exact nonlocal expression
\eqref{Sigma1_RTP_exact} and isolate the singular contribution associated
with these fronts. Let $\Delta=t_f-t$. Recall that the Cole--Hopf fields
$G_\sigma=e^{p_\sigma}$ obey the backward evolution equations, with their
boundary condition imposed at the final time $t_f$. A particle in
orientation state $\sigma=\pm1$ at time $t$ can stay in the same state until $t=t_f$ without
tumbling, with probability $e^{-\gamma\Delta}$. This no-tumble contribution
produces a discontinuity in $G_\sigma(x,t)$ when the corresponding
ballistic trajectory reaches the origin, namely at
\be
x_\sigma^{\rm b}(t)=-\sigma v_0\Delta.
\ee
\begin{figure}[t]
\centering
\includegraphics[width=0.6\columnwidth]{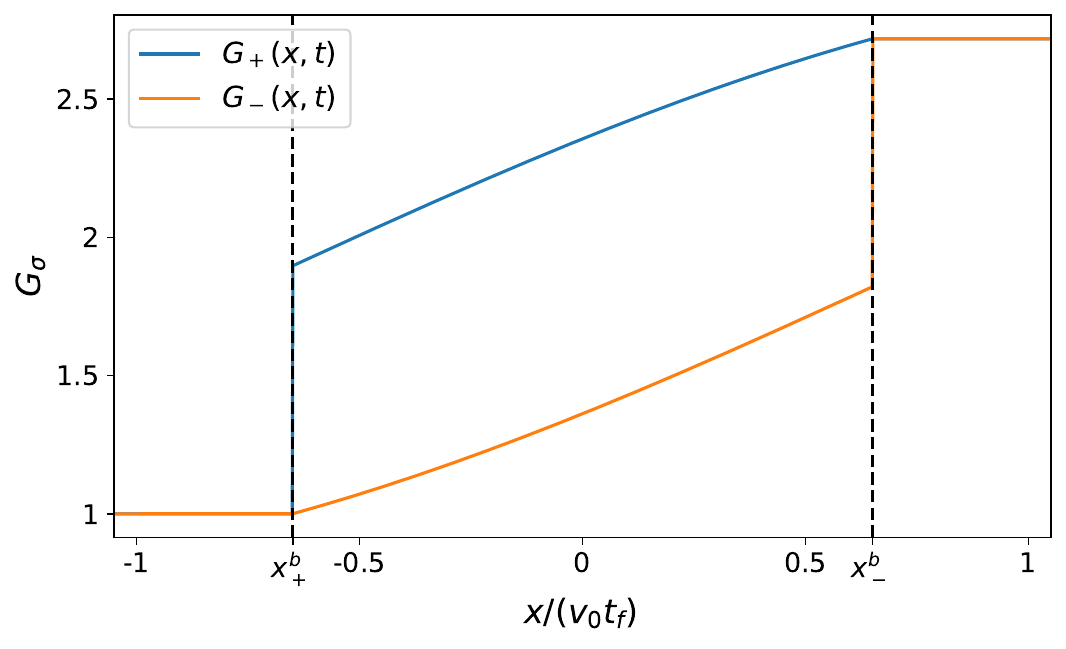}
\caption{
Backward Cole--Hopf fields $G_\pm(x,t)$ for a representative free RTP
saddle in the persistent ballistic regime.
The discontinuities occur at the backward ballistic fronts
$x_\sigma^{\rm b}(t)=-\sigma v_0(t_f-t)$:
$G_+$ jumps at the left front and $G_-$ at the right front.
They correspond to the no-tumble contribution between $t$ and $t_f$ and
produce the Dirac peaks in $\partial_xG_\sigma$ appearing in
Eq.~\eqref{G_ballistic_jump}.
Here $\tau_f=1$, $t/t_f=0.35$, $\lambda=1$, $\rho_0=1$, and $\rho_1=0.5$.
}
\label{fig:G_ballistic_fronts}
\end{figure}

Denoting $[f]_{x_0}=f(x_0^+)-f(x_0^-)$, the jump is
\be
[G_\sigma]_{x_\sigma^{\rm b}}
=
(e^\lambda-1)e^{-\gamma\Delta},
\ee
or equivalently, see Fig. \ref{fig:G_ballistic_fronts}
\be \label{G_ballistic_jump}
\partial_xG_\sigma(x,t)
=
\left(\partial_xG_\sigma\right)_{\rm reg}
+
(e^\lambda-1)e^{-\gamma\Delta}
\delta\left(x-x_\sigma^{\rm b}(t)\right).
\ee
where $(\partial_xG_\sigma)_{\rm reg}$ denotes the part locally regular at
$x_\sigma^{\rm b}(t)$, i.e. without the Dirac mass associated with this backward
front.
This also follows directly from
\eqref{Gs_sol_freeRTP}--\eqref{Gd_sol_freeRTP}: the left backward front
belongs to $G_+$ and the right one to $G_-$.

Away from the isolated crossings with the ballistic discontinuities of
the forward fields, $F_\sigma$ is smooth at $x_\sigma^{\rm b}$. We denote
\be\label{defFb}
F_\sigma^{\rm b}(t)=F_\sigma(x_\sigma^{\rm b}(t),t)\;.
\ee
Since
$\rho_\sigma^{(0)}=F_\sigma G_\sigma$, the corresponding local jump is
\be
[\rho_\sigma^{(0)}]_{x_\sigma^{\rm b}}
=
F_\sigma^{\rm b}(t)
(e^\lambda-1)e^{-\gamma\Delta}.
\ee
Since the forward fields $F_\pm$ also contain ballistic discontinuities at
$x=\pm v_0t$, the density decomposition below is understood locally in a
neighborhood of a given backward front $x_\sigma^{\rm b}$:
\be
\rho_s^{(0)}(x,t)
=
\rho_{s,{\rm reg}}^{(0)}(x,t)
+
[\rho_\sigma^{(0)}]_{x_\sigma^{\rm b}}\,
\theta\left(x-x_\sigma^{\rm b}(t)\right).
\ee
Here ``reg'' means regular at the selected backward front and may still
contain the forward ballistic discontinuities.

By contrast, since $B^{(0)}=F_+\partial_xG_++F_-\partial_xG_-$ contains
no derivatives of $F_\pm$, the forward fronts generate no delta peaks in
$B^{(0)}$. Its only distributional singularities are those at the two
backward fronts, so that, away from their isolated crossings with the
forward fronts,
\be
B^{(0)}(x,t)
=
B_{\rm reg}^{(0)}(x,t)
+
\sum_{\sigma=\pm}
[\rho_\sigma^{(0)}]_{x_\sigma^{\rm b}}\,
\delta\left(x-x_\sigma^{\rm b}(t)\right),
\ee
where $B_{\rm reg}^{(0)}$ is an ordinary piecewise-smooth function and
may have finite jumps at $x=\pm v_0t$.

We now evaluate the leading contribution to \eqref{Sigma1_RTP_exact} by
inserting the above decomposition of $\rho_s^{(0)}$ and $B^{(0)}$ and
retaining, for each $\sigma$, the product of the density step at the
backward front $x_\sigma^{\rm b}$ with the delta peak of $B^{(0)}$ at the
same front. We assume that $W(x)$ is even and regular at the origin.
Since only $W'$ enters the microscopic dynamics, we choose the additive
constant such that $W(x)\to0$ for $|x|\to\infty$. The interaction field
generated by this density step, evaluated at the same backward front, is,
see also Appendix~\ref{app:RTP_ballistic_front0}
\be
\int dy\,W'(x_\sigma^{\rm b}-y)\,
[\rho_\sigma^{(0)}]_{x_\sigma^{\rm b}}\,
\theta(y-x_\sigma^{\rm b})
=
[\rho_\sigma^{(0)}]_{x_\sigma^{\rm b}}
\int_{-\infty}^{0}dr\,W'(r)
=
[\rho_\sigma^{(0)}]_{x_\sigma^{\rm b}}\,W(0).
\ee
Thus the self-interaction of a sharp backward front probes the potential
difference $W(0)-W(\infty)$, rather than the integrated interaction $W_0$
which controls the smooth long-wavelength contribution. All remaining
terms---including smooth contributions, interactions between distinct
backward fronts, and couplings to the forward fronts---are subleading by
$O(a/(v_0t_f))$ in the ballistic scaling regime. Their detailed
analysis is given in Appendix~\ref{app:RTP_ballistic_front0}.

Combining this interaction field with the delta peak in $B^{(0)}$, and
summing the two backward fronts, gives, see Appendix~\ref{app:RTP_ballistic_front0},
\be \label{Sigma1_RTP_ballistic_front}
\Sigma_{1,{\rm fr}}^{\rm an,0}(\lambda)
=
-W(0)(e^\lambda-1)^2
\int_0^{t_f}dt\,
e^{-2\gamma(t_f-t)}
\left[
\left(F_+^{\rm b}(t)\right)^2
+
\left(F_-^{\rm b}(t)\right)^2
\right],
\ee
where the functions $F_\sigma^{\rm b}(t)$ are defined in \eqref{defFb}.
Thus the original space-time nonlocal integral has been reduced, to leading
order in the ballistic regime, to a single time integral.

The expressions of the functions $F_\sigma^{\rm b}$ follow explicitly from
\eqref{Fs_real} and \eqref{Fd_real}. Introducing
$\delta\rho_\lambda=\rho_0-\rho_1e^{-\lambda}$ and
$U_{\rm b}(t)=U(v_0\Delta,t)$, one obtains
\bea
F_+^{\rm b}(t)
&=&
\frac{\rho_0}{2}
-
\frac{\delta\rho_\lambda}{2}U_{\rm b}(t)
+
\frac{\delta\rho_\lambda}{4}
e^{-\gamma t}
I_0\left(\gamma\sqrt{t^2-\Delta^2}\right)
\theta(t-\Delta),
\nn\\
F_-^{\rm b}(t)
&=&
\frac{\rho_1e^{-\lambda}}{2}
+
\frac{\delta\rho_\lambda}{2}U_{\rm b}(t)
-
\frac{\delta\rho_\lambda}{4}
e^{-\gamma t}
I_0\left(\gamma\sqrt{t^2-\Delta^2}\right)
\theta(t-\Delta).
\eea
where $\Delta=t_f-t$. 
Together with \eqref{Urealtime},
Eq.~\eqref{Sigma1_RTP_ballistic_front} is therefore an explicit
one-dimensional integral.

Its ballistic scaling form is most conveniently expressed by setting
$u=t/t_f$. The combination entering $F_\sigma^{\rm b}$ can be written in
terms of the orientation-resolved crossing probability $U_-$ introduced in
\eqref{Upmdef}:
\bea \label{R_RTP_ballistic}
{\cal R}_{\tau_f}(u)
&:=&
U_-\left(v_0t_f(1-u),ut_f\right)
\nn\\
&=&
\frac{\tau_fu}{2}e^{-\tau_fu}
\int_{(1-u)/u}^{1}dq\,
\left[
I_0\left(\tau_fu\sqrt{1-q^2}\right)
+
\sqrt{\frac{1-q}{1+q}}\,
I_1\left(\tau_fu\sqrt{1-q^2}\right)
\right],
\qquad \frac12\leq u\leq1,
\eea
while ${\cal R}_{\tau_f}(u)=0$ for $u<1/2$. The first equality gives a
simple probabilistic interpretation of the scaling function: it is the
crossing probability for a particle initially in the $-$ orientation,
evaluated at the backward ballistic front. The derivation of the second
equality is given in Appendix~\ref{app:RTP_ballistic_front}.

Defining
\be \label{I_RTP_ballistic}
{\cal I}(\tau_f)
=
\int_{1/2}^{1}du\,
e^{-2\tau_f(1-u)}
{\cal R}_{\tau_f}(u)
\left[
1-{\cal R}_{\tau_f}(u)
\right],
\ee
one finds that Eq.~\eqref{Sigma1_RTP_ballistic_front} takes the scaling form
at large $t_f$ and fixed $\tau_f=\gamma t_f$
\be \label{Sigma1_RTP_ballistic_scaling}
\Sigma_{1}^{\rm an,0}(\lambda) \simeq 
\Sigma_{1,{\rm fr}}^{\rm an,0}(\lambda)
= 
-t_f \, W(0)(e^\lambda-1)^2\,
{\cal A}(\tau_f;\lambda,\rho_0,\rho_1),
\ee
where
\be \label{A_RTP_ballistic}
{\cal A}(\tau_f;\lambda,\rho_0,\rho_1)
=
\frac{1-e^{-2\tau_f}}{8\tau_f}
\left(
\rho_0^2+\rho_1^2e^{-2\lambda}
\right)
-
\frac12
\left(
\rho_0-\rho_1e^{-\lambda}
\right)^2
{\cal I}(\tau_f).
\ee
The derivation and the asymptotic analysis of ${\cal I}(\tau_f)$ are given
in Appendix~\ref{app:RTP_ballistic_front}. Thus the leading interaction correction 
to the SCGF is linear in $t_f$ in the strongly
persistent regime and is controlled by $W(0)$ rather than by $W_0$.

\begin{figure}[t]
\centering
\includegraphics[width=0.6\columnwidth]{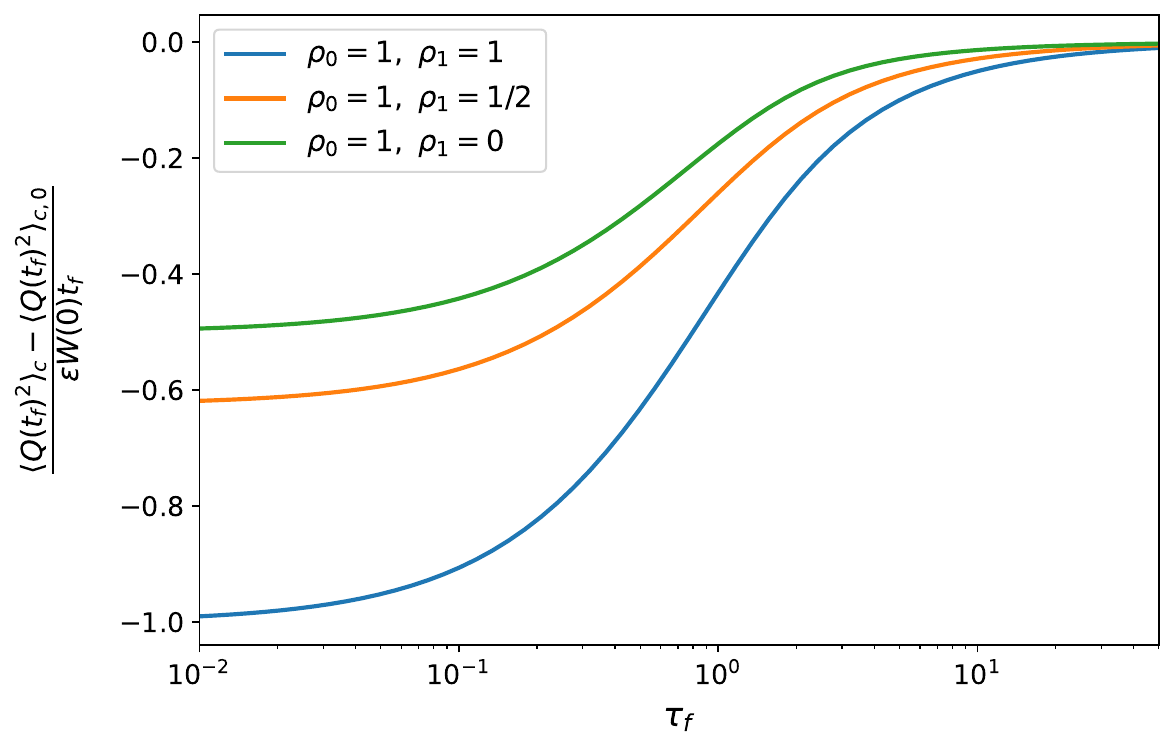}
\caption{
First-order interaction correction to the current variance in the persistent
ballistic regime, normalized by $\varepsilon W(0)t_f$, for fixed
$\rho_0=1$ and $\rho_1=1,1/2,0$.
The plotted quantity is
$\big[\langle Q(t_f)^2\rangle_c-\langle Q(t_f)^2\rangle_{c,0}\big]/
[\varepsilon W(0)t_f]$.
For a repulsive interaction, $\varepsilon W(0)>0$, the correction is
negative for all $\tau_f$, corresponding to a suppression of current
fluctuations. The correction approaches a finite value in the strongly
persistent limit and decays to zero for $\tau_f\gg1$.
}
\label{fig:RTP_variance_correction}
\end{figure}

By expanding the SCGF (\ref{Sigma1_RTP_ballistic_scaling}) in powers of $\lambda$, one can obtain the cumulants of $Q(t_f)$ in the large $t_f$ limit and fixed $\tau_f$. Let us display the first two cumulants in that regime. The first cumulant at large time is not corrected by the interaction to this order, 
\be
\frac{1}{t_f} \left\langle Q(t_f)\right\rangle
\simeq
\frac{v_0}{2}\chi(\tau_f)
\left(
\rho_0-\rho_1
\right) + o(\epsilon).
\ee
while the second cumulant reads, at large $t_f$ 
\be
\frac{1}{t_f}  \left\langle Q(t_f)^2\right\rangle_c \simeq
\frac{v_0}{2}
\chi(\tau_f)
\left(
\rho_0+\rho_1
\right)
-
\varepsilon W(0)
\left[
\frac{1-e^{-2\tau_f}}{4\tau_f}
\left(
\rho_0^2+\rho_1^2
\right)
-
\left(
\rho_0-\rho_1
\right)^2
{\cal I}(\tau_f)
\right] + o(\epsilon) .
\ee
For repulsive interactions the correction is negative and is plotted in Fig. 
\ref{fig:RTP_variance_correction}.

From these results, let us discuss now various limits.
In the strongly persistent limit $\tau_f\ll1$,
\be \label{A_RTP_ballistic_smalltau}
{\cal A}(\tau_f;\lambda,\rho_0,\rho_1)
=
\frac14
\left(
\rho_0^2+\rho_1^2e^{-2\lambda}
\right)
-\tau_f
\left[
\frac14
\left(
\rho_0^2+\rho_1^2e^{-2\lambda}
\right)
+
\frac{
\left(
\rho_0-\rho_1e^{-\lambda}
\right)^2
}{16}
\right]
+
O(\tau_f^2).
\ee
In particular,
\be \label{Sigma1_RTP_ballistic_tau0}
\Sigma_{1,{\rm fr}}^{\rm an,0}(\lambda)
\simeq
-\frac{W(0)t_f}{4}
\left[
\rho_0^2(e^\lambda-1)^2
+
\rho_1^2(e^{-\lambda}-1)^2
\right],
\qquad
\tau_f\ll1.
\ee
For $\tau_f\ll1$, tumbles are rare and the two ballistic fronts contribute
independently, with amplitudes set by $\rho_0$ and $\rho_1$, respectively.

In the opposite limit $\tau_f\gg1$, the front contribution obeys
\be \label{A_RTP_ballistic_largetau}
{\cal A}(\tau_f;\lambda,\rho_0,\rho_1)
=
\frac{
\left(
\rho_0+\rho_1e^{-\lambda}
\right)^2
}{
16\tau_f
}
+
O(\tau_f^{-2}),
\ee
and therefore, at fixed $\gamma$,
\be \label{Sigma1_RTP_ballistic_largetau}
\Sigma_{1,{\rm fr}}^{\rm an,0}(\lambda)
=
-\frac{W(0)}{16\gamma}
(e^\lambda-1)^2
\left(
\rho_0+\rho_1e^{-\lambda}
\right)^2
+
O\left(\frac{1}{\gamma\tau_f}\right).
\ee
Thus the ballistic-front contribution ceases to grow with $t_f$ at fixed
$\gamma$ and eventually becomes subleading to the smooth diffusive
interaction correction, which grows as $\sqrt{t_f}$.

The contributions neglected in retaining only the backward-front
self-interactions are analyzed in Appendix~\ref{app:RTP_ballistic_front0}.
For fixed $\tau_f=O(1)$ they are
$O(\rho^2W_0/v_0)$, whereas
$\Sigma_{1,{\rm fr}}^{\rm an,0}=O(\rho^2W(0)t_f)$.
Thus, for a regular short-range potential with
$W_0=O(aW(0))$, their relative magnitude is
$O(a/(v_0t_f))$, and the backward-front contribution is leading in the
regime \eqref{ballistic_scaling_RTP}. Perturbation theory further requires
$|\varepsilon|\rho|W(0)|/v_0\ll1$. If $W(0)=0$, the front
self-interaction vanishes and the contributions proportional to $W_0$
must instead be retained.

\section{MFT for Fixed boundary conditions on the interval $[0,L]$} \label{sec:MFT_reservoirs}

In this section we study the total stationary current large-deviation problem for open boundaries. We apply the MFT formalism introduced in Section~\ref{sec:MFT_gen} to a different setting, where the particles are confined on an interval $[0,L]$, with reservoirs of particles at the boundaries, with densities $\rho_a$ and $\rho_b$, respectively. This is similar to the setting studied in Refs.~\cite{BodineauDerrida2004,BD2005} for boundary-driven diffusive systems, and in particular for the WASEP and TASEP in Ref.~\cite{BD2005} -- see also Ref.~\cite{SuzukiSasamoto2026} for a recent MFT treatment in which the stochastic exchange of particles with the reservoirs is included explicitly. Here we study RTPs with internal velocities $\pm v_0$, as well as their Brownian counterparts.

For this system, we consider a slightly different observable, namely the current flowing through the system, integrated over both time and space,
\be \label{integrated_current_def2}
\mathcal{Q}_{t_f}(L)
=
\frac{1}{L}
\int_0^{t_f}dt
\int_0^Ldx\,J_s(x,t) \;,
\ee
where the current $J_s$ is defined in \eqref{LJsd}-\eqref{defJsbar}. A similar space-time integrated current has been considered for active lattice gases in Ref.~\cite{Agranov2022}. By the continuity equation, $\mathcal{Q}_{t_f}(L)$ differs from the current integrated through any fixed cross-section only by a contribution depending on the initial and final density profiles. This contribution is subextensive in $t_f$ and therefore does not affect the long-time scaled cumulant generating function. We will calculate the scaled cumulant generating function
\be \label{Sigma_tilde_def}
\tilde\Sigma(\lambda)
=
\lim_{t_f\to+\infty}
\frac{1}{t_f}
\ln\left\langle
e^{\lambda\mathcal{Q}_{t_f}(L)}
\right\rangle \;.
\ee
Equivalently, the probability of observing a time-averaged current
$
j=\frac{\mathcal{Q}_{t_f}(L)}{t_f}
$
takes, at long times, the large-deviation form
\be \label{current_LDP_reservoirs}
P\left(\mathcal{Q}_{t_f}(L)\simeq j t_f\right)
\asymp
e^{-t_f I(j)} \;,
\ee
where the rate function $I(j)$ is related to $\tilde\Sigma(\lambda)$ by the Legendre--Fenchel transform
\be \label{Legendre_current_reservoirs}
I(j)
=
\sup_{\lambda}
\left[
\lambda j-\tilde\Sigma(\lambda)
\right] \;.
\ee

In this section we restrict ourselves to stationary solutions of the saddle-point equations. For passive diffusive systems, this corresponds to the stationary variational realization of the Bodineau--Derrida additivity principle~\cite{BodineauDerrida2004}. In terms of the MFT action ${\cal S}[\rho,J]$, the assumption can be written schematically, for a prescribed time-averaged current $j$, as
\be \label{BD_stationary_principle}
P\left(\mathcal{Q}_{t_f}(L) \simeq j t_f\right)
\asymp
\exp\left[
-\min_{\rho(x)}
{\cal S}\big[
\rho(x,t)=\rho(x),\,J(x,t)=j
\big]
\right],
\ee
where the minimization is performed over stationary density profiles satisfying the reservoir boundary conditions. Equivalently, the stationary approximation to the current rate function is obtained by minimizing the MFT action per unit time over stationary profiles carrying the prescribed constant current.

The assumption that the optimal profile is stationary is not always valid. Time-dependent optimal histories and associated dynamical phase transitions are known to occur in passive diffusive systems
\cite{BertiniCurrent2005,BodineauDerridaDPT2005,ShpielbergDonAkkermans2017,BaekKafriLecomte2017},
including systems driven by boundary reservoirs. Similarly, in active lattice gases a transition to phase-separated traveling-wave optimal profiles has been found within MFT~\cite{Agranov2022}. Since the systems and situations considered here are simpler, we will not consider such possible time-dependent solutions here, and restrict throughout this section to the stationary branch of the saddle-point equations.

As in the previous section, we will first solve these stationary equations for free non-interacting particles (i.e. $V'(x)=0$ and $W'(x)=0$), before extending the results to the presence of interactions. 
In each case we will begin by considering Brownian particles 
as a benchmark.

\subsection{Non-interacting Brownian particles}

Let us first consider non-interacting Brownian particles. 
In this case there is a single density field $\rho(x,t)$ and a single current $J(x,t)$,
satisfying the conservation equation $\partial_t\rho+\partial_x J=0$,
and the deterministic part of the current is $\bar J=-D\partial_x\rho$.
Some details are provided in Appendix~\ref{app:Brownian_boundary_conditions}. 

Introducing an auxiliary field $p(x,t)$ to impose the continuity equation,
the generating function can be written schematically as
\bea \label{LD_Br_reservoirs_current}
\left\langle e^{\lambda\mathcal{Q}_{t_f}(L)}\right\rangle
&\simeq&
{\cal N}
\int{\cal D}\rho\,{\cal D}J\,{\cal D}p\,
P[\rho(x,0)]
\,e^{-\widetilde{\cal S}} \;,
\eea
where we have defined the tilted action
\be \label{Stilde_Br_def}
\widetilde{\cal S}
=
{\cal S}-\lambda\mathcal{Q}_{t_f}(L) 
=
\int_0^{t_f}dt\int_0^Ldx
\left[
\frac{(J+D\partial_x\rho)^2}{4D\rho}
+
p(\partial_t\rho+\partial_xJ)
-
\frac{\lambda}{L}J
\right] \;.
\ee
Here $P[\rho(x,0)]$ denotes the initial distribution. In the long-time limit
$t_f\to\infty$, its contribution $-\ln P[\rho(x,0)]$ to the action is
$O(1)$ in $t_f$ and does not affect the leading stationary saddle-point
equations or the scaled cumulant generating function
$\widetilde\Sigma(\lambda)$. After integrating the term $p\,\partial_xJ$ by parts in space, variation
with respect to the current in the bulk gives
\be \label{saddle_J_Br_reservoirs}
J
=
-D\partial_x\rho
+
2D\rho
\left(
\partial_xp+\frac{\lambda}{L}
\right) \;.
\ee
It is therefore natural to introduce the shifted auxiliary field
$\hat p(x,t)
=
p(x,t)+\frac{\lambda x}{L}$. Before eliminating the current, let us specify the spatial boundary
conditions. The particle reservoirs impose fixed densities,
\be \label{BC_rho_br}
\rho(0,t)=\rho_a
\quad,\quad
\rho(L,t)=\rho_b \;.
\ee
The boundary currents, on the other hand, are unconstrained. As shown
explicitly in Appendix~\ref{app:Brownian_boundary_conditions}, see also 
Refs \cite{KrapivskyVoid,Agranov2016} for related problems, the
corresponding boundary variation of the action then imposes
\be \label{BCphat_br}
p(0,t)=p(L,t)=0
\quad\Leftrightarrow\quad
\hat p(0,t)=0
\quad,\quad
\hat p(L,t)=\lambda \;.
\ee
In particular, the spatial boundary term generated by the integration by
parts of $p\,\partial_xJ$ vanishes on the saddle. Note that the fixed-density
boundary conditions \eqref{BC_rho_br} can also be viewed as the
fast-boundary limit of the more general stochastic-reservoir MFT of Suzuki
and Sasamoto~\cite{SuzukiSasamoto2026}; the relation between the two
descriptions is discussed in
Appendix~\ref{app:Brownian_boundary_conditions}.

Using these boundary conditions and eliminating the bulk current $J$, the
tilted action reduces to
\be \label{S_tilde_Br_reservoirs}
\widetilde{\cal S}
=
\int_0^{t_f}dt\int_0^Ldx\,
\left(
\hat p-\frac{\lambda x}{L}
\right)\partial_t\rho
-
\int_0^{t_f}dt\,{\sf \widetilde H} \;,
\ee
where the Hamiltonian reads
\be \label{H_Br_reservoirs}
{\sf \widetilde H}
=
D\int_0^Ldx\,
\left[
\rho(\partial_x\hat p)^2
-
(\partial_x\rho)(\partial_x\hat p)
\right] \;.
\ee
The corresponding saddle-point equations are
\bea \label{SPeq_Br_reservoirs1}
\partial_t\rho
&=&
\partial_x
\left[
D\partial_x\rho
-
2D\rho\,\partial_x\hat p
\right] \;,
\\
\partial_t\hat p
&=&
-D(\partial_x\hat p)^2
-D\partial_x^2\hat p \;. \label{SPeq_Br_reservoirs2}
\eea

We now restrict ourselves to the stationary branch of the saddle-point
equations, as discussed above in connection with the
Bodineau--Derrida additivity principle. Equations
\eqref{SPeq_Br_reservoirs1}-\eqref{SPeq_Br_reservoirs2} then reduce to
\bea \label{eqstat_brownian1}
0
&=&
\partial_x
\left[
D\partial_x\rho
-
2D\rho\,\partial_x\hat p
\right] \;,
\\
0
&=&
-D(\partial_x\hat p)^2
-D\partial_x^2\hat p \;.
\label{eqstat_brownian2}
\eea

In terms of the Cole--Hopf variables
$F=\rho e^{-\hat p}$, 
$G=e^{\hat p}$, 
these equations simply become
\be
\partial_x^2F=0
\quad,\quad
\partial_x^2G=0 \;.
\ee
The four integration constants are determined by
\eqref{BC_rho_br} and \eqref{BCphat_br}. We thus obtain
\bea \label{SPsolF_Br}
F(x)
&=&
\rho_a
+
\left(
\rho_b e^{-\lambda}-\rho_a
\right)\frac{x}{L} \;,
\\
G(x)
&=&
1+
\left(
e^\lambda-1
\right)\frac{x}{L} \;.
\label{SPsolG_Br}
\eea
Going back to the original variables $\rho=FG$ and
$\hat p=\ln G$, this gives
\bea
\rho(x)
&=&
\rho_a
+
\left(
\rho_a e^\lambda
+
\rho_b e^{-\lambda}
-
2\rho_a
\right)\frac{x}{L}
+
\left(
\rho_a+\rho_b
-\rho_a e^\lambda
-\rho_b e^{-\lambda}
\right)
\left(\frac{x}{L}\right)^2 \;,
\\
\hat p(x)
&=&
\ln\left[
1+
\left(e^\lambda-1\right)\frac{x}{L}
\right] \;.
\eea
For $\lambda=0$, one recovers the usual stationary linear density profile
$
\rho(x)
=
\rho_a
+
(\rho_b-\rho_a)\frac{x}{L}$,
while $\hat p(x)=0$.

Let us now evaluate the scaled cumulant generating function by computing
the tilted action \eqref{S_tilde_Br_reservoirs} at the saddle point. The
term proportional to $\lambda$ in the canonical part of the action is a
temporal boundary contribution,
\be
-\frac{\lambda}{L}
\int_0^{t_f}dt\int_0^Ldx\,x\,\partial_t\rho
=
-\frac{\lambda}{L}
\int_0^Ldx\,x\,
\left[\rho(x,t_f)-\rho(x,0)\right] \;.
\ee
It is therefore subextensive in $t_f$. Likewise, the term
involving $\hat p\,\partial_t\rho$ vanishes on the stationary bulk
solution, while possible initial and final temporal boundary layers
contribute only terms of order $O(1)$ in the long-time limit. Thus, for a
stationary saddle, one has 
\be \label{Sigma_H_relation_Br}
\widetilde{\cal S}^*
=
-t_f\,{\sf\widetilde H}^*
+O(1)
\quad\Rightarrow\quad
\widetilde\Sigma(\lambda)
=
{\sf\widetilde H}^* \;.
\ee

For stationary profiles the Hamiltonian
\eqref{H_Br_reservoirs} takes, in terms of the Cole--Hopf variables,
the particularly simple form
\be
{\sf \widetilde H}
=
-D\int_0^Ldx\,
(\partial_xF)(\partial_xG) \;.
\ee
Evaluating this expression using the saddle-point solution
\eqref{SPsolF_Br}-\eqref{SPsolG_Br}, we obtain
\be \label{Sigma_Br_reservoirs}
\widetilde\Sigma(\lambda)
=
\frac{D}{L}
\left[
\rho_a(e^\lambda-1)
+
\rho_b(e^{-\lambda}-1)
\right]
\;.
\ee
This result agrees with the steady-state fast-boundary limit of the
stochastic-reservoir formulation of Ref.~\cite{SuzukiSasamoto2026}.
The first two scaled cumulants read, at large $t_f$,
\be
\frac{1}{t_f}
\left\langle \mathcal{Q}_{t_f}(L)\right\rangle
\simeq
\widetilde\Sigma'(0)
=
\frac{D}{L}(\rho_a-\rho_b)
\quad,\quad
\frac{1}{t_f}
\left\langle [\mathcal{Q}_{t_f}(L)]^2\right\rangle_c
\simeq
\widetilde\Sigma''(0)
=
\frac{D}{L}(\rho_a+\rho_b) \;.
\ee
The first scaled cumulant recovers Fick's law. More generally,
Eq.~\eqref{Sigma_Br_reservoirs} has the form of the scaled cumulant
generating function of the difference of two independent Poisson currents,
with rates $D\rho_a/L$ and $D\rho_b/L$, respectively.

Finally one finds, upon Legendre transform, that the probability
distribution of the integrated current takes the large-time
large-deviation form
\be \label{LegendreTransformQ}
P(\mathcal{Q}_{t_f}(L) \simeq {\cal Q})
\asymp
\exp\left[
-\frac{\rho_aDt_f}{L}\,
\Psi\left(
\frac{L\mathcal{Q}}
{\rho_aDt_f}
\right)
\right] \;,
\ee
where $q
=
\frac{L\mathcal{Q}}
{D\rho_a t_f}$
is the dimensionless current and
\be \label{LDP_reservoirs_br}
\Psi(q) 
=
q\lambda_q
-
\widetilde{\widetilde\Sigma}(\lambda_q)
\quad,\quad
\widetilde{\widetilde\Sigma}(\lambda)
=
(e^\lambda-1)
+
\frac{\rho_b}{\rho_a}(e^{-\lambda}-1)
\quad,\quad
\lambda_q 
=
\ln\left(
\frac{q}{2}
+
\sqrt{
\frac{q^2}{4}
+
\frac{\rho_b}{\rho_a}
}
\right) \;.
\ee
This is the large deviation rate function of the difference of two Poisson variables.
It is thus exactly the same as \eqref{psi_annealed_br_step} with $Q(t_f) \to {\cal Q}_{t_f}(L)$, $\rho_0 \sqrt{Dt_f/\pi} \to \rho_a D t_f/L$ and $\rho_0,\rho_1\to\rho_a,\rho_b$.

\subsection{Non-interacting RTPs}

We now consider the same problem for non-interacting RTPs, and restrict to
the case without additional thermal diffusion, $D=0$. The dynamics is then
described by the two densities $\rho_\pm(x,t)$, with deterministic
translational currents $J_\pm=\pm v_0\rho_\pm$, together with stochastic
tumbling at rate $\gamma$. We use the MFT for the RTP's introduced in
Section~\ref{sec:MFT_gen}.

For the observable \eqref{integrated_current_def2}, one has
$J_s=J_++J_-$ and therefore
$\frac{\delta\mathcal Q_{t_f}(L)}{\delta J_\pm}=1/L$. Hence the shifted auxiliary
fields defined in \eqref{defhatp} become
\be \label{defhatp_reservoirs}
\hat p_\pm(x,t)
=
p_\pm(x,t)+\frac{\lambda x}{L} \;.
\ee
Moreover, in the limit $D\to0$, the covariance matrix $\mathbf C$ in
\eqref{saddle_point_J} vanishes, so that the saddle-point currents reduce
to their deterministic values,
$J_+^{\rm sp}=v_0\rho_+$ and $J_-^{\rm sp}=-v_0\rho_-$. Thus
$J_s=v_0(\rho_+-\rho_-)$ and
$\mathcal{Q}_{t_f}(L)=(v_0/L)\int_0^{t_f}dt\int_0^Ldx\,(\rho_+-\rho_-)$.

Starting from the general reduced action
\eqref{action_after_JK_general}, including its spatial surface term with $[L_1,L_2]=[0,L]$, we
define the tilted action
$\widetilde{\cal S}={\cal S}-\lambda\mathcal Q_{t_f}(L)$. Substituting the
$D=0$ saddle-point currents and expressing $p_\pm$ in terms of
$\hat p_\pm$, the generating function
\eqref{largedeviationform_general} becomes
\be \label{LD_RTP_reservoirs}
\left\langle e^{\lambda\mathcal{Q}_{t_f}(L)}\right\rangle
\simeq
{\cal N}
\int{\cal D}\rho_+{\cal D}\rho_-{\cal D}p_+{\cal D}p_-\,
P[\vec\rho(x,0)]\,e^{-\widetilde{\cal S}} \;,
\ee
with
\be \label{S_tilde_RTP_reservoirs}
\widetilde{\cal S}
=
\int_0^{t_f}dt\int_0^Ldx\,
\left[
\left(\hat p_+-\frac{\lambda x}{L}\right)\partial_t\rho_+
+
\left(\hat p_--\frac{\lambda x}{L}\right)\partial_t\rho_-
\right]
-\int_0^{t_f}dt\,{\sf\widetilde H}
+\int_0^{t_f}dt\,{\cal B} \;,
\ee
where the bulk Hamiltonian is
\be \label{H_RTP_reservoirs}
{\sf\widetilde H}
=
\int_0^Ldx\,
\Big[
v_0\rho_+\partial_x\hat p_+
-
v_0\rho_-\partial_x\hat p_-
+
\gamma\rho_+
\left(e^{-(\hat p_+-\hat p_-)}-1\right)
+
\gamma\rho_-
\left(e^{\hat p_+-\hat p_-}-1\right)
\Big] \;,
\ee
From \eqref{action_after_JK_general} 
and for $D=0$ the spatial boundary contribution 
reads ${\cal B} = v_0[p_+\rho_+-p_-\rho_-]_0^L$, or equivalently
\be \label{B_RTP_boundaries}
{\cal B}
=
v_0
\left[
\hat p_+\rho_+-\hat p_-\rho_-
\right]_0^L
-
\lambda v_0
\left[
\rho_+(L)-\rho_-(L)
\right] \;.
\ee
At the same time, the explicit bulk term
$-\lambda\mathcal Q_{t_f}(L)$ cancels the terms proportional to $\lambda$
generated when the gradients $\partial_xp_\pm$ appearing in
\eqref{action_after_JK_general} are rewritten as
$\partial_xp_\pm=\partial_x\hat p_\pm-\lambda/L$, leaving the bulk
Hamiltonian \eqref{H_RTP_reservoirs}.
The remaining explicit term proportional to $\lambda$ in
\eqref{S_tilde_RTP_reservoirs},
$-\lambda x\,\partial_t(\rho_++\rho_-)/L$, is a temporal boundary
contribution and is therefore subextensive in $t_f$, as in the Brownian
case. Similarly, $-\ln P[\vec\rho(x,0)]$ does not affect the leading
long-time stationary bulk saddle.

We now specify the spatial boundary conditions. There are several possible
prescriptions for coupling RTPs to particle reservoirs. We choose here the
simplest one: the reservoirs fix the populations entering the interval.
For reservoirs containing equal fractions of the two orientations,
\be \label{BCrhopm}
\rho_+(0,t)=\frac{\rho_a}{2}
\quad,\quad
\rho_-(L,t)=\frac{\rho_b}{2} \;.
\ee
The outgoing densities $\rho_-(0,t)$ and $\rho_+(L,t)$ are not fixed. As
shown explicitly in Appendix~\ref{app:RTP_boundary_conditions}, variation
of the action with respect to these outgoing densities imposes
$p_-(0,t)=0$ and $p_+(L,t)=0$. In terms of the shifted fields,
\be \label{BCphatpm}
\hat p_-(0,t)=0
\quad,\quad
\hat p_+(L,t)=\lambda \;.
\ee
Thus the density boundary conditions are imposed on the incoming
characteristics, while the corresponding auxiliary-field conditions are
imposed on the outgoing ones. In particular, unlike in the Brownian
problem, the auxiliary fields do not vanish on all the boundaries, and the
surface contribution \eqref{B_RTP_boundaries} need not vanish.

We now restrict ourselves to the stationary branch of the saddle-point
equations. The bulk equations follow directly from
\eqref{SPeq_general1}-\eqref{SPeq_general4}, upon setting
$D=0$, $V=W=0$, and replacing $p_\pm$ by $\hat p_\pm$. They read
\bea \label{SPeq_RTPfree_reservoirs1}
0 &=&
-v_0\partial_x\rho_+
-\gamma\rho_+e^{-(\hat p_+-\hat p_-)}
+\gamma\rho_-e^{\hat p_+-\hat p_-} \;,
\\
0 &=&
v_0\partial_x\rho_-
-\gamma\rho_-e^{\hat p_+-\hat p_-}
+\gamma\rho_+e^{-(\hat p_+-\hat p_-)} \;,
\\
0 &=&
-v_0\partial_x\hat p_+
-\gamma\left(e^{-(\hat p_+-\hat p_-)}-1\right) \;,
\\
0 &=&
v_0\partial_x\hat p_-
-\gamma\left(e^{\hat p_+-\hat p_-}-1\right) \;.
\label{SPeq_RTPfree_reservoirs4}
\eea

As in the free-RTP problem considered previously, we introduce the
Cole--Hopf variables $F_\pm=\rho_\pm e^{-\hat p_\pm}$ and
$G_\pm=e^{\hat p_\pm}$, and define
$F_s=F_++F_-$, $F_d=F_+-F_-$ and similarly $G_s,G_d$. The stationary
bulk equations then become
\bea \label{eqFs_freeRTP_boundaries}
0 &=& -v_0\partial_xF_d \;, \qquad
0=-v_0\partial_xF_s-2\gamma F_d \;,
\\
0 &=& v_0\partial_xG_d \;, \qquad
0=v_0\partial_xG_s-2\gamma G_d \;.
\label{eqGd_freeRTP_boundaries}
\eea
Thus $F_d$ and $G_d$ are constant, while
$F_s(x)=F_0-(2\gamma/v_0)F_dx$ and
$G_s(x)=G_0+(2\gamma/v_0)G_dx$. Introducing
\be \label{alpha_RTP_boundaries}
\alpha=\frac{\gamma L}{v_0} \;,
\ee
we have
\be \label{FpmGpm_saddle_RTP_boundaries}
F_\pm(x)
=
\frac{1}{2}
\left[
F_0- 2\alpha F_d\frac{x}{L}\pm F_d
\right] \quad , \quad 
G_\pm(x)
=
\frac{1}{2}
\left[
G_0+ 2\alpha G_d\frac{x}{L}\pm G_d
\right] \;.
\ee

The auxiliary-field boundary conditions \eqref{BCphatpm} give
$G_-(0)=1$ and $G_+(L)=e^\lambda$, or equivalently
$G_0-G_d=2$ and $G_0+(2\alpha+1)G_d=2e^\lambda$. Hence
\bea \label{G0Gd_RTP_boundaries}
G_d = \frac{e^\lambda-1}{\alpha+1} \quad , \quad G_0 = \frac{2\alpha+1+e^\lambda}{\alpha+1} \;.
\eea
In particular,
\be \label{G_boundary_values_RTP}
G_+(0)=\frac{\alpha+e^\lambda}{\alpha+1}
\quad,\quad
G_-(L)=\frac{\alpha e^\lambda+1}{\alpha+1} \;.
\ee

The density boundary conditions \eqref{BCrhopm} give
\be
\frac{1}{4}(F_0+F_d)(G_0+G_d)=\frac{\rho_a}{2}
\quad,\quad
\frac{1}{4}
\left[F_0-(2\alpha+1)F_d\right]
\left[G_0+(2\alpha-1)G_d\right]
=
\frac{\rho_b}{2} \;,
\ee
and therefore
\be \label{F0Fd_RTP_boundaries}
F_d=\frac{\rho_a}{2(\alpha+e^\lambda)} - \frac{\rho_b}{2(\alpha e^\lambda+1)} 
\quad , \quad F_0 = \frac{(2\alpha+1)\rho_a}{2(\alpha+e^\lambda)} + \frac{\rho_b}{2(\alpha e^\lambda+1)} \;.
\ee

The density and polarization profiles follow from
$\rho_\pm=F_\pm G_\pm$. In particular,
\bea
\rho_s(x) &=& \frac{1}{2} \left[
\left(F_0-2\alpha F_d\frac{x}{L}\right)
\left(G_0+2\alpha G_d\frac{x}{L}\right)
+ F_dG_d \right] \;,
\\
\rho_d(x) &=& \frac{F_0G_d+F_dG_0}{2}
=
\frac{\rho_a e^\lambda}{2(\alpha+e^\lambda)}
-
\frac{\rho_b}{2(\alpha e^\lambda+1)} \;.
\label{rhod_RTP_boundaries}
\eea
The polarization is therefore constant, as required by stationarity of the
total current. Since $J_s=v_0\rho_d$,
\be
J_s^*
=
v_0
\left[
\frac{\rho_a e^\lambda}{2(\alpha+e^\lambda)}
-
\frac{\rho_b}{2(\alpha e^\lambda+1)}
\right] \;.
\ee

Let us now evaluate the scaled cumulant generating function
by computing \eqref{S_tilde_RTP_reservoirs} at the saddle point. 
For stationary bulk profiles, the time-derivative terms vanish, while the
initial contribution and possible temporal boundary layers give only
$O(1)$ corrections. Hence
\be \label{Stilde_RTP_boundary_decomp}
\frac{\widetilde{\cal S}^*}{t_f}
=
-{\sf\widetilde H}^*
+
{\cal B}^*
+
O(t_f^{-1})  \quad  \Rightarrow \quad  
\widetilde\Sigma(\lambda)
=
{\sf\widetilde H}^*
-
{\cal B}^* \;.
\ee

Using the boundary conditions \eqref{BCrhopm}-\eqref{BCphatpm}, the boundary
term \eqref{B_RTP_boundaries} becomes
\be \label{B_RTP_boundaries_simplified}
{\cal B}^*
=
-\frac{v_0}{2}
\left[
\rho_a\hat p_+(0)
+
\rho_b\left(\hat p_-(L)-\lambda\right)
\right] \;.
\ee
The bulk Hamiltonian, on the other hand, vanishes identically on the
stationary solution. Indeed, in the Cole--Hopf variables its integrand
reduces to
$ 
\frac{v_0}{2}F_d\partial_xG_s-\gamma F_dG_d=0$
where we used $v_0\partial_xG_s=2\gamma G_d$. Thus
${\sf\widetilde H}^*=0$, and the entire stationary current SCGF comes from
the spatial boundary contribution,
\be
\widetilde\Sigma(\lambda)
=
-{\cal B}^* \;.
\ee
Using \eqref{G_boundary_values_RTP}, together with
$\hat p_+(0)=\ln G_+(0)$ and
$\hat p_-(L)=\ln G_-(L)$,
we finally obtain
\be \label{SCGF_free_RTP_boundaries}
\widetilde\Sigma(\lambda)
=
\frac{v_0}{2}
\left[
\rho_a
\ln\left(
\frac{\alpha+e^\lambda}{\alpha+1}
\right)
+
\rho_b
\ln\left(
\frac{\alpha+e^{-\lambda}}{\alpha+1}
\right)
\right]
 \;.
\ee
Recalling that $\alpha=\gamma L/v_0$ and $D_a=v_0^2/(2\gamma)$, the 
mean current reads, at large $t_f$,
\be \label{mean_current_free_RTP_boundaries}
\frac{1}{t_f} \langle\mathcal{Q}_{t_f}(L)\rangle \simeq
\widetilde\Sigma'(0)
= 
\frac{v_0}{2(\alpha+1)}(\rho_a-\rho_b) = 
\frac{D_a}{L+v_0/\gamma}
(\rho_a-\rho_b) \;.
\ee
Thus the finite persistence length produces an effective boundary
resistance. The second cumulant reads at large time
\be
\frac{1}{t_f} \langle\mathcal{Q}_{t_f}(L)^2\rangle_c
\simeq \widetilde\Sigma''(0) = \frac{v_0\alpha}{2(\alpha+1)^2}(\rho_a+\rho_b) \;.
\ee 
In the diffusive limit $L\gg v_0/\gamma$, Eq.~\eqref{mean_current_free_RTP_boundaries} reduces to
Fick's law, $\widetilde\Sigma'(0)\simeq
D_a(\rho_a-\rho_b)/L$. More generally, expanding
\eqref{SCGF_free_RTP_boundaries} for $\alpha\gg1$ gives
\be
\widetilde\Sigma(\lambda)
=
\frac{D_a}{L}
\left[
\rho_a(e^\lambda-1)
+
\rho_b(e^{-\lambda}-1)
+
O\left(\frac{v_0}{\gamma L}\right) \right] \;,
\ee
and we recover the open Brownian result with $D$ replaced by
$D_a$.

In the opposite persistent limit $\alpha\to0$,
\be
\widetilde\Sigma(\lambda)
\longrightarrow
\frac{v_0}{2}(\rho_a-\rho_b)\lambda \;.
\ee
Thus the current becomes deterministic in this idealized
fixed-incoming-density reservoir model when the system is much shorter than
the persistence length. This property depends on the reservoir prescription:
allowing stochastic injection at the boundaries would introduce an
additional boundary large-deviation cost and modify the finite-$L$ current
fluctuations.

A Legendre-Fenchel transform \eqref{Legendre_current_reservoirs} gives the large deviation result
\be \label{LegendreTransformQRTP}
P(\mathcal{Q}_{t_f}(L) \simeq {\cal Q})
\asymp
\exp\left[
-\frac{\rho_aD_at_f}{L}\,
\Psi\left(
\frac{L {\cal Q}}
{\rho_aD_at_f}
\right)
\right] \; \quad , \quad \alpha=\gamma L/v_0,
\ee
where 
\be
\Psi(q) 
=
q\lambda_q
-
\widetilde{\widetilde\Sigma}(\lambda_q)
\quad,\quad
\widetilde{\widetilde\Sigma}(\lambda)
=
\alpha
\left[
\ln\left(
\frac{\alpha+e^\lambda}{\alpha+1}
\right)
+
\frac{\rho_b}{\rho_a}
\ln\left(
\frac{\alpha+e^{-\lambda}}{\alpha+1}
\right)
\right]
,
\ee
with
\be
\lambda_q 
=
\ln\left(
\frac{\sqrt{4 \alpha^2 (\alpha-q) (\alpha \rho_b/\rho_a+q)+(\alpha (\alpha q+\rho_b/\rho_a-1)+q)^2}+\alpha^2 q+\alpha (\rho_b/\rho_a-1)+q}{2 \alpha (\alpha-q)}
\right) \;.
\ee
This result is valid for $-\alpha\rho_b/\rho_a<q<\alpha$, i.e. $-\frac{\rho_bv_0t_f}{2}<{\cal Q}<\frac{\rho_av_0t_f}{2}$, which is the range of integrated currents available in our idealized reservoir description. In the limit $\alpha \gg 1$, we recover the diffusive result \eqref{LDP_reservoirs_br} with $D=D_a$. 
It is exactly the same as Eq. \eqref{Ian_intro} with the correspondence $Q(t_f) \to {\cal Q}_{t_f}(L)$ and $\mu(t_f) \to \rho_a D t_f/L$ (and $\rho_0,\rho_1\to\rho_a,\rho_b$).

Finally, the result \eqref{SCGF_free_RTP_boundaries} has a simple
single-particle interpretation. Introducing the transmission probability
\be \label{ptr_RTP_boundaries}
p_{\rm tr}
=
\frac{1}{\alpha+1}
=
\frac{1}{1+\gamma L/v_0} \;,
\ee
the SCGF becomes
\be
\widetilde\Sigma(\lambda)
=
\frac{v_0\rho_a}{2}
\ln\left[1+p_{\rm tr}(e^\lambda-1)\right]
+
\frac{v_0\rho_b}{2}
\ln\left[1+p_{\rm tr}(e^{-\lambda}-1)\right] \;.
\ee
Since $\ln[1+p(e^\lambda-1)]$ is the cumulant generating function of a
Bernoulli variable, this has the form of the difference of two independent
binomial transmission processes. Within the present reservoir prescription,
the incoming flux from the left is $v_0\rho_a/2$ and each $+$ particle
crosses the system with probability $p_{\rm tr}$; particles entering from
the right with orientation $-$ behave analogously and contribute the
opposite current.

The quantity $p_{\rm tr}$ is precisely the single-particle transmission
probability of a free RTP. If $q_\pm(x)$ denotes the probability that a
particle starting at $x$ with orientation $\pm$ exits through $x=L$ before
$x=0$, the backward equations are
\be
v_0\partial_xq_+ +\gamma(q_--q_+)=0
\quad,\quad
-v_0\partial_xq_- +\gamma(q_+-q_-)=0 \;,
\ee
with outgoing absorbing conditions $q_-(0)=0$ and $q_+(L)=1$. Their
solution gives
\be
q_+(0)
=
\frac{1}{1+\gamma L/v_0}
=
p_{\rm tr} \;.
\ee
Hence the logarithms in \eqref{SCGF_free_RTP_boundaries} directly encode
the Bernoulli statistics of transmission across the interval. In the
strongly persistent limit $\gamma L/v_0\to0$, one has $p_{\rm tr}\to1$:
every incoming particle crosses the system, and the transmission noise
therefore vanishes.

\subsection{Weakly interacting Brownian particles}
\label{sec:Brownian_interacting_reservoirs}

We now consider weakly interacting Brownian particles on the interval
$[0,L]$. In order to obtain the simplest nontrivial result, we restrict
ourselves in this subsection to equal reservoir densities,
$\rho_a=\rho_b=\rho$, 
and to an even short-ranged pair potential $\varepsilon W(x)$, where
$\varepsilon$ controls the interaction strength. We assume that the
interaction range $a$ is much smaller than the system size $a\ll L$,
and work to first order in $\varepsilon$.
The deterministic current entering the Dean--Kawasaki description is
\be \label{Jbar_Brownian_interacting_res}
\bar J_\varepsilon
=
-D\partial_x\rho
+
\varepsilon\rho\,{\cal H}[\rho],
\qquad
{\cal H}[\rho](x)
=
-\int_0^Ldy\,W'(x-y)\rho(y).
\ee
The density-current MFT action is therefore
\bea \label{S_Brownian_interacting_res}
{\cal S}_\varepsilon[\rho,J,p]
&=&
\int_0^{t_f}dt\int_0^Ldx\,
\Bigg[
\frac{(J-\bar J_\varepsilon)^2}{4D\rho}
+
p\left(\partial_t\rho+\partial_xJ\right)
\Bigg].
\eea
\\

\noindent{\bf Boundary conditions.}
The interaction does not modify the auxiliary-field boundary conditions
associated with the ideal fixed-density reservoirs. As shown in
Appendix~\ref{app:Brownian_boundary_conditions}, the spatial boundary
variation still gives
\be \label{p_BC_Brownian_interacting}
p(0,t)=p(L,t)=0 \quad \Leftrightarrow \quad
\hat p(0,t)=0 \quad , \quad \hat p(L,t)=\lambda,
\ee
where $\hat p(x,t)=p(x,t)+\lambda x/L$.
\\

\noindent {{\bf Initial condition.}}
The interaction also modifies the equilibrium stationary density measure.
The corresponding Gibbs density large-deviation functional is derived in
Appendix~\ref{app:Brownian_interacting_stationary_measure}. This modifies
the initial weight in the path integral, but gives only an $O(1)$
contribution in time. It therefore does not contribute explicitly to the
scaled cumulant generating function in the long-time stationary limit.
\\

\noindent {\bf Short-range weak-interaction expansion.}
We start from the tilted version of the density-current action
\eqref{S_Brownian_interacting_res},
\be
\widetilde{\cal S}_\varepsilon
=
{\cal S}_\varepsilon
-
\frac{\lambda}{L}
\int_0^{t_f}dt\int_0^Ldx\,J \;.
\ee
Using the boundary conditions \eqref{p_BC_Brownian_interacting}, we can
integrate the term $p\,\partial_xJ$ in
\eqref{S_Brownian_interacting_res} by parts without a spatial surface
contribution. Using the shifted field 
$\hat p=p+\lambda x/L$, the tilted action becomes
\be
\widetilde{\cal S}_\varepsilon
=
\int_0^{t_f}dt\int_0^Ldx\,
\left[
\frac{(J-\bar J_\varepsilon)^2}{4D\rho}
+
\left(\hat p-\frac{\lambda x}{L}\right)\partial_t\rho
-
J\partial_x\hat p
\right] \;,
\ee
Variation with respect to the current gives
\be
J^*
=
\bar J_\varepsilon
+
2D\rho\,\partial_x\hat p \;.
\ee
Substituting this saddle-point current into the tilted action gives
\be
\widetilde{\cal S}_\varepsilon
=
\int_0^{t_f}dt\int_0^Ldx\,
\left(\hat p-\frac{\lambda x}{L}\right)\partial_t\rho
-
\int_0^{t_f}dt\,{\sf\widetilde H}_\varepsilon,
\ee
with 
\be \label{H_Brownian_interacting_res}
{\sf\widetilde H}_\varepsilon
=
D\int_0^Ldx\,
\left[
\rho(\partial_x\hat p)^2
-
(\partial_x\rho)(\partial_x\hat p)
\right]
+
\varepsilon
\int_0^Ldx\,
\rho(x)\,{\cal H}[\rho](x)\,
\partial_x\hat p(x) \;.
\ee
This expression is \st{still} valid for an arbitrary interaction potential
$W(x)$. For the time-independent saddle point solution, 
the terms $\partial_t \rho$ do not contribute and the scaled cumulant generating
function is obtained by evaluating this Hamiltonian at its saddle point,
$\widetilde\Sigma(\lambda)={\sf\widetilde H}_\varepsilon^*$.

We now expand
\be
\widetilde\Sigma(\lambda)
=
\widetilde\Sigma_0(\lambda)
+
\varepsilon\widetilde\Sigma_1(\lambda)
+
O(\varepsilon^2).
\ee
where
$\tilde \Sigma_0(\lambda)$ is the SCGF for the non-interacting case given in \eqref{Sigma_Br_reservoirs}. For $\rho_a=\rho_b=\rho$ it can be written as
\be
\widetilde\Sigma_0(\lambda)
=
\frac{2D\rho}{L}(\cosh\lambda-1).
\ee

Since the zeroth-order fields extremize
${\sf\widetilde H}_0$ with fixed boundary conditions, their
$O(\varepsilon)$ corrections do not contribute at first order. Hence
the first-order correction is obtained simply by evaluating the explicit
interaction term in \eqref{H_Brownian_interacting_res} on the
non-interacting saddle,
\be \label{Sigma1_Brownian_res_nonlocal}
\widetilde\Sigma_1(\lambda)
=
\int_0^Ldx\,
\rho^{(0)}(x)\,
{\cal H}[\rho^{(0)}](x)\,
\partial_x\hat p^{(0)}(x) \;.
\ee
Equivalently, using the definition of ${\cal H}$,
\be
\widetilde\Sigma_1(\lambda)
=
-\int_0^Ldx\int_0^Ldy\,
\rho^{(0)}(x)\rho^{(0)}(y)
W'(x-y)\,
\partial_x\hat p^{(0)}(x) \;.
\ee

Let us now specialize to a short-ranged interaction. In the present
perturbative calculation the force field is evaluated on the
non-interacting saddle $\rho^{(0)}$, which, for fixed $\lambda$, varies on
the macroscopic scale $L$. In the hydrodynamic regime $a\ll L$, and
retaining only the bulk contribution while neglecting microscopic
interaction-range boundary layers at the reservoirs, we may therefore
write
\be
{\cal H}[\rho^{(0)}]
=
-W_0\partial_x\rho^{(0)}
+
O\left(W_0a^2\partial_x^3\rho^{(0)}\right),
\qquad
W_0=\int_{-\infty}^{+\infty}dr\,W(r) \;.
\ee
Then Eq.~\eqref{Sigma1_Brownian_res_nonlocal} reduces to
\be \label{Sigma1_Brownian_res_envelope}
\widetilde\Sigma_1(\lambda)
=
-W_0
\int_0^Ldx\,
\rho^{(0)}(x)
\partial_x\rho^{(0)}(x)
\partial_x\hat p^{(0)}(x) \;.
\ee
For equal reservoir densities $\rho_a=\rho_b=\rho$, the free Cole--Hopf solution \eqref{SPsolF_Br}-\eqref{SPsolG_Br} reads
\be
F_0(x)
=
\rho\left[
1+(e^{-\lambda}-1)\frac{x}{L}
\right] \quad , \quad 
G_0(x)
=
1+(e^\lambda-1)\frac{x}{L},
\ee
with $\rho^{(0)}=F_0G_0$ and 
$\hat p^{(0)}=\ln G_0$. Substitution into
\eqref{Sigma1_Brownian_res_envelope} gives
an elementary integral, and one obtains 
\be \label{Sigma1_Brownian_res_final}
\widetilde\Sigma_1(\lambda)
=
-\frac{8\rho^2W_0}{3L}
\sinh^4\left(\frac{\lambda}{2}\right) .
\ee
Hence to first order in the interaction we have obtained 
\be \label{Sigma_Brownian_interacting_res_final}
\widetilde\Sigma(\lambda)
=
\frac{2D\rho}{L}(\cosh\lambda-1)
-
\varepsilon\frac{8\rho^2W_0}{3L}
\sinh^4\left(\frac{\lambda}{2}\right)
+
O(\varepsilon^2).
\ee
This expression is even under $\lambda\to-\lambda$, as expected for equal
reservoir densities. Since the interaction correction starts at order
$\lambda^4$,
there is no correction to the variance at first order,
\be
\frac{1}{t_f} \langle [\mathcal{Q}_{t_f}(L)]^2 \rangle_c \simeq \widetilde\Sigma''(0)
=
\frac{2D\rho}{L}
+
O(\varepsilon^2),
\ee
while the fourth cumulant is corrected as
\be
\frac{1}{t_f} \langle [\mathcal{Q}_{t_f}(L)]^4 \rangle_c \simeq \widetilde\Sigma^{(4)}(0)
=
\frac{2D\rho}{L}
-
\varepsilon\frac{4\rho^2W_0}{L}
+
O(\varepsilon^2).
\ee
\\

\noindent {\bf Comparison with the standard MFT.}
We now compare the above perturbative Dean--Kawasaki calculation with the
standard MFT formulation for diffusive systems
\cite{Derrida2007,BertiniMFT2015}. This approach has recently been
specialized to interacting Brownian particles in
Ref.~\cite{Grabsch2025,Grabsch2026}, where its relation to the Dean--Kawasaki
description was examined explicitly. In the macroscopic MFT approach one
starts directly from the local fluctuating hydrodynamic current
\be
J
=
-D_{\rm coll}(q)\partial_x\,q
+
\sqrt{\sigma(q)}\,\eta \;.
\ee
Here $q(x,t)$ denotes the coarse-grained density field, corresponding to
the field $\rho(x,t)$ used in the Dean--Kawasaki formulation above.
The microscopic interactions are encoded in the two transport
coefficients $D_{\rm coll}(q)$ and $\sigma(q)$.
For overdamped interacting Brownian particles, Ref.~\cite{Grabsch2025,Grabsch2026}
shows that the mobility is independent of the interaction,
\be
\sigma(q)=2Dq,
\ee
in our units, while the collective diffusion coefficient is determined by
equilibrium thermodynamics. More generally, denoting by $\mu_0$ the bare
mobility and by $P(q)$ the equilibrium pressure, the thermodynamic relation
used in Ref.~\cite{Grabsch2025,Grabsch2026} reads
\be
D_{\rm coll}(q)
=
\mu_0 P'(q)
=
\mu_0 q f''(q),
\ee
where $f(q)$ is the equilibrium free-energy density. In our convention
$\mu_0=1$ and the bare diffusion coefficient is $D=k_{\rm B}T$.

The weak-interaction free-energy density associated with the Gibbs measure,
derived in Appendix~\ref{app:Brownian_interacting_stationary_measure}, is
\be \label{freenergy}
f(q)
=
Dq(\ln q-1)
+
\varepsilon\frac{W_0}{2}q^2
+
O(\varepsilon^2).
\ee
Using \eqref{freenergy} therefore gives
\be \label{Dcoll_weak_res}
D_{\rm coll}(q)
=
D+\varepsilon W_0q
+
O(\varepsilon^2).
\ee
By comparison, the short-range expansion performed in our
Dean--Kawasaki calculation gives
${\cal H}[q]\simeq-W_0\partial_xq$, so that
\be
\bar J_\varepsilon
=
-D\partial_xq+\varepsilon q{\cal H}[q]
\simeq
-\left(D+\varepsilon W_0q\right)\partial_xq.
\ee
Thus the local constitutive law obtained by coarse-graining the
Dean--Kawasaki current has precisely the same collective diffusion
coefficient \eqref{Dcoll_weak_res} as the macroscopic MFT description.

Ref.~\cite{Grabsch2025,Grabsch2026} also gives a useful independent check based on the
virial expansion. Introducing the Mayer function
$f_{\rm M}(r)
=
e^{-\beta\varepsilon W(r)}-1$, 
the low-density expansion of the collective diffusion coefficient starts as
\be
\frac{D_{\rm coll}(q)}{D}
=
1
-
q\int dr\,f_{\rm M}(r)
+
O(q^2).
\ee
For a weak interaction,
$f_{\rm M}(r)=-\varepsilon W(r)/D+O(\varepsilon^2)$, and hence, in the weak-interaction and low-density regime,
\be
D_{\rm coll}(q)
=
D+\varepsilon W_0q
+
O(\varepsilon^2,q^2),
\ee
again consistently with \eqref{Dcoll_weak_res}.

We now verify that this correspondence extends to the full stationary
large-deviation functional for the current. Within the stationary realization of the Bodineau--Derrida additivity
principle~\cite{BodineauDerrida2004}, the standard MFT gives
\be \label{BD_variational_Brownian_res}
\widetilde\Sigma(\lambda)
=
\frac{1}{L}
{\sup}_{q,\hat p}
\int_0^1du\,
\left[
\frac{\sigma(q)}{2}(\partial_u\hat p)^2
-
D_{\rm coll}(q)
(\partial_uq)(\partial_u\hat p)
\right],
\ee
where $u=x/L$, with
$q(0)=q(1)=\rho$ and
$\hat p(0)=0$, $\hat p(1)=\lambda$.

Substituting
$\sigma(q)=2Dq$ and
$D_{\rm coll}(q)=D+\varepsilon W_0q+O(\varepsilon^2)$,
and returning to the coordinate $x=Lu$, gives precisely the local
short-range form of the Hamiltonian
\eqref{H_Brownian_interacting_res} obtained from the Dean--Kawasaki
action. The two variational problems therefore coincide to
$O(\varepsilon)$, and the perturbative MFT immediately reproduces
\eqref{Sigma1_Brownian_res_final} and
\eqref{Sigma_Brownian_interacting_res_final}.

This provides a consistency check between the Dean--Kawasaki calculation
and the standard MFT description at the level of the full current
large-deviation functional, and not only of the constitutive current.

\subsection{Weakly interacting RTPs}
\label{sec:RTP_interacting_reservoirs}

We now turn to weakly interacting RTPs on the interval $[0,L]$, again
without additional thermal diffusion, $D=0$. As in the Brownian case, we
restrict ourselves to equal reservoir densities,
$\rho_a=\rho_b=\rho$, and consider an even short-ranged pair potential
$\varepsilon W(x)$, of range $a\ll L$, working to first order in
$\varepsilon$.

Starting from the general currents \eqref{defJbar}, with $V=0$, $D=0$ and
$W\to\varepsilon W$, the translational currents are deterministic and read
\be \label{Jpm_RTP_interacting_res}
J_\pm
=
\left[
\pm v_0+\varepsilon{\cal H}[\rho_s]
\right]\rho_\pm \;.
\ee
where $\rho_s=\rho_++\rho_-$ and
${\cal H}[\rho_s](x)
=
-\int_0^Ldy\,W'(x-y)\rho_s(y)$.
The tumbling events remain Poisson processes with rates
$\gamma\rho_\pm$. 

We consider the same space-time integrated current
\eqref{integrated_current_def2}, with the shifted auxiliary fields
$\hat p_\pm=p_\pm+\lambda x/L$ introduced in \eqref{defhatp_reservoirs}.
Using the same tilted-action construction as in
Eqs.~\eqref{S_tilde_RTP_reservoirs}--\eqref{H_RTP_reservoirs}, the
interaction amounts to replacing
$\widetilde{\cal S}\to\widetilde{\cal S}_\varepsilon$,
${\sf\widetilde H}\to{\sf\widetilde H}_\varepsilon$, and
${\cal B}\to{\cal B}_\varepsilon$, where the latter is the spatial
boundary contribution, whose general form is given below in
\eqref{B_RTP_interacting_res}. Using the deterministic currents
\eqref{Jpm_RTP_interacting_res}, the tilted bulk Hamiltonian becomes
\be \label{H_RTP_interacting_res}
{\sf\widetilde H}_\varepsilon
=
\int_0^Ldx\,
\Bigg[
v_0\rho_+\partial_x\hat p_+
-
v_0\rho_-\partial_x\hat p_-
+
\gamma\rho_+
\left(
e^{-(\hat p_+-\hat p_-)}-1
\right)
+
\gamma\rho_-
\left(
e^{\hat p_+-\hat p_-}-1
\right)
+
\varepsilon{\cal H}[\rho_s]
\left(
\rho_+\partial_x\hat p_+
+
\rho_-\partial_x\hat p_-
\right)
\Bigg] \;.
\ee
Thus the interaction enters the bulk Hamiltonian only through the last term.

There is, however, an additional subtlety which is specific to the RTP
problem. As discussed in the non-interacting case, the stationary current
SCGF contains a spatial boundary contribution because the auxiliary fields
do not vanish at all four ends of the two characteristics. Moreover, once
interactions are present, the precise form of this contribution depends on
how the reservoirs are coupled to the interacting particles. We adopt here
the natural continuation of the ideal reservoir prescription used in the
non-interacting problem: the reservoirs impose the incoming populations
\be \label{BC_RTP_interacting_res}
\rho_+(0,t)=\frac{\rho}{2}
\quad,\quad
\rho_-(L,t)=\frac{\rho}{2},
\qquad
\hat p_-(0,t)=0
\quad,\quad
\hat p_+(L,t)=\lambda,
\ee
and we assume that the pair interaction is a bulk interaction which is
switched off within microscopic contact layers adjacent to the reservoirs.
With this convention, the reservoir coupling and hence the spatial
boundary contribution are unchanged to first order in $\varepsilon$.
We return below to what changes if the interaction is instead allowed to
act all the way to the reservoir contacts.
\\

\noindent {\bf Initial condition.}
As for the Brownian problem, the stationary distribution of the density
fields enters the functional integral through the initial weight
$P[\vec\rho(x,0)]$. There is an important difference, however: interacting
RTPs \st{at $D=0$} constitute an active non-equilibrium system and their
stationary measure is not in general a Gibbs measure. Thus there is no
direct analogue of the equilibrium free-energy functional used for the
interacting Brownian particles.

This stationary measure is nevertheless not needed for the present
long-time calculation. Whatever its density large-deviation functional
${\cal V}^{\rm RTP}_\varepsilon[\rho_+,\rho_-]$, it contributes only at
the initial time. For fixed $L$,
\be
\lim_{t_f\to+\infty}
\frac{
{\cal V}^{\rm RTP}_\varepsilon
[\rho_+(x,0),\rho_-(x,0)]
}{t_f}
=
0 \;.
\ee
It therefore does not contribute explicitly to the leading stationary
scaled cumulant generating function.
\\

\noindent{\bf Weak-interaction expansion.}
For stationary profiles the tilted action has, as in
\eqref{Stilde_RTP_boundary_decomp}, the structure
\be
\frac{\widetilde{\cal S}^*}{t_f}
=
-{\sf\widetilde H}^*_{\varepsilon}
+
{\cal B}^*_{\varepsilon}
+
O(t_f^{-1}),
\qquad
\widetilde\Sigma(\lambda)
=
{\sf\widetilde H}^*_{\varepsilon}
-
{\cal B}^*_{\varepsilon} \;.
\ee
Under the ideal-contact description specified above,
${\cal B}_\varepsilon$ has no explicit $O(\varepsilon)$ correction. We
therefore expand
\be
\widetilde\Sigma(\lambda)
=
\widetilde\Sigma_0(\lambda)
+
\varepsilon\widetilde\Sigma_1(\lambda)
+
O(\varepsilon^2) \;.
\ee
For equal reservoir densities the non-interacting result
\eqref{SCGF_free_RTP_boundaries} becomes
\be \label{Sigma0_RTP_interacting_res}
\widetilde\Sigma_0(\lambda)
=
\frac{v_0\rho}{2}
\ln\left[
\frac{
(\alpha+e^\lambda)(\alpha+e^{-\lambda})
}{
(\alpha+1)^2
}
\right],
\qquad
\alpha=\frac{\gamma L}{v_0} \;.
\ee

Since the zeroth-order fields extremize the complete free variational
problem, including its spatial boundary contribution, their
$O(\varepsilon)$ corrections do not contribute to the SCGF at first
order. Hence the first-order correction is obtained by evaluating only the
explicit interaction term in \eqref{H_RTP_interacting_res} on the free
RTP saddle,
\be \label{Sigma1_RTP_res_nonlocal}
\widetilde\Sigma_1(\lambda)
=
\int_0^Ldx\,
{\cal H}[\rho_s^{(0)}](x)
\left[
\rho_+^{(0)}(x)\partial_x\hat p_+^{(0)}(x)
+
\rho_-^{(0)}(x)\partial_x\hat p_-^{(0)}(x)
\right] \;.
\ee
This formula is still valid for an arbitrary interaction potential in the
bulk.

Let us now specialize to a short-range interaction. The force field is
evaluated on the non-interacting stationary saddle, whose profiles vary on
the macroscopic scale $L$. Away from the microscopic contact layers, the
same gradient expansion as in the Brownian case gives
\be
{\cal H}[\rho_s^{(0)}]
=
-W_0\partial_x\rho_s^{(0)}
+
O\left(
W_0a^2\partial_x^3\rho_s^{(0)}
\right),
\qquad
W_0
=
\int_{-\infty}^{+\infty}dr\,W(r) \;.
\ee
Equation~\eqref{Sigma1_RTP_res_nonlocal} therefore reduces to
\be \label{Sigma1_RTP_res_local}
\widetilde\Sigma_1(\lambda)
=
-W_0
\int_0^Ldx\,
\partial_x\rho_s^{(0)}
\left[
\rho_+^{(0)}\partial_x\hat p_+^{(0)}
+
\rho_-^{(0)}\partial_x\hat p_-^{(0)}
\right] \;.
\ee

The integral can be evaluated explicitly using the free Cole--Hopf
solution obtained above. Introducing $u=x/L$, we recall from
\eqref{FpmGpm_saddle_RTP_boundaries} that
\be
F_s^{(0)}(u)
=
F_0-2\alpha F_du,
\qquad
G_s^{(0)}(u)
=
G_0+2\alpha G_du,
\ee
where, for equal reservoir densities,
\bea
G_d
&=&
\frac{(e^\lambda-1)}{\alpha+1}
\quad,\quad
G_0
=
\frac{2\alpha+1+e^\lambda}{\alpha+1},
\\
F_d
&=&
\frac{\rho}{2}
\left[
\frac{1}{\alpha+e^\lambda}
-
\frac{1}{\alpha e^\lambda+1}
\right]
\quad , \quad F_0 =
\frac{\rho}{2}
\left[
\frac{2\alpha+1}{\alpha+e^\lambda}
+
\frac{1}{\alpha e^\lambda+1}
\right] \;.
\eea
Using
$F_\pm=(F_s\pm F_d)/2$,
$G_\pm=(G_s\pm G_d)/2$,
$\rho_\pm=F_\pm G_\pm$ and
$\hat p_\pm=\ln G_\pm$, one finds the useful identities
\be
\rho_s^{(0)}
=
\frac{
F_s^{(0)}G_s^{(0)}+F_dG_d
}{2} \quad , \quad 
\partial_x\rho_s^{(0)}
=
\frac{\alpha}{L}
\left[
G_dF_s^{(0)}
-
F_dG_s^{(0)}
\right] \;,
\ee
and
\be
\rho_+^{(0)}\partial_x\hat p_+^{(0)}
+
\rho_-^{(0)}\partial_x\hat p_-^{(0)}
=
\frac{\alpha G_d}{L}F_s^{(0)} \;.
\ee
The remaining integral in
\eqref{Sigma1_RTP_res_local} is therefore elementary. It is convenient
during the calculation to introduce
$s_\lambda=\sinh^2(\lambda/2)=(\cosh(\lambda)-1)/2$ and $K_\lambda=(\alpha+1)^2+4\alpha s_\lambda$.
One obtains
\be \label{Sigma1_RTP_res_final_alpha}
\widetilde\Sigma_1(\lambda)
=
-\frac{8W_0\rho^2\alpha^2}{3L}\,
\frac{
s_\lambda
\left[
3(\alpha+1)^3
+
\left(
\alpha^4+\alpha^3+10\alpha^2+9\alpha+3
\right)s_\lambda
\right]
}{
(\alpha+1)^2 K_\lambda^2
} \;.
\ee

Combining this with the free result gives the stationary SCGF to first
order in the interaction. Eliminating the shorthand quantities from the final expression, we obtain
\bea \label{Sigma_RTP_interacting_res_final}
\widetilde\Sigma(\lambda)
&=&
\frac{v_0\rho}{2}
\ln\left[\frac{1+\alpha^2+2\alpha\cosh\lambda}{
(1+\alpha)^2}
\right]
\\
&&
-
\varepsilon\frac{2W_0\rho^2\alpha^2}{3L}
(\cosh\lambda-1) \, 
\frac{6(1+\alpha)^3
+
\left[
\alpha^4+\alpha^3+10\alpha^2+9\alpha+3
\right]
(\cosh\lambda-1)
}{(1+\alpha)^2
\left[
1+\alpha^2+2\alpha\cosh\lambda
\right]^2
}
+
O(\varepsilon^2) \;. \nn 
\eea
This expression is even under $\lambda\to-\lambda$, as expected for equal
reservoir densities. In contrast with the interacting Brownian result
\eqref{Sigma_Brownian_interacting_res_final}, the interaction correction
now starts at order $\lambda^2$ for finite persistence. The variance is
therefore modified already at first order,
\be \label{variance_RTP_interacting_res}
\frac{1}{t_f}
\left\langle
[{\cal Q}_{t_f}(L)]^2
\right\rangle_c
\simeq
\widetilde\Sigma''(0)
=
\frac{
v_0\rho\alpha
}{
(1+\alpha)^2
}
-
\varepsilon
\frac{4W_0\rho^2}{L}
\frac{
\alpha^2
}{
(1+\alpha)^3
}
+
O(\varepsilon^2) \;.
\ee
Similarly, the fourth cumulant reads
\be \label{fourth_cumulant_RTP_interacting_res}
\frac{1}{t_f}
\left\langle
[{\cal Q}_{t_f}(L)]^4
\right\rangle_c
\simeq
\widetilde\Sigma^{(4)}(0)
=
v_0\rho
\frac{
\alpha
\left[
\alpha^2-4\alpha+1\right]
}{
(1+\alpha)^4
}
-
\varepsilon
\frac{4W_0\rho^2}{L}
\frac{
\alpha^2
\left[
\alpha^4+2\alpha^3-11\alpha^2-12\alpha+4
\right]
}{
(1+\alpha)^6
}
+
O(\varepsilon^2) \;.
\ee

A useful check is provided by the diffusive limit
$L\gg v_0/\gamma$ {(i.e. $\alpha\gg 1$)}. Introducing again
$D_a=v_0^2/(2\gamma)$, expansion of
\eqref{Sigma_RTP_interacting_res_final} gives
\be \label{diffusive_limit_RTP_interacting_res}
\widetilde\Sigma(\lambda)
=
\frac{2D_a\rho}{L}
(\cosh\lambda-1)
-
\varepsilon
\frac{8\rho^2W_0}{3L}
\sinh^4\left(\frac{\lambda}{2}\right)
+
O\left(
\varepsilon^2,\frac{v_0}{\gamma L}
\right) \;.
\ee
We thus recover precisely the weakly interacting Brownian result
\eqref{Sigma_Brownian_interacting_res_final}, with the thermal diffusion
constant $D$ replaced by the active diffusion constant $D_a$. In
particular, the $O(\varepsilon)$ correction to the variance in
\eqref{variance_RTP_interacting_res} vanishes in this limit, consistently
with the Brownian result.

In the opposite strongly persistent limit,
$\alpha=\gamma L/v_0\to0$, both the free SCGF and its $O(\varepsilon)$ correction
vanish for equal reservoir densities within the present ideal
fixed-incoming-density reservoir model.
\\

\noindent{\bf Dependence on the reservoir contact model.}
The result \eqref{Sigma_RTP_interacting_res_final} assumes that the
interaction is switched off in microscopic contact layers adjacent to the
reservoirs. It is useful to make explicit why this assumption matters. If
the pair force is instead allowed to act up to the boundaries, the
deterministic currents entering the spatial surface term are the full
currents \eqref{Jpm_RTP_interacting_res}. Before imposing the boundary
conditions, the stationary boundary contribution can be written as
\be \label{B_RTP_interacting_res}
{\cal B}_\varepsilon
=
\left[
\left(
\hat p_+-\frac{\lambda x}{L}
\right)J_+
+
\left(
\hat p_--\frac{\lambda x}{L}
\right)J_-
\right]_{0}^{L} \;.
\ee

Its explicit first-order contribution, evaluated on the free saddle, gives
an additional correction to the SCGF, 
\be \label{Sigma1_RTP_contact}
\widetilde\Sigma_{1,\mathrm{contact}}(\lambda)
=
\frac{\rho}{2}
\left[
{\cal H}[\rho_s^{(0)}](0)\,\hat p_+^{(0)}(0)
-
{\cal H}[\rho_s^{(0)}](L)\,
\left(\hat p_-^{(0)}(L)-\lambda\right)
\right] \;,
\ee
where, from \eqref{G_boundary_values_RTP} with $\hat p_\pm=\ln G_\pm$,
\be
\hat p_+^{(0)}(0)
=
\ln\left(
\frac{\alpha+e^\lambda}{\alpha+1}
\right),
\qquad
\hat p_-^{(0)}(L)-\lambda
=
\ln\left(
\frac{\alpha+e^{-\lambda}}{\alpha+1}
\right) \;.
\ee
The endpoint values ${\cal H}[\rho_s^{(0)}](0)$ and
${\cal H}[\rho_s^{(0)}](L)$ cannot in general be obtained from the bulk
approximation ${\cal H}[\rho_s^{(0)}]\simeq
-W_0\partial_x\rho_s^{(0)}$.
They depend on the microscopic
continuation of the interaction through the reservoir contact region:
for example, on whether reservoir particles exert forces on particles
inside the interval, or whether the interaction is suppressed as the
boundary is approached. In such a model the complete first-order
correction has the form
\be
\widetilde\Sigma_1
=
\widetilde\Sigma_{1,\mathrm{bulk}}
+
\widetilde\Sigma_{1,\mathrm{contact}} \;,
\ee
with the bulk contribution given by
\eqref{Sigma1_RTP_res_final_alpha}. Thus, unlike the bulk correction, the
finite-$L$ contact contribution is not universal without a more detailed
microscopic specification of the interacting reservoirs. The ideal-contact
prescription adopted above removes this non-universal contribution and
isolates the bulk weak-interaction correction.
\\

{Comparison with multi-state MFT}. It is useful to place the above Dean--Kawasaki calculation in the more
general framework of multi-component macroscopic fluctuation theory.
The hydrodynamics of particles carrying an internal ``color'' was
introduced in the seminal work of Quastel~\cite{QuastelColor1992}, and
the corresponding dynamical large-deviation theory was subsequently
developed for exclusion processes in
Ref.~\cite{QuastelRezakhanlouVaradhan1999}. Multi-species diffusive
systems have since been studied in a variety of settings, including open
systems within MFT~\cite{Vanicat2017} and two-species cross-diffusion
models with a non-diagonal mobility matrix
\cite{MasonJackBruna2023}. The extension to active particles with an
internal orientation has also been developed at the hydrodynamic level
\cite{Erignoux2021,MasonErignouxJackBruna2023}. In parallel, large
deviations for systems combining conservative transport with local
reaction or conversion processes were studied in
Refs.~\cite{JonaLandimVares1993,BodineauLagouge2010,
BodineauLagouge2012}. The latter ingredient is essential for RTPs,
since tumbling corresponds to the stochastic interconversion of the two
internal states. A particularly close formulation to the present one is
the MFT of active lattice gases developed in
Ref.~\cite{Agranov2022}, where Gaussian conservative currents coexist
with Poissonian tumbling events.

The generic structure can be summarized as follows. For two coarse-grained densities $q_\pm$ (corresponding to our
$\rho_\pm$), conservative currents $j_\pm$, and local conversion
currents $k_{+-}$ and $k_{-+}$, the continuity equations read
\bea \label{continuity_two_state_MFT}
\partial_tq_+ +\partial_xj_+
&=&
-k_{+-}+k_{-+},
\\
\partial_tq_- +\partial_xj_-
&=&
-k_{-+}+k_{+-}.
\nn
\eea
For a diffusive multi-component system with deterministic current
$\bar{\boldsymbol j}(\boldsymbol q)$, mobility matrix
$\boldsymbol{\sigma}(\boldsymbol q)$, and local transition rates
$r_{+-}(\boldsymbol q)$ and $r_{-+}(\boldsymbol q)$, the corresponding
dynamical cost has the general form
\bea \label{LD_two_state_MFT}
{\cal I}
&=&
\int dt\,dx\,
\Bigg[
\frac12
\left(
\boldsymbol j-\bar{\boldsymbol j}
\right)^T
\boldsymbol{\sigma}^{-1}
\left(
\boldsymbol j-\bar{\boldsymbol j}
\right)
\nn\\
&&\hspace{0.8cm}
+
k_{+-}\ln\frac{k_{+-}}{r_{+-}}
-k_{+-}+r_{+-}
+
k_{-+}\ln\frac{k_{-+}}{r_{-+}}
-k_{-+}+r_{-+}
\Bigg] .
\eea
The first term is the usual multi-component MFT cost for conservative
currents, while the last two terms are the Poisson large-deviation costs
associated with the conversion processes
\cite{JonaLandimVares1993,BodineauLagouge2010,
BodineauLagouge2012,Agranov2022}. Introducing conjugate fields $p_\pm$
and eliminating the conservative and conversion currents gives the
Hamiltonian structure
\bea \label{H_two_state_MFT}
{\sf H}
&=&
\int dx\,
\Bigg[
\bar{\boldsymbol j}(\boldsymbol q)\cdot
\partial_x\boldsymbol p
+
\frac12
(\partial_x\boldsymbol p)^T
\boldsymbol{\sigma}(\boldsymbol q)
(\partial_x\boldsymbol p)
\nn\\
&&\hspace{0.8cm}
+
r_{+-}(\boldsymbol q)
\left(e^{p_--p_+}-1\right)
+
r_{-+}(\boldsymbol q)
\left(e^{p_+-p_-}-1\right)
\Bigg].
\eea

The present $D=0$ RTP model is a particularly simple limiting case of
this general two-state structure. Since the translational dynamics is
deterministic, the conservative mobility matrix vanishes,
$\boldsymbol{\sigma}=0$, while
\be
\bar j_\pm
=
\left[
\pm v_0+\varepsilon{\cal H}[q_s]
\right]q_\pm,
\qquad
r_{+-}=\gamma q_+,
\qquad
r_{-+}=\gamma q_-,
\qquad
q_s=q_++q_-.
\ee
Substitution into \eqref{H_two_state_MFT} gives
\bea
{\sf H}
&=&
\int dx\,
\Bigg[
v_0q_+\partial_xp_+
-
v_0q_-\partial_xp_-
+
\varepsilon{\cal H}[q_s]
\left(
q_+\partial_xp_+
+
q_-\partial_xp_-
\right)
\nn\\
&&\hspace{0.8cm}
+
\gamma q_+
\left(e^{-(p_+-p_-)}-1\right)
+
\gamma q_-
\left(e^{p_+-p_-}-1\right)
\Bigg],
\eea
which is precisely the bulk Hamiltonian
\eqref{H_RTP_interacting_res} obtained above from the
Dean--Kawasaki construction, after replacing $p_\pm$ by the shifted
fields $\hat p_\pm$. Thus the Dean--Kawasaki approach reproduces the
standard multi-state MFT structure: the weak interaction modifies the
deterministic constitutive currents, whereas the Poissonian tumbling
sector remains unchanged.

This comparison is somewhat different from the Brownian one discussed
above. For interacting Brownian particles, equilibrium thermodynamics
provides an independent relation between the collective diffusion
coefficient and the free-energy density. At finite persistence there is
in general no analogous equilibrium relation for RTPs, and the
constitutive currents of the two-state MFT have to be supplied from the
underlying microscopic dynamics. Consequently, inserting the currents
above into the general multi-state MFT is primarily a consistency check
at the level of the full fluctuation functional, rather than an
independent determination of the interaction correction.

There is nevertheless an important independent limiting check. On scales
large compared with the persistence length $v_0/\gamma$, the polarization
can be eliminated and the two-state theory reduces to a one-field
diffusive MFT. To leading order,
\be
J_s
\simeq
-\left[
D_a+\varepsilon W_0q_s
\right]\partial_xq_s,
\qquad
D_a=\frac{v_0^2}{2\gamma},
\ee
while the effective mobility becomes $2D_aq_s$. The standard scalar MFT
is therefore recovered with
$D_{\rm coll}(q_s)=D_a+\varepsilon W_0q_s+O(\varepsilon^2)$.
This explains directly why the large-persistence-length ratio limit of
\eqref{Sigma_RTP_interacting_res_final} reduces to
\eqref{Sigma_Brownian_interacting_res_final} with $D$ replaced by $D_a$.
At finite persistence, by contrast, the full two-state description is
essential and produces, in particular, the $O(\varepsilon)$ correction
to the current variance found in
\eqref{variance_RTP_interacting_res}.

\section{Conclusion}

In this work, we studied the fluctuations in a gas of interacting RTPs in the continuum, using the Dean-Kawasaki (DK) equations as a starting point. In the first part of the paper, we obtained the lowest order density correlations and the variance of the integrated current through the origin by linearizing the DK equations around a homogeneous stationary state. We then derived a continuum MFT for interacting RTPs, starting from the DK equations but retaining the Poisson statistics of the tumbling events. This MFT framework is a useful tool to study the large deviations of the current and its higher order moments. 
In the absence of interactions, a generalized Cole–Hopf transformation allows us to solve the MFT equations and recover the known current distributions for a step initial density \cite{FreeRTPs2020,JoseRossoRamola2023,JoseRossoRamola2024}. Beyond these distributions, we have obtained explicit optimal density histories for both RTPs and Brownian particles, including the regime of large quenched currents. These profiles illustrate what MFT can reveal about how a rare current fluctuation develops in time. We have also computed the first-order corrections to the current generating functions due to a weak, pairwise, short-range interaction, both on the infinite line and on an interval connected to reservoirs. Our treatment of the interactions is expected to be valid in the limit where the interaction potential varies slowly compared to interparticle distances, and allows for particle crossings.
Although we were able to obtain nontrivial results in perturbation theory for short-range interactions, the method would need to be improved 
to be able to access the large deviations for long-range interactions 
beyond the second cumulant. Indeed in that case the first-order interaction correction grows faster with time than the non-interacting contribution.

Our work raises several interesting questions. Since we have obtained the exact MFT saddle-point equations in the presence of interactions, it would be natural to attempt to solve them numerically. This would allow in principle to access the large deviations in a regime of stronger interactions and to treat the case of long-range interactions, e.g.  
the regularized Riesz gas. For the system connected to reservoirs this would allow us to test the additivity principle~\cite{BodineauDerrida2004}, i.e. whether the saddle point solution develops time-dependence in some regime of parameters. 
A related question concerns the effect of interactions on the optimal density histories. In particular, does an arbitrarily weak repulsion change the accumulation near the origin found for large quenched currents (see Fig. \ref{fig:ballistic_outer_history}), and on which spatial and temporal scales? The non-interacting profiles obtained here provide a starting point for this analysis. 
Finally, for active particles the microscopic description of the reservoir is expected to play a crucial role which would deserve further investigation. Beyond these specific questions, the formalism developed here could also be applied to a variety of other settings, for instance to particles in an external potential, to driven particles, and to different observables such as the number of particles in an interval.

\vspace*{0.5cm}
\noindent{\bf Acknowledgements} P.L.D. and G.S. acknowledge support from ANR Grant No.~ANR-23-CE30-0020-01 EDIPS. L.T. acknowledges support from the National Science Foundation through the Physics Frontier Center for Living Systems (PHY-2317138). The authors thank the Kavli Institute for
Theoretical Physics (KITP) for hospitality during the program ``New Trends
in Non-equilibrium Dynamics,'' where part of this work was carried out.
This research was supported in part by grant NSF PHY-2309135 to the Kavli
Institute for Theoretical Physics (KITP). The authors acknowledge use of generative AI to assist in derivations and manuscript writing. All AI output was checked by the authors, and they take full responsibility for the results presented in this work.

\vspace*{0.5cm}
\noindent{\bf Note added} While completing this work we learned about an upcoming related work
[K. Sharma, S. Jangid, A. Pattanayak, T. Arnoulx de Pirey, J. Klamser and T. Sadhu, {\it Current Fluctuations in Active Ideal Gases from Phase-Space Hydrodynamics}, in preparation].
The connection between our two works will be discussed in a future version of this paper.

\appendix

\section{Optimal density profiles in the quenched ballistic regime}
\label{app:density_RTP}

In this Appendix we derive the density profiles summarized in Sec.~\ref{sec:intro_results}. We consider non-interacting RTPs with quenched initial density $r(x)=\rho_0\theta(-x)$ (i.e. we take $\rho_1=0$) and annealed initial orientations. We focus on the extreme large deviation regime described by \eqref{Pqu_ldf}, where $\gamma t_f\to\infty$ with
\be
q=\frac{Q}{\rho_0v_0t_f}=O(1),\qquad 0<q<1\;.
\ee
The goal of this Appendix is to reconstruct the optimal density profile corresponding to a given value of the current $Q$. For this we use the relation
\be \label{rho_reconstruction_app}
\rho_s(x,t)=\frac12\left[F_s(x,t)G_s(x,t)+F_d(x,t)G_d(x,t)\right] \;,
\ee
where the exact expressions of $F_s$, $G_s$ and $G_d$ are given in \eqref{Fs_quenched} and
\eqref{Gs_sol_freeRTP}--\eqref{Gd_sol_freeRTP}, while $F_d$ can be obtained from \eqref{eqFs_freeRTP}. These quantities depend on the current $Q$ through the parameter $\lambda$. 
Using the saddle-point relation \eqref{SigmaPrime},
$Q^*(t_f)=\Sigma_{\rm qu}'(\lambda)$, together with
Eqs.~\eqref{sigma_tf_sq}--\eqref{def_W}, gives, for $0<\alpha<1$,
\be \label{alpha_q_app1}
\lambda=\alpha\gamma t_f, \qquad q={\cal W}'(\alpha)=\sqrt{\alpha(2-\alpha)}
\quad \Rightarrow \quad
\alpha=1-\sqrt{1-q^2}\;.
\ee

The calculation is organized as follows. We first determine the macroscopic profile at time $t=\tau t_f$, with $0<\tau<1$ fixed, on the ballistic spatial
scale $x=O(v_0t_f)$. We then resolve the final-time layer $t=t_f-\kappa/\gamma$ with $\kappa=O(1)$, on the scale $x=O(v_0/\gamma)$. For fixed $0<\tau<1$, the macroscopic result does not resolve the smaller diffusive
boundary layer around $x=0$, whose width is $O(\sqrt{D_at_f})$. We calculate this profile separately and use it to match the final-time regime. We finally integrate the density on $x>0$ to obtain the time-dependent integrated current $Q(t)$ in the two regimes.

\subsection{Profile at fixed $0<\tau=t/t_f<1$}

For $y=-uv_0t_f$ with $0<u<1$, the function $U$ has the  large-time behavior at exponential order given in \eqref{ldf_rho1.2}, $U(-y,t_f)\asymp e^{-\gamma t_f(1-\sqrt{1-u^2})}$.
Using Eq.~\eqref{alpha_q_app1}, we find that $e^\lambda U(-y,t_f)$ tends to infinity for $0<u<q$ and to zero for $u>q$. Consequently, we obtain from \eqref{Fs_quenched} and
\eqref{Gs_sol_freeRTP}
\be \label{h:split}
F_s(x,\tau t_f)\simeq
\rho_0\int_{-\infty}^{-qv_0t_f}dy\,P_s(x,\tau t_f|y)
+\rho_0e^{-\lambda}\int_{-qv_0t_f}^0dy\,
\frac{P_s(x,\tau t_f|y)}{U(-y,t_f)}\;,
\ee
where we recall that the propagator $P_s$ is given in \eqref{freeRTP_propagator}.
As a function of the initial position $y$, the propagator $P_s(x,\tau t_f|y)$ is peaked around $y=x$. In the limit $\gamma t_f\to\infty$ considered here, the first integral can thus be approximated to 1 for $x<-qv_0 t_f$, while it is exponentially small for $x>-qv_0 t_f$.
In the second integral, the exponential part of the ratio is
\be \label{ibis:rate}
\frac{P_s(x,\tau t_f|y)}{U(-y,t_f)}
\asymp e^{-\gamma t_f\Phi(y)},\qquad
\Phi(y)=\sqrt{1-\frac{y^2}{v_0^2t_f^2}}
-\sqrt{\tau^2-\frac{(x-y)^2}{v_0^2t_f^2}}-(1-\tau)\;.
\ee
The saddle-point equation $\Phi'(y_*)=0$ reads
\be \label{h:saddle}
\frac{x-y_*}{\sqrt{v_0^2\tau^2t_f^2-(x-y_*)^2}}
+\frac{y_*}{\sqrt{v_0^2t_f^2-y_*^2}}=0 \quad \Rightarrow \quad y_* = \frac{x}{1-\tau}\;.
\ee
This saddle lies in the integration interval when
$-q(1-\tau)v_0t_f<x<0$. In this range, we have
\be
\Phi(y_*) = -(1-\tau)\left(1-\sqrt{1-\frac{x^2}{v_0^2t_f^2(1-\tau)^2}}\right) .
\ee
Using the asymptotic behavior of $U$ from (\ref{ldf_rho1.2}) in the expression of $G_s$ given in (\ref{Gs_sol_freeRTP}), we find that, for
$-q(1-\tau)v_0t_f<x<0$, 
$
e^\lambda U(-x,(1-\tau)t_f)\gg 1$. Indeed, the exponent is minimal at $x=-q(1-\tau)v_0t_f$, where it is equal to $\alpha\tau>0$
and therefore 
\bea
G_s(x,\tau t_f)\simeq
2e^\lambda U(-x,(1-\tau)t_f) \;.
\label{asympt_Gs_App}
\eea
{For $-q(1-\tau)v_0t_f<x<0$, the exponential contributions to $F_s$ and $G_s$ thus cancel out and one expects the density $\rho_s$ to be nontrivial (see below).}
For $x<-qv_0t_f$, the first integral in Eq.~(\ref{h:split}) gives $F_s\simeq\rho_0$. In the same region, $e^\lambda U(-x,(1-\tau)t_f)$ is exponentially small whenever $U$ is nonzero (and $U=0$ outside the corresponding ballistic cone), so that $G_s\simeq 2$. Besides, Eq.~(\ref{Gd_sol_freeRTP}) also shows that $G_d$ is exponentially small. Hence $\rho_s(x,\tau t_f)\simeq\rho_0$ in this region. In the remaining intervals $-qv_0t_f<x<-q(1-\tau)v_0t_f$ and $x>0$, the term $F_sG_s/2$ is exponentially small.

To obtain the density in the interval $-q(1-\tau)v_0t_f<x<0$, one must also keep the
prefactors in front of the exponential contributions studied so far. Writing $u_*=-y_*/(v_0t_f)$ and $s_* = \sqrt{1-u_*^2}$, and using the Bessel asymptotics
$I_0(z)\simeq I_1(z)\simeq e^z/\sqrt{2\pi z}$ in \eqref{freeRTP_propagator} and \eqref{Urealtime}, we obtain
\be \label{h:prefactors}
P_s(x,\tau t_f|y) \simeq \frac{\gamma}{v_0} \frac{1+s_*}{2s_*\sqrt{2\pi\gamma t_f \tau s_*}}
 e^{-\gamma t_f \tau(1-s_*)} , \qquad U(-y_*,t_f)\simeq
\frac{1+s_*}{2u_*\sqrt{2\pi\gamma t_fs_*}}
 e^{-\gamma t_f(1-s_*)}.
\ee
Using also $\Phi''(y_*)=\frac{1-\tau}{v_0^2\tau t_f^2s_*^3}$, we can perform the Gaussian integration around the saddle-point $y_*$ to obtain (combining with \eqref{asympt_Gs_App})
\be \label{ibis:product1}
\frac{F_sG_s}{2}\simeq\frac{\rho_0(1+s_*)}{2(1-\tau)}.
\ee
We now need to consider the term $F_d G_d/2$. 
From \eqref{Gd_sol_freeRTP} and from the relation $\partial_t F_s=-v_0\partial_xF_d$, we see that $F_d$ and $G_d$ have the same exponential dependence as $F_s$ and $G_s$, respectively. The term $F_dG_d/2$ therefore gives a non-negligible contribution only for
$-q(1-\tau)v_0t_f<x<0.$
On this interval, we find that
\be 
\frac{F_d}{F_s}\simeq\frac{G_d}{G_s}\simeq\frac{u_*}{1+s_*} \quad \Rightarrow \quad \frac{F_dG_d}{2} \simeq\frac{\rho_0(1-s_*)}{2(1-\tau)}.
\ee
Putting everything together, we thus obtain, in the limit $\gamma t_f\to\infty$,
\be \label{h:macro}
\frac{\rho_s(x,\tau t_f)}{\rho_0}\longrightarrow
\begin{cases}
1,&x<-qv_0t_f,\\
0,&-qv_0t_f<x<-q(1-\tau)v_0t_f,\\
(1-\tau)^{-1},&-q(1-\tau)v_0t_f<x<0,\\
0,&x>0.
\end{cases}
\ee
This limit holds at fixed $x/(v_0t_f)$ away from the three boundaries
$x=-qv_0t_f$, $x=-q(1-\tau)v_0t_f$ and $x=0$ separating the four intervals
in Eq.~\eqref{h:macro}. The interval adjacent to the origin has mass $Q$ and width
$qv_0(t_f-t)$. When $t_f-t=O(\gamma^{-1})$, its width is of order
$v_0/\gamma$, which is the spatial scale considered next.

\subsection{Profile close to the final time} \label{app:density_diffusivescale}

We now set $t=t_f-\kappa/\gamma$ and $x=(v_0/\gamma)\xi$, keeping
$\kappa$ and $\xi$ fixed as $\gamma t_f\to\infty$.
In the second integral of Eq.~\eqref{h:split}, we set
$y=-uv_0t_f$ and write $s(u)=\sqrt{1-u^2}$ (here we omit the part of the density corresponding to $x<-qv_0t_f$, and thus we can neglect the first integral). The modified Bessel functions
$I_0$ and $I_1$ in the continuous part of the propagator
\eqref{freeRTP_propagator} have the common argument
$\sqrt{\gamma^2t^2-(\gamma(x-y)/v_0)^2}$. Using
$t=t_f-\kappa/\gamma$, $x=(v_0/\gamma)\xi$ and $y=-uv_0t_f$, its large-$t_f$
expansion is
\be \label{ibis:expansion}
\sqrt{\gamma^2t^2-(\gamma x/v_0+u\gamma t_f)^2}
=\gamma t_fs(u)-\frac{\kappa+u\xi}{s(u)}+o(1)\;.
\ee
Keeping the prefactors in the propagator and in $U$, we obtain
\be \label{ibis:ratio}
\frac{P_s(x,t|-uv_0t_f)}{U(uv_0t_f,t_f)}
\simeq\frac{\gamma}{v_0}\frac{u}{s(u)}
 e^{-\kappa[1/s(u)-1]-\xi u/s(u)}\;.
\ee
For each fixed $u$, the equation for $F_d$ gives the amplitude ratio
$F_d/F_s=u/[1+s(u)]$. The backward fields satisfy
\be \label{ibis:backward}
\frac{G_s}{2e^\lambda}\longrightarrow
U\!\left(-\frac{v_0\xi}{\gamma},\frac{\kappa}{\gamma}\right),
\qquad
\frac{G_d}{e^\lambda}\longrightarrow
 e^{-\kappa}I_0\!\left(\sqrt{\kappa^2-\xi^2}\right)
 \theta(\kappa-|\xi|)\;.
\ee
Substituting into Eq.~\eqref{rho_reconstruction_app} then gives
$\rho_s(x,t)\simeq\rho_0\gamma t_f{\cal F}_q(\xi,\kappa)$, with
\be \label{h:terminal}
{\cal F}_q(\xi,\kappa)=\int_0^qdu\,
 e^{-\kappa[1/s(u)-1]-\xi u/s(u)}
\Bigg[\frac{u}{s(u)}
 U\!\left(-\frac{v_0\xi}{\gamma},\frac{\kappa}{\gamma}\right)+\frac12\left(\frac1{s(u)}-1\right)e^{-\kappa}
 I_0\!\left(\sqrt{\kappa^2-\xi^2}\right)
 \theta(\kappa-|\xi|)\Bigg]\;.
\ee
The terms containing $I_0$ are understood to vanish for $|\xi|>\kappa$.
The contribution from any fixed $u>q$ is exponentially small; the
shrinking neighborhoods of $u=0$ and $u=q$ give subleading corrections. The function $F_q(\xi,\kappa)$ has discontinuities at $\xi=\pm\kappa$, originating from the ballistic edges of the RTP propagator. We give explicitly the jump at $\xi=\kappa$. This discontinuity comes from both the function $U$ in the first term and the $\theta$ function in the second term of Eq. (\ref{h:terminal}). One finds
\be \label{Fq_jump_plus_app}
{\cal F}_q(\kappa^+,\kappa)-{\cal F}_q(\kappa^-,\kappa)
=\frac{e^{-\kappa}}{2}\int_0^qdu\,
\left[1-\sqrt{\frac{1-u}{1+u}}\right]\exp\left\{-\kappa\left[
\sqrt{\frac{1+u}{1-u}}-1\right]\right\}>0\;.
\ee

At $\kappa=0$, i.e. at time $t=t_f$, the expression \eqref{h:terminal} reduces to
\be \label{h:final}
{\cal F}_q(\xi,0)=\theta(\xi)\int_0^qdu\,
\frac{u}{\sqrt{1-u^2}}e^{-\xi u/\sqrt{1-u^2}},
\qquad
\int_0^{\infty}d\xi\,{\cal F}_q(\xi,0)=q\;.
\ee
The second equality follows by integrating over $\xi$ first and gives $\int_0^\infty dx\,\rho_s(x,t_f)=Q$, as one should expect.
For $\xi<-\kappa$, i.e. $x<-v_0(t_f-t)$, one has ${\cal F}_q(\xi,\kappa)=0$ since particles there would not have time to cross the origin before time $t_f$.
For large positive $\xi$ at fixed $q$ and $\kappa$, only
$u=O(\xi^{-1})$ contributes, so that we get the asymptotic behavior
\be
{\cal F}_q(\xi,\kappa)\underset{\xi\to+\infty}{\simeq}\int_0^\infty du\,u e^{-\xi u}=\frac{1}{\xi^2} \;,
\ee
as announced in Eq. (\ref{Fq_tail_intro_new}) in the main text.

\subsection{Boundary layer for the density near the origin}

We now want to resolve more closely the behavior of the density near the origin at all times. At fixed $0<\tau<1$, we keep
$X=x/\sqrt{4D_at_f}$ fixed, where $D_a=v_0^2/(2\gamma)$.
The propagator $P_s$ is Gaussian on this scale. Substituting
$y=-2\sqrt{D_at_f}\,Y$ into Eq.~\eqref{Fs_quenched} and
\eqref{Gs_sol_freeRTP} thus gives
\be \label{h:diff_fields}
F_s\simeq\frac{2\rho_0e^{-\lambda}}{\sqrt{\pi\tau}}
\int_0^\infty dY\,\frac{e^{-(X+Y)^2/\tau}}{\operatorname{erfc}(Y)},
\qquad
G_s\simeq e^\lambda\operatorname{erfc}\!\left(-\frac{X}{\sqrt{1-\tau}}\right)\;.
\ee
Here $F_dG_d/(F_sG_s)=O((\gamma t_f)^{-1})$, so the leading density is
\be \label{h:R}
\rho_s(X\sqrt{4D_at_f},\tau t_f)\simeq\rho_0{\cal R}(X,\tau)\;, \qquad
{\cal R}(X,\tau)=
\frac{\operatorname{erfc}(-X/\sqrt{1-\tau})}{\sqrt{\pi\tau}}
\int_0^\infty dY\,\frac{e^{-(X+Y)^2/\tau}}{\operatorname{erfc}(Y)}\;.
\ee
For $X\to+\infty$, the integral is restricted to $Y=O(X^{-1})$.
Replacing $\operatorname{erfc}(Y)$ by one therefore gives
${\cal R}(X,\tau)\simeq\sqrt{\tau/\pi}\,e^{-X^2/\tau}/X$, which describes the decay of the density for $x>0$.
\\

\noindent {\bf Matching with the final time regime.} To compare with the final-time result, set $\epsilon=1-\tau$ and keep
$z=X/\sqrt\epsilon$ fixed. The relevant values of $Y$ are now of order
$\epsilon^{-1/2}$. Using
$\operatorname{erfc}(Y)\simeq e^{-Y^2}/(\sqrt\pi Y)$ and setting
$Y=w/\sqrt\epsilon$, we find
\be \label{h:matching}
\begin{split}
\epsilon{\cal R}(X,1-\epsilon)
&\simeq\operatorname{erfc}(-z)\int_0^\infty dw\,w e^{-w^2-2zw}\\
&=\operatorname{erfc}(-z)
\left[\frac12-\frac{\sqrt\pi}{2}z e^{z^2}\operatorname{erfc}(z)\right]\;.
\end{split}
\ee
The variables in the two regimes are related by
$\epsilon=\kappa/(\gamma t_f)$ and
$z=\xi/\sqrt{2\kappa}$. Thus, for large $\kappa$ keeping $z$ fixed one finds
\be \label{h:Fmatch}
{\cal F}_q(\xi,\kappa)\simeq\frac{1}{\kappa}
\operatorname{erfc}(-z)
\left[\frac12-\frac{\sqrt\pi}{2}z e^{z^2}\operatorname{erfc}(z)\right],
\qquad z=\frac{\xi}{\sqrt{2\kappa}}\;.
\ee
This matches the expression obtained from Eq.~\eqref{h:terminal} by expanding
the argument of the exponential as $-\kappa u^2/2-\xi u$, using the diffusive approximation of $P_s$, and extending the integral to infinity.
For fixed $\xi/\sqrt\kappa$, this integral is dominated by $u=O(\kappa^{-1/2})$.
Its upper limit may thus be extended to infinity if $q^2\kappa\gg1$.
The matching therefore requires
$1\ll\kappa\ll\gamma t_f$ and $q^2\kappa\gg1$.

\subsection{Integrated current for all time $t \leq t_f$}

Since $x>0$ is empty initially, one has at all times $Q(t)=\int_0^\infty dx\,\rho_s(x,t)$.
The two density scalings above give
\bea \label{h:current_diff}
&&Q(\tau t_f)\simeq\rho_0\sqrt{D_at_f}\,{\cal J}_{\rm diff}(\tau),
\qquad {\cal J}_{\rm diff}(\tau)=2\int_0^\infty dX\,{\cal R}(X,\tau),\\
&&Q(t_f-\kappa/\gamma)\simeq\rho_0v_0t_f\,{\cal J}_{\rm bal}(\kappa),
\qquad {\cal J}_{\rm bal}(\kappa)=\int_0^\infty d\xi\,{\cal F}_q(\xi,\kappa)\;. \label{h:current_bal}
\eea
In particular, from \eqref{h:final} we have ${\cal J}_{\rm bal}(0)=q$. In the matching $\tau\to 1^-$ for \eqref{h:current_diff} and $1\ll\kappa\ll\gamma t_f$ with $q^2\kappa\gg1$ for \eqref{h:current_bal}, both expressions can be evaluated using Eq.~\eqref{h:matching} and the integral
\be \label{h:integral}
\int_0^\infty dz\,\operatorname{erfc}(-z)
\left[\frac12-\frac{\sqrt\pi}{2}z e^{z^2}\operatorname{erfc}(z)\right]
=\frac{\pi+2}{4\sqrt\pi}\;,
\ee
which follows by writing the bracket as
$-(\sqrt\pi/4)\,d[e^{z^2}\operatorname{erfc}(z)]/dz$ and integrating by parts. Hence we obtain in this regime
\be \label{h:current_match}
{\cal J}_{\rm diff}(\tau)\simeq
\frac{\pi+2}{2\sqrt{\pi(1-\tau)}},\qquad
{\cal J}_{\rm bal}(\kappa)\simeq
\frac{\pi+2}{2\sqrt{2\pi\kappa}}\;,
\ee
which give the same $Q(t)$ with $1-\tau=\kappa/(\gamma t_f)$.

\section{Optimal density profiles for large quenched Brownian currents}
\label{app:density_br}

In this Appendix, we perform the same calculations as in Appendix~\ref{app:density_RTP} but in the case of Brownian particles. We consider non-interacting Brownian particles with diffusion coefficient $D$ and quenched initial density $r(x)=\rho_0\theta(-x)$ (i.e. we take $\rho_1=0$). We consider large deviations of the current $Q/(\rho_0\sqrt{Dt_f})\gg1$ (we do not impose a
specific scaling of $Q$ with $t_f$), and we would like to reconstruct the corresponding optimal density profile.

The exact solution is obtained through $\rho=FG$, with $F$ and $G$ given in \eqref{F_Brownian_quenched} and \eqref{G_Brownian_annealed}. Using ${\rm erfc}(-x)=1+{\rm erf}(x)$, these can be written
\be \label{i:exact}
F(x,t)=\rho_0\int_{-\infty}^0dy\,
\frac{P_{\rm B}(x,t|y)}{1+(e^\lambda-1)U_{\rm B}(y,t_f)},
\qquad G(x,t)=1+(e^\lambda-1)U_{\rm B}(x,t_f-t)\;,
\ee
where
\be \label{i:PU}
P_{\rm B}(x,t|y)=\frac{e^{-(x-y)^2/(4Dt)}}{\sqrt{4\pi Dt}},
\qquad U_{\rm B}(y,t)=\frac12\operatorname{erfc}\!\left(-\frac{y}{\sqrt{4Dt}}\right)\;.
\ee
The saddle-point relation $Q^*(t_f)=\Sigma_{\rm qu}'(\lambda)$, together with the large $\lambda$ approximation of \eqref{sigma_free_brownian_quenched}, $\Sigma_{\rm qu}(\lambda)
\simeq \frac{4}{3} \rho_0 \sqrt{Dt_f}\,\lambda^{3/2}$, gives
\be
\lambda \simeq \frac{Q^2}{4\rho_0^2Dt_f}.
\ee
As in the RTP case, we first determine the macroscopic profile at time $t=\tau t_f$, with $0<\tau<1$ fixed, on the spatial scale $x=O(Q/\rho_0)$, before resolving the smaller diffusive boundary layer around $x=0$ and the final-time boundary layer. We end this appendix once again by calculating the time-dependent integrated current $Q(t)$ in the two regimes.

\subsection{Profile at fixed $0<\tau=t/t_f<1$}

The calculation of the macroscopic profile is very similar to the RTP case. For $y<0$ and $|y|\gg\sqrt{Dt_f}$,
$U_{\rm B}(y,t_f)\simeq-\sqrt{Dt_f}\,e^{-y^2/(4Dt_f)}/(\sqrt\pi \,y)$.
Thus the denominator in Eq.~\eqref{i:exact} is dominated by its second term for $-Q/\rho_0<y<0$, and is approximately equal to 1 for $y<-Q/\rho_0$, to leading order on the scale $Q/\rho_0$. This gives
\be \label{i:split}
F(x,\tau t_f)\simeq
\rho_0\int_{-\infty}^{-Q/\rho_0}dy\,P_{\rm B}(x,\tau t_f|y)
+\rho_0e^{-\lambda}\int_{-Q/\rho_0}^0dy\,
\frac{P_{\rm B}(x,\tau t_f|y)}{U_{\rm B}(y,t_f)}\;.
\ee
Since the Gaussian propagator $P_{\rm B}(x,\tau t_f|y)$ is peaked around $y=x$, we can approximate it to 1 for $x<-Q/\rho_0$ and neglect it for $x>-Q/\rho_0$.
In the second integral, we can again use $|y|\gg\sqrt{Dt_f}$ to write
\be \label{i:ratio}
\frac{P_{\rm B}(x,\tau t_f|y)}{U_{\rm B}(y,t_f)}
\simeq\frac{-y}{2Dt_f\sqrt\tau}
\exp\!\left[-\frac{(x-y)^2}{4D\tau t_f}+\frac{y^2}{4Dt_f}\right]\;.
\ee
The argument of the exponential in {\magenta Eq.~\eqref{i:ratio}} is maximal at $y_*=x/(1-\tau)$.
For $-(1-\tau)Q/\rho_0<x<0$, this point belongs to the integration
interval. Performing the Gaussian integral gives in this interval
\be\label{i:F}
F(x,\tau t_f)\simeq\rho_0e^{-\lambda}
\frac{\sqrt\pi(-y_*)}{\sqrt{Dt_f(1-\tau)}}
 e^{x^2/[4D(1-\tau)t_f]}\;.
\ee
When $x$ is outside of the interval $(-(1-\tau)Q/\rho_0,0)$ the contribution of the second integral is exponentially smaller.

In the interval relevant for the saddle, $-(1-\tau)Q/\rho_0<x<0$, the large-current limit gives $e^\lambda U_B(x,(1-\tau)t_f)\gg 1$. Using the large-negative-$x$ asymptotics of $U_B$, we thus obtain
\bea
G(x,\tau t_f)
\simeq
e^\lambda\,
\frac{\sqrt{D(1-\tau)t_f}}{\sqrt{\pi}(-x)}
\exp\left[-\frac{x^2}{4D(1-\tau)t_f}\right],
\qquad
-\frac{(1-\tau)Q}{\rho_0}<x<0 .
\label{I7}
\eea
{For $-(1-\tau)Q/\rho_0<x<0$, multiplying (\ref{i:F}) with (\ref{I7}) thus gives $\rho(x,\tau t_f)\simeq \rho_0/(1-\tau)$.} For $x<-Q/\rho_0$, the first integral in Eq.~(\ref{i:split}) gives $F\simeq\rho_0$. Moreover, $e^\lambda U_B(x,(1-\tau)t_f)\ll1$ in this region, so that $G\simeq1$ and therefore $\rho(x,\tau t_f)\simeq\rho_0$. {Finally,} for $-Q/\rho_0<x<-(1-\tau)Q/\rho_0$ {and} for $x>0$ the density is exponentially small on the scale $Q/\rho_0$.
Putting everything together,
\be \label{i:macro}
\frac{\rho(x,\tau t_f)}{\rho_0}\longrightarrow
\begin{cases}
1,&x<-Q/\rho_0,\\
0,&-Q/\rho_0<x<-(1-\tau)Q/\rho_0,\\
(1-\tau)^{-1},&-(1-\tau)Q/\rho_0<x<0,\\
0,&x>0.
\end{cases}
\ee
This limit is understood at fixed $x\rho_0/Q$, away from the three boundaries
$x=-Q/\rho_0$, $x=-(1-\tau)Q/\rho_0$ and $x=0$ separating the four intervals
in Eq.~\eqref{i:macro}. The nonzero interval adjacent to the origin has mass $Q$. This result is the same as the RTP result \eqref{h:macro}, but without requiring a ballistic scaling of the current $Q$.

Near the origin, we instead keep $X=x/\sqrt{4Dt_f}$ fixed.
Taking $y=-2\sqrt{Dt_f}\,Y$ in Eq.~\eqref{i:exact} gives
\be \label{i:diff_fields}
F\simeq\frac{2\rho_0e^{-\lambda}}{\sqrt{\pi\tau}}
\int_0^\infty dY\,\frac{e^{-(X+Y)^2/\tau}}{\operatorname{erfc}(Y)},
\qquad G\simeq\frac{e^\lambda}{2}
\operatorname{erfc}\!\left(-\frac{X}{\sqrt{1-\tau}}\right)\;.
\ee
Hence $\rho(x,\tau t_f)\simeq\rho_0{\cal R}(X,\tau)$, with exactly the same function ${\cal R}(X,\tau)$ as in Eq.~\eqref{h:R}. In particular, at fixed $\tau<1$,
${\cal R}(X,\tau)\simeq\sqrt{\tau/\pi}\,e^{-X^2/\tau}/X$ as
$X\to+\infty$.

\subsection{Profile close to the final time}

The two lengths $(1-\tau)Q/\rho_0$ and $\sqrt{D(t_f-t)}$ become
comparable when $(1-\tau)Q^2/(\rho_0^2Dt_f)=O(1)$.
We therefore introduce
\be \label{i:variables}
\kappa_{\rm B}=\frac{Q^2(t_f-t)}{4\rho_0^2Dt_f^2},
\qquad \xi_{\rm B}=\frac{xQ}{4\rho_0Dt_f}\;,
\ee
and keep these variables fixed in the large-current limit.
In the second integral of Eq.~\eqref{i:split}, set
$y=-(Q/\rho_0)u$, with $0<u<1$. The argument of the exponential in
Eq.~\eqref{i:ratio} then becomes
\be \label{jbis:exponent}
-\frac{(x-y)^2}{4Dt}+\frac{y^2}{4Dt_f}
=-\kappa_{\rm B}u^2-2\xi_{\rm B}u
+O\!\left(\frac{\rho_0^2Dt_f}{Q^2}\right)\;.
\ee
Keeping the prefactor gives
\be \label{i:terminal_fields}
F(x,t)\simeq\frac{Q^2e^{-\lambda}}{2\rho_0Dt_f}
\int_0^1du\,u e^{-\kappa_{\rm B}u^2-2\xi_{\rm B}u},
\qquad
G(x,t)\simeq\frac{e^\lambda}{2}
\operatorname{erfc}\!\left(-\frac{\xi_{\rm B}}{\sqrt{\kappa_{\rm B}}}\right).
\ee
The contribution of the first integral in Eq.~\eqref{i:split}, which comes from initial positions $y<-Q/\rho_0$, is subleading on this scale. This remains true in the neighborhood of the boundary $y=-Q/\rho_0$ separating the two integration intervals in
Eq.~\eqref{i:split}. We therefore obtain
\be \label{i:B}
\rho(x,t)\simeq\frac{Q^2}{4\rho_0Dt_f}
{\cal B}(\xi_{\rm B},\kappa_{\rm B}),\qquad
{\cal B}(\xi,\kappa)=\operatorname{erfc}\!\left(-\frac{\xi}{\sqrt\kappa}\right)
\int_0^1du\,u e^{-\kappa u^2-2\xi u}\;.
\ee
At the final time, integration over $u$ yields
\be \label{i:final}
{\cal B}(\xi,0)=\theta(\xi)
\frac{1-(1+2\xi)e^{-2\xi}}{2\xi^2},\qquad
{\cal B}(0^+,0)=1,\qquad
\int_0^\infty d\xi\,{\cal B}(\xi,0)=1\;.
\ee
The normalization imposes $\int_0^\infty dx\,\rho(x,t_f)=Q$, as one should expect. Since ${\cal B}(\xi,0)\simeq1/(2\xi^2)$ for large positive $\xi$, we have
\be \label{i:tail}
\rho(x,t_f)\simeq\frac{2\rho_0Dt_f}{x^2},
\qquad \frac{\rho_0Dt_f}{Q}\ll x\ll\sqrt{Dt_f}\;.
\ee

To match this result with the finite time scaling, let us take $1\ll\kappa_{\rm B}\ll Q^2/(4\rho_0^2Dt_f)$ and keep
$z=\xi_{\rm B}/\sqrt{\kappa_{\rm B}}$ fixed. Substituting
$u=w/\sqrt{\kappa_{\rm B}}$ in Eq.~\eqref{i:B} and extending the
upper limit to infinity gives
\be \label{i:matching}
{\cal B}(\xi_{\rm B},\kappa_{\rm B})\simeq
\frac{1}{\kappa_{\rm B}}\operatorname{erfc}(-z)
\left[\frac12-\frac{\sqrt\pi}{2}z e^{z^2}\operatorname{erfc}(z)\right]\;,
\ee
which is compatible with the limit $\tau\to1^-$ of the function $\mathcal{R}(X,\tau)$ given in \eqref{h:matching}, with $1-\tau=4\rho_0^2Dt_f\kappa_{\rm B}/Q^2$ and
$X/\sqrt{1-\tau}=z$.

\subsection{Integrated current for all time $t \leq t_f$}

Integrating the density over the positive half-line gives
\bea \label{i:current_diff}
&&Q(\tau t_f)\simeq\rho_0\sqrt{Dt_f}\,{\cal J}_{\rm diff}(\tau),
\qquad {\cal J}_{\rm diff}(\tau)=2\int_0^\infty dX\,{\cal R}(X,\tau),\\
&&Q\!\left(t_f-\frac{4\rho_0^2Dt_f^2}{Q^2}\kappa_{\rm B}\right)
\simeq Q\,{\cal J}_{\rm B}(\kappa_{\rm B}),
\qquad {\cal J}_{\rm B}(\kappa)=\int_0^\infty d\xi\,{\cal B}(\xi,\kappa)\;,\label{i:current_B}
\eea
where ${\cal J}_{\rm diff}$ is the same as in Eq.~\eqref{h:current_diff}, and
Eq.~\eqref{i:final} gives ${\cal J}_{\rm B}(0)=1$.
For large $\kappa_{\rm B}$, substituting
$\xi_{\rm B}=\sqrt{\kappa_{\rm B}}\,z$ and using
Eqs.~\eqref{i:matching} and \eqref{h:integral} yields
\be \label{jbis:current_match}
{\cal J}_{\rm B}(\kappa_{\rm B})\simeq
\frac{\pi+2}{4\sqrt{\pi\kappa_{\rm B}}}\;.
\ee
This gives the same expression for $Q(t)$ as the limit $\tau\to 1^-$ of Eq.~\eqref{i:current_diff}, already given in Eq.~\eqref{h:current_match}, with
$1-\tau=4\rho_0^2Dt_f\kappa_{\rm B}/Q^2$.

\section{Two-point correlations in Fourier space and time}
\label{app:two_point_fourier_time}

In Section~\ref{sec:correlations}, we derived the two-point correlation functions of the density in Fourier space and real time. Here we give their expression in Fourier space and Fourier time. We again work directly in the thermodynamic limit and use the Fourier convention
\be
\tilde f(k,\omega)
=
\int dx\,dt \; e^{i(\omega t-kx)} f(x,t) \;.
\ee
The linearized Dean-Kawasaki equations \eqref{dean_delta2} then read
\bea \label{dean_fourier2} 
-i\omega \delta \tilde\rho_s(k,\omega) &=& - ik v_0 \delta \tilde \rho_d(k,\omega) - k \rho h(k) \delta \tilde\rho_s(k,\omega) \;,
\\
-i\omega \delta \tilde \rho_d(k,\omega) &=& - ik v_0 \delta \tilde \rho_s(k,\omega) - 2\gamma \delta \tilde \rho_d(k,\omega) + 2 \sqrt{\gamma \rho} \, \tilde \zeta_K(k,\omega) \;, \nonumber
\eea
with
\be \label{zetacorr_k_omega}
\langle \tilde \zeta_K(k,\omega) \tilde \zeta_K(k',\omega')^*\rangle = (2 \pi)^2 \delta(k-k') \delta(\omega-\omega') \;.
\ee
We henceforth consider $k\neq0$, the zero mode being fixed by particle-number conservation. The first equation yields the relation
\be \label{rhosdrel_fourier} 
\delta \tilde\rho_d(k,\omega) = \frac{i\omega-\rho k h(k)}{iv_0k} \delta \tilde\rho_s(k,\omega) \;.
\ee
Inserting in the second equation we get
\be \label{rhosd_fourier} 
\delta \tilde\rho_s(k,\omega) = \frac{2iv_0k\sqrt{\gamma \rho}}{Z(k,\omega)} \; \tilde \zeta_K(k,\omega) \;,
\ee
where we have introduced
\be
Z(k,\omega) = \omega^2 -v_0^2k^2  -2\gamma\rho kh(k) + i\omega(2\gamma+\rho kh(k)) \;.
\ee
From these two relations, as well as the noise correlations \eqref{zetacorr_k_omega}, we easily obtain the two-point correlation functions,
\bea \label{corr_fourier_def}
&&\left\langle \delta \tilde\rho_s(k,\omega) \delta \tilde\rho_s^*(k',\omega') \right\rangle
= (2 \pi)^2 \delta(k-k') \delta(\omega-\omega') C_{ss}(k,\omega) \;,
\\
&&\left\langle \delta \tilde\rho_d(k,\omega) \delta \tilde\rho_d^*(k',\omega') \right\rangle
= (2 \pi)^2 \delta(k-k') \delta(\omega-\omega') C_{dd}(k,\omega) \;, \nn
\\
&&\left\langle \delta \tilde\rho_s(k,\omega) \delta \tilde\rho_d^*(k',\omega') \right\rangle
= (2 \pi)^2 \delta(k-k') \delta(\omega-\omega') C_{sd}(k,\omega) \;, \nn
\eea
where
\bea
C_{ss}(k,\omega) &=& \frac{ 4 \gamma \rho v_0^2 k^2}{ |Z(k,\omega)|^2 } \;, \\
C_{dd}(k,\omega) &=& \frac{ 4 \gamma \rho (\omega^2 + \rho^2 k^2 h(k)^2) }{ |Z(k,\omega)|^2 } \;, \nn \\
C_{sd}(k,\omega) &=& \frac{4 \gamma \rho v_0k(\omega-i\rho k h(k))}{ |Z(k,\omega)|^2 } \;. \nn
\eea

\section{Evaluation of the first-order interaction correction for Brownian particles}
\label{app:Brownian_first_order}

In this Appendix we evaluate explicitly the space-time integral entering the first-order
interaction correction to the annealed current cumulant generating function for weakly
interacting Brownian particles. For a sufficiently short-ranged even interaction, the
gradient expansion performed in Section~\ref{sec:step_interaction} gives
\be \label{app:Sigma1_def}
\Sigma_1^{\rm an,0}(\lambda)
\simeq
-W_0\,{\cal I}(\lambda),
\qquad
W_0=\int_{-\infty}^{+\infty}dr\,W(r),
\ee
where
\be \label{app:I_def}
{\cal I}(\lambda)
=
\int_0^{t_f}dt
\int_{-\infty}^{+\infty}dx\,
F_0(x,t)\partial_xG_0(x,t)
\partial_x\left[F_0(x,t)G_0(x,t)\right].
\ee
The free annealed saddle-point fields can be written as
\be \label{app:FG_abcd}
F_0(x,t)
=
a+b\,{\rm erf}\left(\frac{x}{\sqrt{4Dt}}\right),
\qquad
G_0(x,t)
=
c+d\,{\rm erf}\left(\frac{x}{\sqrt{4D(t_f-t)}}\right),
\ee
where
\be \label{app:abcd}
a=\frac{\rho_0+\rho_1e^{-\lambda}}{2},
\qquad
b=\frac{\rho_1e^{-\lambda}-\rho_0}{2},
\qquad
c=\frac{1+e^\lambda}{2},
\qquad
d=\frac{e^\lambda-1}{2}.
\ee

We introduce the dimensionless variables $\tau=\frac{t}{t_f}$, 
$z=\frac{x}{\sqrt{Dt_f}}$ and define
\be
E(z,\tau)
=
{\rm erf}\left(\frac{z}{2\sqrt{\tau}}\right),
\qquad
H(z,\tau)
=
{\rm erf}\left(\frac{z}{2\sqrt{1-\tau}}\right).
\ee
Using $F_0\partial_xG_0\,\partial_x(F_0G_0)
=
F_0G_0\,\partial_xF_0\,\partial_xG_0
+
F_0^2(\partial_xG_0)^2$, together with
$dt\,dx
=
t_f\sqrt{Dt_f}\,d\tau\,dz$,
we obtain
\bea \label{app:I_rescaled}
{\cal I}(\lambda)
=
\frac{1}{\pi}\sqrt{\frac{t_f}{D}}
\int_0^1d\tau\int_{-\infty}^{+\infty}dz
\Bigg[
&&
\frac{bd}{\sqrt{\tau(1-\tau)}}
e^{-\frac{z^2}{4\tau(1-\tau)}}
(a+bE)(c+dH)
+
\frac{d^2}{1-\tau}
e^{-\frac{z^2}{2(1-\tau)}}
(a+bE)^2
\Bigg].
\eea

Since $E(z,\tau)$ and $H(z,\tau)$ are odd functions of $z$, while the Gaussian
factors are even, all terms containing an odd number of error functions vanish upon
integration over $z$. Hence only four contributions remain,
\be \label{app:I_four_terms}
{\cal I}(\lambda)
=
\sqrt{\frac{t_f}{D}}
\left[
acbd\,{\cal J}_0
+
b^2d^2\,{\cal J}_{EH}
+
a^2d^2\,{\cal K}_0
+
b^2d^2\,{\cal K}_{E^2}
\right],
\ee
where
\bea
{\cal J}_0
&=&
\frac{1}{\pi}
\int_0^1\frac{d\tau}{\sqrt{\tau(1-\tau)}}
\int_{-\infty}^{+\infty}dz\,
e^{-\frac{z^2}{4\tau(1-\tau)}},
\label{app:J0_def}
\\
{\cal J}_{EH}
&=&
\frac{1}{\pi}
\int_0^1\frac{d\tau}{\sqrt{\tau(1-\tau)}}
\int_{-\infty}^{+\infty}dz\,
e^{-\frac{z^2}{4\tau(1-\tau)}}
E(z,\tau)H(z,\tau),
\label{app:JEH_def}
\\
{\cal K}_0
&=&
\frac{1}{\pi}
\int_0^1\frac{d\tau}{1-\tau}
\int_{-\infty}^{+\infty}dz\,
e^{-\frac{z^2}{2(1-\tau)}},
\label{app:K0_def}
\\
{\cal K}_{E^2}
&=&
\frac{1}{\pi}
\int_0^1\frac{d\tau}{1-\tau}
\int_{-\infty}^{+\infty}dz\,
e^{-\frac{z^2}{2(1-\tau)}}
E(z,\tau)^2.
\label{app:KE2_def}
\eea
The two integrals without error functions are elementary
and give ${\cal J}_0=\frac{2}{\sqrt{\pi}}$ and 
${\cal K}_0=
\frac{2\sqrt2}{\sqrt{\pi}}$. We now evaluate the two nontrivial integrals involving products of error functions.
We use the identity
\be \label{app:erf_identity}
{\cal K}(A,B,C)
:= \int_{-\infty}^{+\infty}dz\,
e^{-A^2z^2}
{\rm erf}(Bz){\rm erf}(Cz)
=
\frac{2}{A\sqrt{\pi}}
\arctan\left[
\frac{BC}
{A\sqrt{A^2+B^2+C^2}}
\right],
\ee
valid for $A>0$. This is obtained by differentiating with respect to $C$ and using integration by parts, which gives
\be
\partial_C{\cal K}
=
\frac{2}{\sqrt{\pi}}
\int_{-\infty}^{+\infty}dz\,
z\,e^{-(A^2+C^2)z^2}{\rm erf}(Bz) = 
\frac{2B}
{\sqrt{\pi}(A^2+C^2)
\sqrt{A^2+B^2+C^2}}.
\ee
Integrating over $C$, together with ${\cal K}(A,B,0)=0$, gives
\eqref{app:erf_identity}.

We first consider ${\cal J}_{EH}$. In this case
\be
A=\frac{1}{2\sqrt{\tau(1-\tau)}},
\qquad
B=\frac{1}{2\sqrt{\tau}},
\qquad
C=\frac{1}{2\sqrt{1-\tau}},
\ee
and equation~\eqref{app:erf_identity} gives
\be \label{app:JEH_tau}
{\cal J}_{EH}
=
\frac{4}{\pi^{3/2}}
\int_0^1d\tau\,
\arctan\sqrt{\frac{\tau(1-\tau)}{2}} = \frac{6-4\sqrt2}{\sqrt{\pi}}.
\ee
We now calculate ${\cal K}_{E^2}$. In this case \eqref{app:erf_identity} applies with
\be
A=\frac{1}{\sqrt{2(1-\tau)}},
\qquad
B=C=\frac{1}{2\sqrt{\tau}},
\ee
which gives
\be \label{app:KE2_tau}
{\cal K}_{E^2}
=
\frac{2\sqrt2}{\pi^{3/2}}
\int_0^1\frac{d\tau}{\sqrt{1-\tau}}
\arctan\left(\frac{1-\tau}{2\sqrt{\tau}}\right) = 
\frac{6\sqrt2-8}{\sqrt{\pi}}.
\ee
Collecting the four contributions, 
we obtain
\be
{\cal I}(\lambda)
=
2\sqrt{\frac{t_f}{\pi D}}
\left[
acbd
+
\sqrt2\,a^2d^2
+
(\sqrt2-1)b^2d^2
\right].
\label{app:I_final}
\ee
Using \eqref{app:Sigma1_def}, this gives
\be \label{app:Sigma1_final}
\Sigma_1^{\rm an,0}(\lambda)
=
-\frac{W_0}{D}
\sqrt{\frac{Dt_f}{\pi}}\,
\Phi(\lambda;\rho_0,\rho_1)
+
o(\sqrt{t_f}),
\ee
where
\be \label{app:Phi_compact}
\Phi(\lambda;\rho_0,\rho_1)
=
2\left[
acbd
+
\sqrt2\,a^2d^2
+
(\sqrt2-1)b^2d^2
\right].
\ee
Substituting the expressions \eqref{app:abcd} for $a,b,c,d$ gives the result
\eqref{defPhi} in the main text. 

\section{Extraction of the leading ballistic-front contribution}
\label{app:RTP_ballistic_front0}

We give here the details of the reduction of the exact first-order expression
\eqref{Sigma1_RTP_exact} to the ballistic-front result
\eqref{Sigma1_RTP_ballistic_front}. Away from the isolated times at which
fronts cross, it is useful to decompose the density as
\be
\rho_s^{(0)}
=
\rho_{\rm sm}^{(0)}
+
\rho_{\rm b}^{(0)}
+
\rho_{\rm f}^{(0)},
\qquad
\rho_{\rm b}^{(0)}
=
\sum_{\sigma=\pm}
R_\sigma(t)\,
\theta\left(x-x_\sigma^{\rm b}(t)\right),
\qquad
\rho_{\rm f}^{(0)}
=
\sum_{\sigma=\pm}
S_{\sigma}(t)\,
\theta\left(x-x_\sigma^{\rm f}(t)\right),
\ee
where
\be
R_\sigma(t)
=
[\rho_\sigma^{(0)}]_{x_\sigma^{\rm b}}
=
F_\sigma^{\rm b}(t)
(e^\lambda-1)e^{-\gamma(t_f-t)},
\qquad
x_\sigma^{\rm b}
=
-\sigma v_0(t_f-t),
\qquad
x_\sigma^{\rm f}
=
\sigma v_0t.
\ee
Here $S_\sigma(t)$ denotes the jump of the total density at the corresponding
forward front. Its explicit expression will not be needed below. Both
$R_\sigma$ and $S_\sigma$ are $O(\rho)$.

By contrast, $B^{(0)}$ has delta singularities only at the backward fronts,
\be
B^{(0)}(x,t)
=
B_{\rm reg}^{(0)}(x,t)
+
\sum_{\sigma=\pm}
R_\sigma(t)\,
\delta\left(x-x_\sigma^{\rm b}(t)\right).
\ee
The forward fronts produce finite jumps in $B_{\rm reg}^{(0)}$, but no
delta peaks, since no derivative acts on $F_\pm$ in
$B^{(0)}=F_+\partial_xG_++F_-\partial_xG_-$.

Inserting these decompositions into \eqref{Sigma1_RTP_exact} generates six
classes of terms,
\be
\Sigma_1^{\rm an,0}
=
\Sigma_{{\rm sm},{\rm reg}}
+
\Sigma_{{\rm b},{\rm reg}}
+
\Sigma_{{\rm f},{\rm reg}}
+
\Sigma_{{\rm sm},\delta}
+
\Sigma_{{\rm b},\delta}
+
\Sigma_{{\rm f},\delta},
\ee
according to which part of $\rho_s^{(0)}$ is paired with either
$B_{\rm reg}^{(0)}$ or the backward-front delta peaks.

We first consider $\Sigma_{{\rm b},\delta}$, which contains the leading
contribution. Using $W(\infty)=0$,
\be
\int dy\,
W'(x_\sigma^{\rm b}-y)
\theta\left(y-x_{\sigma'}^{\rm b}\right)
=
W\left(x_\sigma^{\rm b}-x_{\sigma'}^{\rm b}\right),
\ee
and therefore
\bea
\Sigma_{{\rm b},\delta}
&=&
-\sum_{\sigma,\sigma'=\pm}
\int_0^{t_f}dt\,
R_\sigma(t)R_{\sigma'}(t)
W\left(x_\sigma^{\rm b}-x_{\sigma'}^{\rm b}\right)
\nn\\
&=&
-W(0)\int_0^{t_f}dt\,
\left[R_+^2(t)+R_-^2(t)\right]
-
2\int_0^{t_f}dt\,
R_+(t)R_-(t)
W\left(2v_0(t_f-t)\right).
\eea

The first term is the self-interaction of each backward front. Since it is
present throughout the full observation time, it is
$O(\rho^2W(0)t_f)$. Using
$F_\sigma^{\rm b}(t)\equiv
F_\sigma(x_\sigma^{\rm b}(t),t)$ and the definition of $R_\sigma(t)$,
we obtain
\be
\Sigma_{1,{\rm fr}}^{\rm an,0}(\lambda)
=
-W(0)(e^\lambda-1)^2
\int_0^{t_f}dt\,
e^{-2\gamma(t_f-t)}
\left[
\left(F_+^{\rm b}(t)\right)^2
+
\left(F_-^{\rm b}(t)\right)^2
\right].
\ee
The second term describes the interaction between the two distinct
backward fronts. Their separation is $2v_0(t_f-t)$, so for a short-ranged
potential it contributes only during the final interval
$t_f-t=O(a/v_0)$. It is therefore
$O(\rho^2W_0/v_0)$.

The remaining terms are of the same subleading order. Indeed, in the
smooth regions the fields vary over the ballistic scale $v_0t_f$, so that
$B_{\rm reg}^{(0)}=O(\rho/(v_0t_f))$ and
$\partial_x\rho_{\rm sm}^{(0)}=O(\rho/(v_0t_f))$. Hence
$\Sigma_{{\rm sm},{\rm reg}}=O(\rho^2W_0/v_0)$.
Similarly, the interaction field generated by the smooth density is
$O(\rho W_0/(v_0t_f))$, and its evaluation on a backward delta peak gives
$\Sigma_{{\rm sm},\delta}=O(\rho^2W_0/v_0)$.
The terms $\Sigma_{{\rm b},{\rm reg}}$ and
$\Sigma_{{\rm f},{\rm reg}}$ are also
$O(\rho^2W_0/v_0)$. Indeed, the interaction field generated by a density
step is localized over a distance $O(a)$ around the corresponding front,
while $B_{\rm reg}^{(0)}=O(\rho/(v_0t_f))$. Integration over space and time
therefore again gives $O(\rho^2W_0/v_0)$.
Finally, $\Sigma_{{\rm f},\delta}$ couples a forward density front to a
backward delta peak. Such a term contains
$W(x_\sigma^{\rm b}-x_{\sigma'}^{\rm f})$ and is appreciable only when the two
fronts approach within $O(a)$. The pairs with $\sigma'=-\sigma$ cross at
$t=t_f/2$, giving a time window $O(a/v_0)$, while the other pairs remain
separated by a distance of order $v_0t_f$. Thus
$\Sigma_{{\rm f},\delta}=O(\rho^2W_0/v_0)$.

Consequently, for a regular short-ranged potential with
$W_0=O(aW(0))$, one has
\be
\frac{
\Sigma_1^{\rm an,0}
-
\Sigma_{1,{\rm fr}}^{\rm an,0}
}{
\Sigma_{1,{\rm fr}}^{\rm an,0}
}
=
O\left(\frac{a}{v_0t_f}\right),
\ee
for fixed $\tau_f=\gamma t_f=O(1)$. Thus the only contribution of order
$\rho^2W(0)t_f$ is the self-interaction of each backward ballistic front. 
All smooth, forward--backward, and distinct backward-front contributions
are thus smaller.

\section{Ballistic scaling function for weakly interacting RTPs}
\label{app:RTP_ballistic_front}

In this Appendix we derive the scaling function entering
\eqref{Sigma1_RTP_ballistic_scaling} from the front expression
\eqref{Sigma1_RTP_ballistic_front}, and obtain its small- and large-$\tau_f$
asymptotics. The derivation and justification of the ballistic-front contribution are
given in Appendix~\ref{app:RTP_ballistic_front0} and will not be repeated here.

Set $u=t/t_f$, $\tau_f=\gamma t_f$ and $\Delta=t_f(1-u)$. Using the
expressions for $F_\pm^{\rm b}$ given in the main text, it is convenient to
introduce
\be
{\cal R}_{\tau_f}(u)
=
U\left(v_0t_f(1-u),ut_f\right)
-
\frac12 e^{-\tau_fu}
I_0\left(\tau_f\sqrt{2u-1}\right)
\theta\left(u-\frac12\right).
\ee
Then
\be
F_+^{\rm b}
=
\frac12
\left[
\rho_0-
\left(\rho_0-\rho_1e^{-\lambda}\right)
{\cal R}_{\tau_f}(u)
\right],
\qquad
F_-^{\rm b}
=
\frac12
\left[
\rho_1e^{-\lambda}
+
\left(\rho_0-\rho_1e^{-\lambda}\right)
{\cal R}_{\tau_f}(u)
\right].
\ee

The relation \eqref{Upm_realtime} immediately identifies this combination
as
\be
{\cal R}_{\tau_f}(u)
=
U_-\left(v_0t_f(1-u),ut_f\right).
\ee
In particular, ${\cal R}_{\tau_f}(u)=0$ for $u<1/2$, since a particle
starting at the corresponding backward front cannot reach the origin by
time $t$.

We now derive the single-integral representation used in
\eqref{R_RTP_ballistic}. For $u>1/2$, set
$q_0=(1-u)/u$ and $z(q)=\tau_fu\sqrt{1-q^2}$. From
\eqref{Urealtime},
\be
{\cal R}_{\tau_f}(u)
=
\frac{e^{-\tau_fu}}{2}
\Bigg[
1-I_0\left(\tau_f\sqrt{2u-1}\right)
+\tau_fu
\int_{q_0}^{1}dq\,
\left\{
I_0(z(q))
+
\frac{I_1(z(q))}{\sqrt{1-q^2}}
\right\}
\Bigg].
\ee
Since $z(q_0)=\tau_f\sqrt{2u-1}$ and
$\partial_q I_0(z(q))
=-\tau_fu\,q\,I_1(z(q))/\sqrt{1-q^2}$, one has
\be
1-I_0\left(\tau_f\sqrt{2u-1}\right)
=
-\tau_fu
\int_{q_0}^{1}dq\,
\frac{q}{\sqrt{1-q^2}}I_1(z(q)).
\ee
Combining the two terms yields
\be
{\cal R}_{\tau_f}(u)
=
\frac{\tau_fu}{2}e^{-\tau_fu}
\int_{(1-u)/u}^{1}dq\,
\left[
I_0\left(\tau_fu\sqrt{1-q^2}\right)
+
\sqrt{\frac{1-q}{1+q}}\,
I_1\left(\tau_fu\sqrt{1-q^2}\right)
\right],
\ee
which is Eq.~\eqref{R_RTP_ballistic}.

\begin{figure}[t]
\centering
\includegraphics[width=0.6\columnwidth]{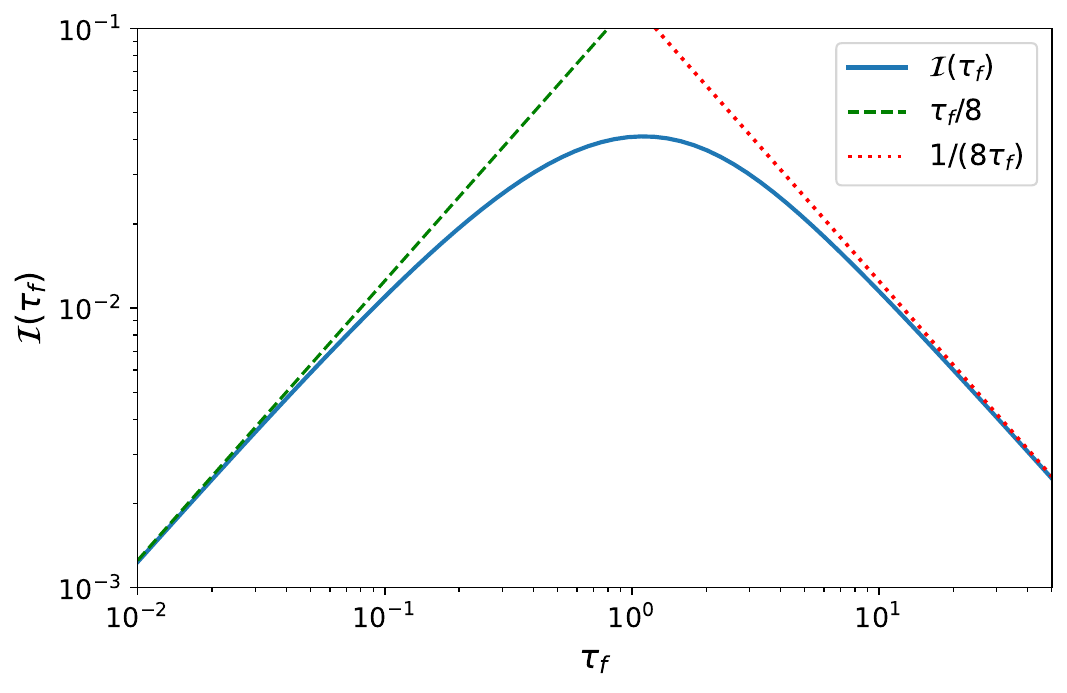}
\caption{
Plot of the scaling function ${\cal I}(\tau_f)$ defined in
Eq.~\eqref{I_RTP_ballistic}, together with its asymptotic behaviors.
For $\tau_f\ll1$, ${\cal I}(\tau_f)\simeq\tau_f/8$, while for
$\tau_f\gg1$, ${\cal I}(\tau_f)\simeq1/(8\tau_f)$.
The numerical evaluation displays the crossover between the strongly
persistent and diffusive regimes.
}
\label{fig:I_RTP_ballistic}
\end{figure}

The combination entering the front contribution simplifies to
\be
\left(F_+^{\rm b}\right)^2
+
\left(F_-^{\rm b}\right)^2
=
\frac14
\Big[
\rho_0^2+\rho_1^2e^{-2\lambda}
-
2\left(\rho_0-\rho_1e^{-\lambda}\right)^2
{\cal R}_{\tau_f}(u)
\left(1-{\cal R}_{\tau_f}(u)\right)
\Big].
\ee
Substituting this result into \eqref{Sigma1_RTP_ballistic_front}, using
$dt=t_fdu$, and performing the elementary integral
$\int_0^1du\,e^{-2\tau_f(1-u)}
=(1-e^{-2\tau_f})/(2\tau_f)$ gives directly
\eqref{Sigma1_RTP_ballistic_scaling}--\eqref{A_RTP_ballistic}, with the
remaining nontrivial integral ${\cal I}(\tau_f)$ defined in
\eqref{I_RTP_ballistic}. This function is plotted in Fig. 
\ref{fig:I_RTP_ballistic}. We now study its asymptotics.

We first consider the strongly persistent limit $\tau_f\ll1$. Expanding
the single-integral representation of ${\cal R}_{\tau_f}$ gives, for
$u>1/2$,
\be
{\cal R}_{\tau_f}(u)
=
\tau_f\left(u-\frac12\right)
+
O(\tau_f^2),
\ee
while it vanishes for $u<1/2$. Hence
\be
{\cal I}(\tau_f)
=
\frac{\tau_f}{8}
+
O(\tau_f^2).
\ee
Together with
$(1-e^{-2\tau_f})/(8\tau_f)
=1/4-\tau_f/4+O(\tau_f^2)$, this gives
\eqref{A_RTP_ballistic_smalltau} and therefore
\eqref{Sigma1_RTP_ballistic_tau0}.

We next consider $\tau_f\gg1$. The exponential factor in
\eqref{I_RTP_ballistic} restricts the integral to a boundary layer near
$u=1$. Writing $s=\tau_f(1-u)$, with $s=O(1)$, the orientation-resolved
crossing probability satisfies
\be
{\cal R}_{\tau_f}\left(1-\frac{s}{\tau_f}\right)
=
\frac12+O(\tau_f^{-1/2}).
\ee
This follows, for example, from the diffusive form
$U(x,t)\simeq \frac12{\rm erfc}(x/\sqrt{4D_at})$, together with
\eqref{Upm_realtime}: in the present boundary layer
$x/\sqrt{4D_at}=s/\sqrt{2\tau_f}$, while the orientation-dependent
difference $U_--U$ is $O(\tau_f^{-1/2})$. Therefore
${\cal R}_{\tau_f}(1-{\cal R}_{\tau_f})
=1/4+O(\tau_f^{-1})$, and
\be
{\cal I}(\tau_f)
=
\frac{1}{8\tau_f}
+
O(\tau_f^{-2}).
\ee
Using this result in \eqref{A_RTP_ballistic} gives
\be
{\cal A}(\tau_f;\lambda,\rho_0,\rho_1)
=
\frac{
\left(\rho_0+\rho_1e^{-\lambda}\right)^2
}{16\tau_f}
+
O(\tau_f^{-2}),
\ee
which is \eqref{A_RTP_ballistic_largetau}.

The hierarchy of the terms neglected in retaining only the backward-front
self-interaction was established in Appendix~\ref{app:RTP_ballistic_front0}.
For fixed $\tau_f=O(1)$, all these contributions are smaller than
$\Sigma_{1,{\rm fr}}^{\rm an,0}$ by $O(a/(v_0t_f))$ for a regular
short-range potential with $W_0=O(aW(0))$.

\section{Brownian MFT: saddle-point equations and boundary conditions}
\label{app:Brownian_boundary_conditions}

In this Appendix we recall the MFT for Brownian particles, starting from
the density-current formulation. We derive the bulk and boundary
saddle-point equations on the interval $[0,L]$. The latter are of particular
interest for the problem studied in Section~\ref{sec:MFT_reservoirs}, where
reservoirs impose fixed densities at the two boundaries.

The density and current obey the continuity equation
$\partial_t\rho+\partial_xJ=0$. For independent Brownian particles, the
deterministic part of the current is $\bar J=-D\partial_x\rho$, and the
dynamical large-deviation cost of a history $(\rho,J)$ is
\be
{\cal S}[\rho,J]
=
\int_0^{t_f}dt\int_0^Ldx\,
\frac{\left(J+D\partial_x\rho\right)^2}{4D\rho},
\ee
subject to the continuity equation. Introducing the auxiliary field $p$ as
a Lagrange multiplier gives
\be \label{S_Brownian_boundary_app}
{\cal S}[\rho,J,p]
=
\int_0^{t_f}dt\int_0^Ldx\,
\left[
\frac{\left(J+D\partial_x\rho\right)^2}{4D\rho}
+
p\left(\partial_t\rho+\partial_xJ\right)
\right] \;,
\ee
where the three fields $\rho$, $J$ and $p$ are now varied independently.

Let us first write the total variation of
\eqref{S_Brownian_boundary_app}. Denoting
$A(x,t)=J(x,t)+D\partial_x\rho(x,t)$, integration by parts in space and
time gives
\bea \label{deltaS_Brownian_full}
\delta{\cal S}
&=&
\int_0^{t_f}dt\int_0^Ldx\,
\Bigg\{
\left[
-\partial_t p
-\partial_x\left(\frac{A}{2\rho}\right)
-\frac{A^2}{4D\rho^2}
\right]\delta\rho
+
\left[
\frac{A}{2D\rho}
-\partial_xp
\right]\delta J
+
\left[
\partial_t\rho+\partial_xJ
\right]\delta p
\Bigg\}
\nn\\
&&
+
\int_0^Ldx\,
\left[p\,\delta\rho\right]_{t=0}^{t=t_f}
+
\int_0^{t_f}dt\,
\left[
\frac{A}{2\rho}\,\delta\rho
+
p\,\delta J
\right]_{x=0}^{x=L} \;.
\eea
The first term is the bulk variation, while the last two terms are,
respectively, the temporal and spatial boundary contributions. On the
infinite line only the spatial contribution disappears under the usual
decay conditions; the temporal endpoint term remains and determines the
initial and final saddle-point conditions.

For the current generating function considered in
Section~\ref{sec:MFT_reservoirs}, the relevant action is the tilted action
$\widetilde{\cal S}={\cal S}-\lambda\mathcal{Q}_{t_f}$, whose total
variation is therefore
\be \label{deltaStilde_Brownian_full}
\delta\widetilde{\cal S}
=
\delta{\cal S}
-
\frac{\lambda}{L}
\int_0^{t_f}dt\int_0^Ldx\,\delta J \,.
\ee
Thus the current tilt modifies only the bulk coefficient of $\delta J$;
in particular it does not modify the spatial boundary variation.
In terms of the shifted auxiliary field
$\hat p(x,t)
=
p(x,t)+\frac{\lambda x}{L}$, 
the bulk variation with respect to $J$ gives
$
J
=
-D\partial_x\rho
+
2D\rho\,\partial_x\hat p$,
which is Eq.~\eqref{saddle_J_Br_reservoirs} of the main text. 
Variation with respect to $p$ gives the continuity equation. Combining it
with the saddle-point current above gives equation
\eqref{SPeq_Br_reservoirs1}. Substituting the same saddle-point current in
the coefficient of $\delta\rho$, and noting that
$\partial_t p=\partial_t\hat p$, gives equation
\eqref{SPeq_Br_reservoirs2}.

We now focus on the spatial boundary variation. The reservoirs considered
in Section~\ref{sec:MFT_reservoirs} impose Dirichlet boundary conditions
on the density,
\be
\rho(0,t)=\rho_a
\quad,\quad
\rho(L,t)=\rho_b \;,
\ee
so that the allowed variations satisfy
\be
\delta\rho(0,t)=\delta\rho(L,t)=0 \;.
\ee
The contribution of the spatial boundary conditions in \eqref{deltaStilde_Brownian_full} therefore
reduces to
\be
\delta\widetilde{\cal S}_{\rm bdry}
=
\int_0^{t_f}dt\,
\left[
p(L,t)\delta J(L,t)
-
p(0,t)\delta J(0,t)
\right] \;.
\ee
The boundary currents are not fixed by the reservoirs and can
therefore be varied independently. Stationarity then requires
\be \label{p_boundary_Brownian_app}
p(0,t)=p(L,t)=0 \quad \Rightarrow \quad \hat p(0,t)=0
\quad,\quad
\hat p(L,t)=\lambda \;.
\ee
which reproduces the boundary conditions \eqref{BCphat_br} of the main
text. With these conditions the spatial surface
term generated by integrating $p\,\partial_xJ$ by parts vanishes, and the
bulk current can then be eliminated as done in the main text.
\\

\noindent{\bf Stochastic reservoirs}. Ref.~\cite{SuzukiSasamoto2026}
presents a more general treatment in which the exchange of particles with
the reservoirs is itself described as a stochastic process. In that case,
in addition to the tilted bulk action considered above, one includes a
reservoir contribution,
\be
\widetilde{\cal S}_{\rm tot}
=
\widetilde{\cal S}_{\rm bulk}
+
\widetilde{\cal S}_{\rm res},
\ee
with
\bea
\widetilde{\cal S}_{\rm res}
=
-\int_0^{t_f}dt\,\Big\{&
C_L^+(\rho(0),\rho_a)
\left[e^{\hat p(0)}-1\right]
+
C_L^-(\rho(0),\rho_a)
\left[e^{-\hat p(0)}-1\right]
\nn\\
&+
C_R^+(\rho(L),\rho_b)
\left[e^{\hat p(L)-\lambda}-1\right]
+
C_R^-(\rho(L),\rho_b)
\left[e^{-\hat p(L)+\lambda}-1\right]
\Big\},
\eea
where $C_{L,R}^+$ and $C_{L,R}^-$ are the macroscopic injection and
removal rates. The factors of $\lambda$ at the right boundary result from
our convention $\hat p=p+\lambda x/L$ for the spatially integrated current;
this is equivalent, up to a constant shift of the conjugate field, to the
convention of Ref.~\cite{SuzukiSasamoto2026}, where the current through the
left reservoir is counted.

It is important to distinguish $\widetilde{\cal S}_{\rm res}$ from the
surface term appearing in the variation of the bulk action. The former is a
genuine physical large-deviation cost for particle injection and removal,
whereas the latter is only a boundary term generated by the variation of
the bulk action. At finite reservoir coupling the boundary densities are
therefore allowed to fluctuate, and the boundary conditions follow from
the combined variation of
$\widetilde{\cal S}_{\rm bulk}+\widetilde{\cal S}_{\rm res}$.

Our prescription corresponds to the fast-reservoir limit of this more
general theory. More precisely, suppose that the reservoir injection and
removal rates are chosen so that, in the absence of a current bias, they
drive the boundary densities toward $\rho_a$ at the left boundary and
$\rho_b$ at the right boundary. Writing
$C_{L,R}^{\pm}=\Gamma\widetilde C_{L,R}^{\pm}$ and taking
$\Gamma\to\infty$, the boundary degrees of freedom then relax
instantaneously to these reservoir-imposed densities, so that
$\rho(0,t)=\rho_a$ and $\rho(L,t)=\rho_b$. Hence
$\delta\rho(0,t)=\delta\rho(L,t)=0$, and the remaining spatial variation of
the bulk action gives $p(0,t)=p(L,t)=0$, or equivalently
$\hat p(0,t)=0$ and $\hat p(L,t)=\lambda$, precisely the boundary
conditions used above.
\\

\noindent{\bf Adding interactions.} Let us finally verify that a pair interaction does not modify the
auxiliary-field boundary conditions derived above. For the weakly
interacting Brownian particles considered in
Section~\ref{sec:Brownian_interacting_reservoirs}, the deterministic
current is given by \eqref{Jbar_Brownian_interacting_res} and the
density-current action by \eqref{S_Brownian_interacting_res}. The variation of the interaction force field is
\be
\delta{\cal H}[\rho](x)
=
-\int_0^Ldy\,W'(x-y)\delta\rho(y).
\ee
This term does not contain a derivative of $\delta\rho$. Hence the only
spatial derivative of $\delta\rho$ in the variation of
\eqref{S_Brownian_interacting_res} comes from the diffusive contribution
$D\partial_x\rho$. After integration by parts, the spatial boundary term is
\be \label{boundary_variation_Brownian_interacting}
\delta{\cal S}_{\varepsilon,\mathrm{bdry}}
=
\int_0^{t_f}dt\,
\left[
\frac{
J+D\partial_x\rho-\varepsilon\rho{\cal H}[\rho]
}{2\rho}\,
\delta\rho
+
p\,\delta J
\right]_{x=0}^{x=L}.
\ee
The reservoirs impose the Dirichlet boundary conditions
$\rho(0,t)=\rho(L,t)=\rho$, so that
$\delta\rho(0,t)=\delta\rho(L,t)=0$. Since the instantaneous boundary
currents are not fixed, $\delta J(0,t)$ and $\delta J(L,t)$ are arbitrary.
Equation~\eqref{boundary_variation_Brownian_interacting} therefore implies
$p(0,t)=p(L,t)=0$, or equivalently
$\hat p(0,t)=0$ and $\hat p(L,t)=\lambda$. This reproduces
\eqref{p_BC_Brownian_interacting} and shows that the interaction does not
modify the auxiliary-field boundary condition associated with ideal
fixed-density reservoirs.

\section{Boundary conditions for the auxiliary fields in the RTP MFT}
\label{app:RTP_boundary_conditions}

In this Appendix we derive the boundary conditions obeyed by the auxiliary
fields for free RTPs on the interval $[0,L]$ in the absence of additional
thermal diffusion, $D=0$. This case differs qualitatively from the Brownian
case discussed in Appendix~\ref{app:Brownian_boundary_conditions}, since
the translational currents are fixed by the densities,
$J_+=v_0\rho_+$ and $J_-=-v_0\rho_-$, and are therefore not independent
fluctuating variables. The independent dynamical variables are thus
$\rho_\pm$, $p_\pm$ and the two tumbling currents $K_\pm$.
The continuity equations read
\bea
\partial_t\rho_+
+v_0\partial_x\rho_+
+K_+-K_-
&=&0,
\\
\partial_t\rho_-
-v_0\partial_x\rho_-
+K_--K_+
&=&0.
\eea
The tumbling events are Poisson processes with local large-deviation costs
\be
{\cal L}_{K_\pm}
=
K_\pm\ln\left(\frac{K_\pm}{\gamma\rho_\pm}\right)
-K_\pm+\gamma\rho_\pm \;.
\ee
The corresponding untilted MSR action is
\be \label{S_RTP_boundary_app}
{\cal S}
=
\int_0^{t_f}dt\int_0^Ldx\,
\Big[
{\cal L}_{K_+}+{\cal L}_{K_-}
+p_+
\left(
\partial_t\rho_+
+v_0\partial_x\rho_+
+K_+-K_-
\right)
+p_-
\left(
\partial_t\rho_-
-v_0\partial_x\rho_-
+K_--K_+
\right)
\Big] \;.
\ee

Varying all the independent fields and integrating by parts in space and time gives
\bea \label{deltaS_RTP_full}
\delta{\cal S}
&=&
\int_0^{t_f}dt\int_0^Ldx\,
\Bigg\{
\left[
\gamma-\frac{K_+}{\rho_+}
-\partial_t p_+
-v_0\partial_xp_+
\right]\delta\rho_+
+
\left[
\gamma-\frac{K_-}{\rho_-}
-\partial_t p_-
+v_0\partial_xp_-
\right]\delta\rho_-
\nn\\
&&
+
\left[
\ln\left(\frac{K_+}{\gamma\rho_+}\right)
+p_+-p_-
\right]\delta K_+
+
\left[
\ln\left(\frac{K_-}{\gamma\rho_-}\right)
+p_--p_+
\right]\delta K_-
\nn\\
&&
+
\left(
\partial_t\rho_+
+v_0\partial_x\rho_+
+K_+-K_-
\right)\delta p_+
+
\left(
\partial_t\rho_-
-v_0\partial_x\rho_-
+K_--K_+
\right)\delta p_-
\Bigg\}
\nn\\
&&
+
\int_0^Ldx\,
\left[
p_+\delta\rho_+
+p_-\delta\rho_-
\right]_{t=0}^{t=t_f}
+
v_0\int_0^{t_f}dt\,
\left[
p_+\delta\rho_+
-p_-\delta\rho_-
\right]_{x=0}^{x=L} \;.
\eea

For the current generating function considered in
Section~\ref{sec:MFT_reservoirs}, the tilted action is
$\widetilde{\cal S}={\cal S}-\lambda{\cal Q}_{t_f}(L)$, with
${\cal Q}_{t_f}(L)
=
\frac{v_0}{L}
\int_0^{t_f}dt\int_0^Ldx\,
(\rho_+-\rho_-)$. 
Its variation is therefore
\be \label{deltaStilde_RTP_full}
\delta\widetilde{\cal S}
=
\delta{\cal S}
-
\frac{\lambda v_0}{L}
\int_0^{t_f}dt\int_0^Ldx\,
(\delta\rho_+-\delta\rho_-) \;.
\ee
Thus the tilt modifies only the bulk coefficients of
$\delta\rho_\pm$ and leaves the spatial boundary contribution in
\eqref{deltaS_RTP_full} unchanged.

Since the tilt does not depend on $K_\pm$, their saddle-point values remain
\be
K_+
=
\gamma\rho_+e^{-(p_+-p_-)},
\qquad
K_-
=
\gamma\rho_-e^{p_+-p_-} \;.
\ee
The variations with respect to $\rho_\pm$ then give, after introducing the shifted fields
$\hat p_\pm(x,t)
=
p_\pm(x,t)+\frac{\lambda x}{L}$,
\bea
\partial_t\hat p_+
&=&
-v_0\partial_x\hat p_+
-\gamma\left(e^{-(\hat p_+-\hat p_-)}-1\right),
\\
\partial_t\hat p_-
&=&
v_0\partial_x\hat p_-
-\gamma\left(e^{\hat p_+-\hat p_-}-1\right),
\eea
while the variations with respect to $p_\pm$ give
\bea
\partial_t\rho_+
&=&
-v_0\partial_x\rho_+
-\gamma\rho_+e^{-(\hat p_+-\hat p_-)}
+\gamma\rho_-e^{\hat p_+-\hat p_-},
\\
\partial_t\rho_-
&=&
v_0\partial_x\rho_-
-\gamma\rho_-e^{\hat p_+-\hat p_-}
+\gamma\rho_+e^{-(\hat p_+-\hat p_-)}.
\eea
Setting the time derivatives to zero reproduces the stationary bulk
equations \eqref{SPeq_RTPfree_reservoirs1}--\eqref{SPeq_RTPfree_reservoirs4}
used in Section~\ref{sec:MFT_reservoirs}.

We now turn to the spatial boundary variation. Since the additional term
$-\lambda{\cal Q}_{t_f}(L)$ in the tilted action contains no spatial
derivatives of $\rho_\pm$, its variation generates no spatial boundary
term. Therefore the spatial boundary contribution is the same as in
\eqref{deltaS_RTP_full},
\bea \label{deltaS_RTP_boundary}
\delta\widetilde{\cal S}_{\rm bdry}
&=&
v_0\int_0^{t_f}dt\,
\Big[
p_+(L,t)\delta\rho_+(L,t)
-p_-(L,t)\delta\rho_-(L,t)
\nn\\
&&\hspace{2.5cm}
-p_+(0,t)\delta\rho_+(0,t)
+p_-(0,t)\delta\rho_-(0,t)
\Big].
\eea
For the reservoir prescription considered in
Section~\ref{sec:MFT_reservoirs}, the incoming populations are fixed,
\be \label{RTP_incoming_BC_app}
\rho_+(0,t)=\frac{\rho_a}{2},
\qquad
\rho_-(L,t)=\frac{\rho_b}{2} \;,
\ee
and hence
$\delta\rho_+(0,t)=0$ and $\delta\rho_-(L,t)=0$.
By contrast, the outgoing populations $\rho_-(0,t)$ and $\rho_+(L,t)$
are not fixed, so their variations are arbitrary. Stationarity of the action therefore
requires
\be \label{p_RTP_boundary_app}
p_-(0,t)=0,
\qquad
p_+(L,t)=0  \quad \Rightarrow \quad \hat p_-(0,t)=0,
\qquad
\hat p_+(L,t)=\lambda \, ,
\ee
as displayed in the main text.

\section{Stationary measure for weakly interacting Brownian particles}
\label{app:Brownian_interacting_stationary_measure}

In this Appendix we discuss the stationary density measure for the weakly
interacting Brownian system considered in
Section~\ref{sec:Brownian_interacting_reservoirs}. This is needed to
determine whether the interaction-induced modification of the initial
measure contributes to the long-time current large deviations.

For the equilibrium system at uniform density $\rho$, the interaction also
modifies the stationary density measure. The latter is the Gibbs measure
associated with the pair potential $\varepsilon W$, and the corresponding
stationary density large-deviation functional (also called the
quasi-potential) is the equilibrium free-energy cost relative to the
homogeneous profile. To first order in $\varepsilon$, it can be written as
\bea
{\cal V}_\varepsilon[\rho]
&=&
\int_0^Ldx\,
\left[
\rho(x)\ln\frac{\rho(x)}{\rho}
-\rho(x)+\rho
\right] +
\frac{\varepsilon}{2D}
\int_0^Ldx\int_0^Ldy\,
[\rho(x)-\rho]\,
W(x-y)\,
[\rho(y)-\rho]
+
O(\varepsilon^2) \;.
\eea
In the short-range hydrodynamic regime, where the density varies slowly
over the range of $W$, the stationary density large-deviation functional
reduces to the local form
\be \label{V_Brownian_interacting_res}
{\cal V}_\varepsilon[\rho]
=
\int_0^Ldx\,
\left[
\rho(x)\ln\frac{\rho(x)}{\rho}
-\rho(x)+\rho
\right]
+
\varepsilon\frac{W_0}{2D}
\int_0^Ldx\,[\rho(x)-\rho]^2
+
O(\varepsilon^2) \;.
\ee
where
$W_0=\int_{-\infty}^{+\infty}dr\,W(r)$.

This contribution must in principle be included in the path integral,
\be
\left\langle e^{\lambda{\cal Q}_{t_f}(L)}\right\rangle
\simeq
\int{\cal D}\rho\,{\cal D}J\,{\cal D}p\,
\exp\left[
-{\cal V}_\varepsilon[\rho(x,0)]
+\lambda{\cal Q}_{t_f}(L)
-{\cal S}_\varepsilon
\right].
\ee
However, ${\cal V}_\varepsilon=O(L)$ is an initial-time contribution,
whereas the action of a stationary optimal profile is proportional to
$t_f$. Hence, in the long-time stationary limit,
$\lim_{t_f\to+\infty}
{\cal V}_\varepsilon[\rho(x,0)]/t_f=0$.
The change of stationary measure therefore does not contribute explicitly
to the scaled cumulant generating function considered in
Section~\ref{sec:Brownian_interacting_reservoirs}.

\end{document}